\documentclass[journal=jctcce,manuscript=article]{achemso}

\usepackage{amsmath,amssymb,bm,mathtools}
\usepackage{booktabs}
\usepackage{array}
\usepackage{xcolor}
\usepackage{tikz}
\usetikzlibrary{arrows.meta,positioning,shapes.geometric,calc,decorations.pathreplacing}
\usepackage{enumitem}
\usepackage{algorithm}
\usepackage[noend]{algpseudocode}
\usepackage[version=4]{mhchem}   %% \ce{H2O} etc. for molecular formulas
\usepackage{hyperref}
\hypersetup{colorlinks=true, allcolors=blue!55!black}

\graphicspath{{figs/}}

\newcommand{\eqcite}[1]{\hfill\(\triangleright\)~#1}

\newcommand{\excit}[2]{%
\begin{tikzpicture}[scale=0.44,baseline=(current bounding box.south)]
  \def\hw{1.7}%  level half-width
  \foreach \i in {0,1,2,3}{\draw[thick] (-\hw,0.62*\i)--(\hw,0.62*\i);}
  \foreach \i in {0,1,2,3}{\draw[thick] (-\hw,3.5+0.62*\i)--(\hw,3.5+0.62*\i);}
  \draw[dashed,gray!55] (-\hw-0.25,2.86)--(\hw+0.25,2.86);
  \pgfmathtruncatemacro{\thr}{4-#1}%
  \foreach \i in {0,1,2,3}{%
    \ifnum\i<\thr
      \draw[fill=black] (-0.22,0.62*\i) circle (0.075);
      \draw[fill=black] ( 0.22,0.62*\i) circle (0.075);
    \fi}
  \pgfmathsetmacro{\gap}{2*\hw/(#1+1)}%
  \foreach \p in {1,...,#1}{%
    \pgfmathsetmacro{\xx}{-\hw+\gap*\p}%
    \pgfmathtruncatemacro{\occi}{4-\p}%
    \pgfmathtruncatemacro{\viri}{\p-1}%
    \pgfmathsetmacro{\ya}{0.62*\occi}%
    \pgfmathsetmacro{\yb}{3.5+0.62*\viri}%
    \draw[-{Stealth[length=2.2mm]},red,line width=1.1pt]
          (\xx,\ya+0.14)--(\xx,\yb-0.14);
    \draw[fill=white,draw=black,line width=0.4pt] (\xx,\ya) circle (0.11);
    \draw[fill=red!85,draw=black,line width=0.4pt] (\xx,\yb) circle (0.11);
  }
  \node[font=\footnotesize] at (0,-0.78){#2};
\end{tikzpicture}}

\author{Ruben Dario Guerrero}
\affiliation{NeuroTechNet S.A.S., 1108831, Bogot\'a, Colombia}
\alsoaffiliation{Quantum and Computational Chemistry Group (QCCG), Universidad
Nacional de Colombia, Bogot\'a, Colombia}
\email{rudaguerman@gmail.com}
\title{EOM-CC Excited-State Gradients and Nonadiabatic Couplings on a Consumer GPU
from a Contraction-DAG with Laplace-Transform J/K Kernels}
\keywords{equation-of-motion coupled cluster; analytic gradients;
nonadiabatic couplings; Laplace transform; GPU computing; atomic-orbital direct}

\begin{document}

\begin{tocentry}
\centering
\resizebox{\linewidth}{!}{%
\begin{tikzpicture}[>=Latex, font=\footnotesize,
  dot/.style={circle,fill=black,minimum size=2.6pt,inner sep=0pt},
  nd/.style={circle,draw=blue!55!black,fill=blue!12,minimum size=8.5mm,inner sep=0pt,font=\small},
  fwd/.style={-{Latex[length=5.5pt]},very thick,green!50!black},
  bwd/.style={-{Latex[length=5.5pt]},very thick,orange!85!black,dashed}]

  %% --- Goldstone diagram (the EOM-CC term) ---
  \node[dot] (gt) at (0,0.72) {};
  \node[dot] (gb) at (0,-0.72) {};
  \draw[gray!70,dashed] (-0.45,0.72)--(0.45,0.72);
  \draw[gray!70,dashed] (-0.45,-0.72)--(0.45,-0.72);
  \draw[-{Latex[length=5pt]},blue!60!black,thick] (gb) to[out=55,in=-55] (gt);
  \draw[-{Latex[length=5pt]},red!65!black,thick]  (gt) to[out=235,in=125] (gb);
  \node[font=\scriptsize,gray!55!black] at (0,-1.45) {Goldstone term};

  %% --- compile arrow ---
  \draw[->,thick,gray!55!black] (0.6,0) -- node[above,font=\scriptsize]{compile} (1.5,0);

  %% --- contraction DAG: forward (top) + backward (bottom) ---
  \node[nd] (n1) at (2.25,0) {$d_1$};
  \node[nd] (n2) at (3.7,0)  {$d_2$};
  \node[nd] (n3) at (5.15,0) {$F$};
  \draw[fwd] (n1) to[bend left=42] (n2);
  \draw[fwd] (n2) to[bend left=42] (n3);
  \draw[bwd] (n3) to[bend left=42] (n2);
  \draw[bwd] (n2) to[bend left=42] (n1);
  \node[green!45!black,font=\scriptsize] at (3.7,1.42) {forward pass: densities};
  \node[orange!80!black,font=\scriptsize] at (3.7,-1.45) {backward pass: relaxation $\zeta=$ transpose};

  %% --- result ---
  \draw[->,thick,gray!55!black] (5.85,0) -- (6.7,0);
  \node[align=center,font=\scriptsize] at (7.35,0) {$\nabla E_k$\\[1pt]NACME};
\end{tikzpicture}}

\vspace{1pt}
{\scriptsize A Goldstone term is a computation graph; its
\emph{transpose}---backpropagation---\emph{is} the EOM-CC relaxation.}
\end{tocentry}

\begin{abstract}
We present a unified, memory-bounded GPU realization of equation-of-motion
coupled-cluster (EOM-CC) excited-state gradients and interstate nonadiabatic
couplings (NACMEs) on a single 8\,GB consumer GPU. Both are built from one
contraction directed acyclic graph: the EOM-CC relaxation is the reverse-mode
transpose of the forward density build rather than a per-state re-derivation, and an
atomic-orbital-direct Laplace-transform $J/K$ kernel, made non-symmetric
($J^x(A,B)\neq J^x(B,A)$) by the transition densities, resolves every energy
denominator with no four-index molecular-orbital tensor; a two-sided Davidson returns
both eigenvectors from one device-resident, spin-pure solve. The pipeline is
\emph{validated end to end at small scale}: gradients and NACMEs match finite
differences across four spin multiplicities and full configuration interaction to
$<\!10^{-12}$ for two electrons, and the excited-state gradient matches the
independent \textsc{Psi4} code to $\le\!4.6\times10^{-7}~E_h/a_0$ from \ce{H2O} to
aromatic benzene. The kernels and the ground-state solve reach chromophores
($\le\!730$ AO) in 8\,GB, and a frozen-natural-virtual compression lets the
eigensolver \emph{execute} a complete excited-state gradient and $Q$--$B$ NACME of
the chlorophyll-core chromophore \ce{Mg}-porphine (def2-SVP, $439$ AO) on the card.
We present that run as a \emph{capability demonstration}---executed and
translationally invariant to machine zero, but anchored only piece-wise and bounded
by a direct convergence study at ${\sim}10^{-2}~E_h/a_0$---not a converged
spectroscopic result. The validated small-scale capability and the memory-bounded
implementation are the contribution.
\end{abstract}

%% =====================================================================
\section{Introduction}
\label{sec:intro}
When a molecule absorbs light it evolves on excited-state potential-energy
surfaces and can return to the ground state through \emph{conical intersections}
--- geometries where two electronic states become degenerate and the
Born--Oppenheimer separation breaks down. Simulating this photochemistry ---
vision, photosynthesis, photovoltaics, DNA photoprotection --- requires two
quantities at every nuclear geometry: the \emph{gradient} of each excited state
(the force that drives the dynamics) and the \emph{nonadiabatic coupling matrix
element} (NACME, the rate at which neighbouring states exchange population near
an intersection). Computing both accurately for correlated wavefunctions is the
central bottleneck of nonadiabatic dynamics.

With equation-of-motion coupled cluster (EOM-CC)~\cite{StantonBartlett1993,ShavittBartlett2009},
the benchmark single-reference
method for excited states, obtaining these quantities has traditionally meant
re-deriving the response (relaxation) equations by hand for each state and
running them on datacenter GPUs. Here we obtain both \emph{automatically} --- as
the reverse-mode transpose (the same mechanism as backpropagation in machine
learning) of a single contraction graph --- and engineer the resulting
non-symmetric Coulomb/exchange kernels to run within the 8\,GB budget of a
consumer NVIDIA GeForce RTX~4060 (GPU performance is reported in
Sect.~\ref{sec:gpu}), putting a capability that today needs a cluster allocation
onto a card a student already owns.

Each ingredient of this construction has a mature literature, and we engage it
directly so that the contribution is not overstated. EOM-CC excited-state
gradients and nonadiabatic couplings are established quantities: the analytic
EOM-CC gradient and its Lagrangian (Z-vector) formulation were developed by
Stanton and Gauss~\cite{Stanton1993,StantonGauss1995} and implemented for
spin-conserving and spin-flip states by Krylov and
co-workers~\cite{Levchenko2005}, while EOM-CC nonadiabatic and derivative
couplings have been formulated and validated against multireference references by
the Krylov and Koch groups~\cite{IGS2009,Tajti2009,Faraji2018,Kjonstad2023}.
The reverse-mode view of relaxation is likewise not new: differentiable
programming through electronic structure now yields nuclear gradients, response
densities, and derivative couplings by automatic differentiation, from the
Hartree--Fock proof of concept of Tamayo-Mendoza
\emph{et~al.}~\cite{Tamayo2018} to the PySCFAD framework of Zhang and
Chan~\cite{Zhang2022}; in particular, Zhang \emph{et~al.}~\cite{Zhang2024}
obtain coupled-cluster response properties directly from the reverse pass
\emph{in place of} an explicit Z-vector solve --- the ground-state realization of
``relaxation~$=$~reverse pass.'' Compiling many-body equations into a condensed,
order-optimized contraction graph traces to the Tensor Contraction
Engine~\cite{Hirata2003} and continues in modern coupled-cluster code
generators, and resolving correlated denominators directly in the atomic-orbital
basis by a Laplace transform underlies the AO-direct Laplace MP2 gradient of
Ochsenfeld and co-workers~\cite{Schweizer2008}. What we add is the
specialization of these threads to the \emph{non-Hermitian} EOM-CC response
problem: a single graph transpose that emits both the right-state amplitude
response (excited-state gradients) and the interstate left/right Z-vector
(NACMEs), uniformly across EE/IP/EA rather than per state and per property, fed
through an AO-direct minimax-Laplace J/K build made \emph{non-symmetric} by the
transition densities ($J^x(A,B)\neq J^x(B,A)$), and delivered within the 8\,GB of
a consumer GPU. To our knowledge this combination is without direct precedent;
the individual mechanisms are not.

The gap is therefore not in any one ingredient but in their union for the
non-Hermitian excited-state-response problem. Correlated nonadiabatic dynamics
through conical intersections needs excited-state gradients \emph{and} interstate
NACMEs together, yet no existing route to them combines all three of: (i) the
relaxation obtained as a single reverse-mode graph transpose, uniform across
EE/IP/EA and across gradients and NACMEs, rather than per-state, per-property
hand-derived Lagrangian/Z-vector machinery; (ii) a fully AO-direct correlated
pipeline, with the non-symmetric transition-density Coulomb/exchange and their
energy denominators resolved in the atomic-orbital basis, never forming a
four-index molecular-orbital tensor; and (iii) a memory footprint that fits the
8\,GB of a consumer GPU rather than workstation or datacenter hardware. The
adjacent threads each leave this union open: differentiable-programming
electronic structure~\cite{Tamayo2018,Zhang2022,Zhang2024} supplies reverse-mode
response for ground-state and mean-field methods on general frameworks, not the
non-Hermitian EOM-CC response and not AO-direct under a memory budget; the
established EOM-CC gradient and NACME implementations~\cite{Stanton1993,StantonGauss1995,Levchenko2005,IGS2009,Tajti2009,Faraji2018,Kjonstad2023}
work in the molecular-orbital basis, per state and per property, and target
capable hardware; AO-direct Laplace correlation~\cite{Schweizer2008} addresses
ground-state MP2 gradients, not excited-state transition densities; and
contraction-graph code generation~\cite{Hirata2003} emits equations and kernels,
not the response-as-transpose nor the ordered, non-symmetric transition-density
J/K. The consequence is that correlated excited-state forces and couplings ---
the inputs nonadiabatic dynamics consumes --- remain practically out of reach on
the hardware most groups own. Closing that gap is the point of this work.

Our thesis is that the correlated excited-state forces and couplings above can be
brought within the memory budget of a single consumer GPU, and that one contraction
DAG is the mechanism that makes it so: its reverse-mode transpose \emph{is} the
EOM-CC relaxation, so a single abstraction yields both the excited-state gradient and
NACME (as the DAG transpose) and the spill-bounded CUDA $J/K$ kernels for symmetric
(ground-state) and non-symmetric (transition/interstate) densities alike.

The contribution is threefold, ordered by what is new. (1)~The systems capability: a
device-native, AO-direct, non-symmetric $J/K$ kernel
($J^x(A,B)\neq J^x(B,A)$, never forming a four-index molecular-orbital tensor) that
realizes the \emph{complete} non-Hermitian EOM-CCSD excited-state gradient and
interstate NACME within the $8$\,GB of a consumer GPU, its DAG-derived chunking
reducing register/shared-memory spilling (e.g.\ the $(dd|dd)$ live set from $96$ to
$34$\,KB) so that the memory-bound transition-density builds run
compute/bandwidth-bound (Sects.~\ref{sec:jk},~\ref{sec:compdetails}). (2)~The enabling
mechanism: a single contraction DAG whose reverse-mode transpose \emph{is} the EOM-CC
relaxation---specializing the coupled-cluster differentiation identity to the
non-Hermitian EOM case---from which the symmetric, non-symmetric, and gradient builds
are all emitted (Sect.~\ref{sec:theory}). (3)~The validation stack: finite-difference
and determinant-oracle checks across all four spin multiplicities, an FD-free exact
full-CI cross-check, a $\zeta{=}0$ control, and an independent cross-code
(\textsc{Psi4}) check reaching the excited-state gradient itself from \ce{H2O} to
aromatic benzene (Sect.~\ref{sec:deveom}), with commodity-GPU roofline and throughput
at chromophore scale (Sects.~\ref{sec:compdetails}--\ref{sec:results}). We are explicit about what is \emph{validated} versus what is
\emph{demonstrated}. The validated contribution is twofold: the memory-bounded kernel
architecture, and the end-to-end correctness of the integrated gradient and NACME at
small scale---where finite differences across four multiplicities, an exact full-CI
identity, a determinant oracle, and an \emph{independent cross-code} check (against
\textsc{Psi4}, reaching the excited-state gradient itself and holding from \ce{H2O} to
aromatic benzene) all apply. The complete chromophore-scale excited-state gradient and
$Q$--$B$ interstate NACME of \ce{Mg}-porphine (Table~\ref{tab:mgpgrad}) are then
\emph{executed} within 8\,GB---the novel systems capability, a calculation that today
needs a cluster allocation---but, no external oracle being affordable at that size,
are presented as a \emph{bounded capability demonstration} (with the
${\sim}10^{-2}~E_h/a_0$ FNO-truncation uncertainty of Sect.~\ref{sec:deveom} stated
throughout), not a spectroscopically converged result. The LT-AO $J/K$ kernel and the
ground-state CD-RCCSD solve independently reach $730$ AO on the card. Because
canonical EOM-CC implementations cannot run EOM-CCSD at chromophore scale on commodity
hardware, no external \emph{large-scale} comparison exists; at that scale the method
is supported by finite differences, internal consistency, and the kernel-identity
argument together with the small-scale and benzene-scale cross-code agreement above.
Section~\ref{sec:theory} develops the theory---the EOM-CC gradient functional
(Sect.~\ref{sec:eom}), the LT-AO J/K kernel that resolves every energy
denominator (Sect.~\ref{sec:lt}), the relaxation as the DAG transpose
(Sect.~\ref{sec:dag}), and the non-symmetric build the transition densities force
(Sect.~\ref{sec:jk}); Sect.~\ref{sec:compdetails} gives the computational
details, and Sect.~\ref{sec:results} the validation and GPU results.

%% =====================================================================
\section{Theory}
\label{sec:theory}
The theory is organized around a single observation: every analytic first
derivative in this framework is the contraction of one- and two-particle
(transition) density matrices with skeleton (derivative) integrals through one
Coulomb/exchange (J/K) kernel, with every energy denominator resolved by a
Laplace quadrature into a short grid of atomic-orbital (AO) density builds. We
develop this in four steps. Section~\ref{sec:eom} defines the EOM-CC
excited-state densities and the gradient functional they enter.
Section~\ref{sec:lt} develops the AO-direct Laplace-transform (LT-AO) J/K kernel
that resolves every denominator and assembles the gradient---the formalism that
keeps the calculation within an 8\,GB memory budget.
Section~\ref{sec:dag} shows that the relaxation (response) equations are the
transpose of the forward density-build graph, generated mechanically rather than
re-derived per state. Section~\ref{sec:jk} treats the non-symmetric J/K build
that the EOM transition densities force.

\paragraph{Notation and conventions.}
Indices $i,j,k,l$ label spin orbitals occupied in the Hartree--Fock reference
$|0\rangle\equiv|\Phi_0\rangle$, $a,b,c,d$ the unoccupied (virtual) orbitals,
$p,q,r,s$ either, and $\mu,\nu,\kappa,\rho$ the atomic-orbital (AO) basis;
$C_{\mu p}$ are the molecular-orbital (MO) coefficients and $\varepsilon_p$ the
orbital energies. We work in second quantization, with $\hat p^{\dagger}$ and
$\hat p$ the creation and annihilation operators for spin orbital $p$, and
$h_{pq}$ and $(pq|rs)$ (chemist notation) the one- and two-electron MO integrals.
Excitations are generated by the operators
\begin{equation}
\hat\tau^a_i=\hat a^{\dagger}\hat i,\qquad
\hat\tau^{ab}_{ij}=\hat a^{\dagger}\hat b^{\dagger}\hat j\,\hat i,\qquad\dots
\label{eq:tauops}
\end{equation}
(each creates particles in virtual orbitals and holes in occupied ones); the full
set is written $\{\hat\tau_\mu\}$, and $|\mu\rangle\equiv\hat\tau_\mu|0\rangle$ are
the excited determinants. The cluster operator is
$\hat T=\sum_{ia}t^a_i\,\hat\tau^a_i
+\tfrac14\sum_{ijab}t^{ab}_{ij}\,\hat\tau^{ab}_{ij}+\cdots$, with amplitudes
$t\equiv\{t_\mu\}$ fixed by the coupled-cluster equations
$\langle\mu|\bar H|0\rangle=0$; the EOM operators $\hat R_k,\hat L_k$
(Eq.~\eqref{eq:eomansatz}) are expanded in the same manifold $\{\hat\tau_\mu\}$,
with $r_0$ the weight on $|0\rangle$. A superscript $x$ denotes a skeleton
derivative with respect to a nuclear coordinate $x$ at fixed MO coefficients ---
the core-Hamiltonian derivative $h^{x}_{\mu\nu}$, the overlap derivative
$S^{x}_{\mu\nu}$, and the integral derivative $(\mu\nu|\kappa\rho)^{x}$ --- while
orbital relaxation is restored separately through the coupled-perturbed
Hartree--Fock (CPHF) coefficients $U^{x}$ (Eq.~\eqref{eq:master}). One- and
two-particle density matrices are
$\gamma_{pq}=\langle\hat p^{\dagger}\hat q\rangle$ and
$\Gamma_{pqrs}=\langle\hat p^{\dagger}\hat r^{\dagger}\hat s\,\hat q\rangle$;
their interstate (transition) forms are defined in Eq.~\eqref{eq:tdm}.

\emph{Convention for the density contractions and J/K builds.} The wavefunction
operators above are written in spin orbitals, but the one- and two-particle
density contractions, the cumulant collapse, and the Coulomb/exchange (J/K) builds
and gradient assembly that the GPU kernels implement
(Eqs.~\eqref{eq:Ffunc}, \eqref{eq:master}, \eqref{eq:laplacefock},
\eqref{eq:ltgrad}, \eqref{eq:cumulant}, and \eqref{eq:cumcollapse})
are stated in the \emph{spatial} closed-shell (RCCSD) convention that the
production code uses---in which the separable two-particle density carries the
spatial exchange coefficient $\tfrac12$ and the closed-shell two-electron build is
$2J-K$ on the per-spatial-orbital density, equivalently $J-\tfrac12 K$ on the
spin-summed total density (the two related by $2J-K=2(J-\tfrac12 K)$). Throughout,
$J^x$ and $K^x$ denote the \emph{single} bare builds of Eq.~\eqref{eq:jkderiv},
which carry no embedded prefactor; the exchange coefficient $\tfrac12$, the
closed-shell spin sum (the half-vs-total-density factor of two), and the bra--ket
pair-interchange $(pq|rs)=(rs|pq)$ are written explicitly wherever they occur and
are never folded into the build. The
spin-orbital expressions enter only the reference verification engine
(Sect.~\ref{sec:verif}), which uses a separate spin-orbital implementation.

%% ---------------------------------------------------------------------
\subsection{EOM-CC excited states, response densities, and the gradient
functional}
\label{sec:eom}
We begin with the object being differentiated: the EOM-CC excited-state energy,
its left and right eigenvectors, and the non-symmetric densities that the
gradient contracts.

\paragraph{Setup.}
EOM-CC parameterizes an excited state by acting on the coupled-cluster ground
state with a linear, CI-like excitation operator
$\hat R_k=r_0+\sum_{ia}r^a_i\hat\tau^a_i+\tfrac14\sum r^{ab}_{ij}\hat\tau^{ab}_{ij}
+\cdots$~\cite{StantonBartlett1993},
\begin{equation}
|\Psi_k\rangle=\hat R_k\,|\Psi_0\rangle,\qquad
|\Psi_0\rangle=e^{\hat T}|\Phi_0\rangle,
\label{eq:eomansatz}
\end{equation}
where $|\Psi_0\rangle$ is the CC ground state built from the reference
$|\Phi_0\rangle$. Inserting this ansatz into the Schr\"odinger equation gives
$\hat H\,e^{\hat T}\hat R_k|\Phi_0\rangle=E_k\,e^{\hat T}\hat
R_k|\Phi_0\rangle$, with $E_k$ the total energy of state~$k$. Because $\hat R_k$ and
$\hat T$ are both pure excitation operators they commute, $[\hat T,\hat R_k]=0$,
so $\hat R_k$ slides through $e^{\hat T}$ and the latter factors out. Left-%
multiplying by $e^{-\hat T}$ then yields an eigenvalue problem,
\begin{equation}
\big(e^{-\hat T}\hat H\,e^{\hat T}-E_k\big)\hat R_k|0\rangle=0,
\label{eq:eomeig}
\end{equation}
in which the bare Hamiltonian is replaced by the similarity-transformed operator
$\bar H\equiv e^{-\hat T}\hat H\,e^{\hat T}$, whose eigenvalue $E_k=E_0+\omega_k$
is the total energy of state~$k$ ($E_0$ the coupled-cluster ground-state energy,
the $\hat R_0=1$ root; $\omega_k$ the excitation energy). This is the central
object of EOM-CC: a similarity transform leaves the spectrum unchanged, so the
construction is formally exact; in practice $\hat T$ and $\hat R_k$ are truncated
to a common excitation level (singles through quadruples here, EOM-CCSDTQ).

\begin{figure}[htbp]
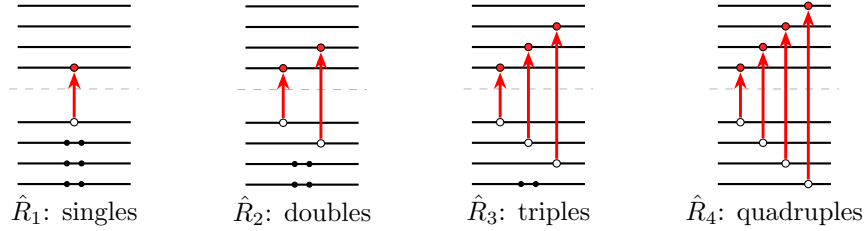
\centering
\setlength{\tabcolsep}{14pt}
\begin{tabular}{cccc}
\excit{1}{$\hat R_1$: singles} & \excit{2}{$\hat R_2$: doubles} &
\excit{3}{$\hat R_3$: triples} & \excit{4}{$\hat R_4$: quadruples}\\
\end{tabular}
\caption{EOM-CCSDTQ excitation manifolds. The right (and left) eigenvectors of
$\bar H$ span singles through quadruples,
$\hat R_k=r_0+\sum_{ia}r^a_i\,\hat\tau^a_i
 +\tfrac14\sum r^{ab}_{ij}\hat\tau^{ab}_{ij}+\cdots$; the dashed line is the
Fermi level, filled circles are reference electrons, and red arrows promote $n$
electrons from occupied to virtual orbitals. The non-symmetric transition
densities $\gamma^{AB},\Gamma^{AB}$ of Eq.~\eqref{eq:tdm} are formed from these
manifolds. The quadruples manifold is the reach of the symbolic
reference-equation generator (\texttt{p}$^{\dagger}$\texttt{q}) and the determinant
oracle; the device pipeline and every validated gradient and NACME in this work are
EOM-CCSD.}
\label{fig:manifolds}
\end{figure}

The price of the transform is that $\bar H$ is \emph{not} Hermitian---the wave
operator $e^{\hat T}$ is non-unitary. Its left and right eigenvectors are
therefore \emph{distinct} objects (unlike ordinary quantum mechanics, where one
bra/ket pair suffices), and they form a biorthonormal pair
$\langle L_j|R_k\rangle=\delta_{jk}$ rather than an orthonormal one. That
non-Hermiticity, and the $L\neq R$ structure it forces, is the root of nearly
everything ``non-symmetric'' in this paper. Collecting the two-sided eigenproblem,
\begin{equation}
\bar H = e^{-\hat T}\hat H\,e^{\hat T},\qquad
\bar H\,\hat R_k|0\rangle=(E_0+\omega_k)\hat R_k|0\rangle,\quad
\langle 0|\hat L_k\,\bar H=(E_0+\omega_k)\langle 0|\hat L_k,
\label{eq:hbar}
\end{equation}
biorthonormalized as $\langle L_j|R_k\rangle=\delta_{jk}$. In the $k$-fold
excitation basis $\{\hat\tau_\mu\}$, $\bar H$ is represented by the (non-symmetric)
\emph{EOM Jacobian}
\begin{equation}
A_{\mu\nu}=\langle\mu|\,[\,\bar H,\hat\tau_\nu\,]\,|0\rangle,
\label{eq:jacobian}
\end{equation}
whose right and left eigenvectors are $R_k,L_k$ with eigenvalue $\omega_k$. $A$ is
the operator that the relaxation step below transposes. For EOM-CCSDTQ the
operators $\hat R_k,\hat L_k$ span the singles through quadruples excitation
manifolds (Fig.~\ref{fig:manifolds}).

\paragraph{Why the densities are non-symmetric.}
Because $\bar H$ (hence $A$) is non-Hermitian, the left and right eigenvectors
differ ($L_k\neq R_k$), so the one- and two-particle (transition) densities
\begin{equation}
\gamma^{AB}_{pq}=\langle L_A|\,p^\dagger q\,|R_B\rangle,\qquad
\Gamma^{AB}_{pqrs}=\langle L_A|\,p^\dagger r^\dagger s\,q\,|R_B\rangle
\label{eq:tdm}
\end{equation}
are \emph{non-symmetric}: $\gamma^{AB}\neq(\gamma^{AB})^{\!\top}$. Operationally
this forbids the usual $A\!\leftrightarrow\!B$ permutation symmetry of the
Coulomb/exchange build, forcing the \emph{ordered} kernel
$J^x(A,B)\neq J^x(B,A)$ of Sect.~\ref{sec:jk} below --- the GPU kernel at the heart of
this work. The common shortcut $L_k=R_k$ symmetrizes the densities and corrupts
the gradient; we instead retain the true biorthonormal triple $(\omega_k,R_k,L_k)$.
We obtain it from a \emph{native} two-sided non-Hermitian Davidson that advances the
right and left Krylov subspaces of $\bar H$ together and returns both eigenvectors
from a \emph{single} solve. Because the pair is non-Hermitian, ordinary
Gram--Schmidt fails: we use biorthogonal Gram--Schmidt (deflating each $L_k$ against
the converged right set), holding the right vector at unit $L_2$ norm and rescaling
only the dual left vector so that $\langle L_k|R_k\rangle=1$; this fixes the relative
normalization of the biorthonormal pair on which every transition density (and hence
the gradient) depends. The solver is device-resident: each matvec $A\,r$ and its
transpose $A^{\!\top}w$ are evaluated by the same GPU kernels that assemble the
gradient, so one solve yields $(\omega_k,R_k,L_k)$ directly on the card. This
\emph{eliminates} the two-step seed---a right vector from a borrowed dense Davidson
followed by a GMRES inverse-iteration left vector---used to bootstrap earlier
validation: on \ce{H2O}/STO-3G the native pair reproduces that borrowed pair to
$\Delta\omega=4.5\times10^{-9}$, with right- and left-vector overlaps of unity and
biorthonormality error $\lVert L^{\top}R-I\rVert$ at $10^{-16}$.

The solver is multiplicity-general. For a closed-shell reference it targets a chosen
total spin through a matrix-free spin-sector projector
$P_S=\prod_{S'\neq S}(\hat S^2-\sigma')/(\sigma-\sigma')$ with
$\sigma=S(S+1)$; because $[\bar H,\hat S^2]=0$ the projection commutes through the
iteration and costs \emph{no} additional $\bar H$ matvec, separating singlets from
triplets in a single run. For an open-shell reference the spin-orbital
singles-and-doubles manifold is not closed under $\hat S^2$~\cite{Pauncz1979}, so no
determinant-space projector can reach a spin-pure state; we instead iterate in a
spin-adapted (genealogical) configuration-state-function (CSF) basis on a
semicanonical ROHF reference, which renders doublets and quartets spin-pure
\emph{by construction} (Sect.~\ref{sec:openshell}). The validated device path is
exercised at small scale---\ce{H2O}/STO-3G for the closed-shell device gates, and
gate-scale ($n_{\mathrm{so}}\le18$) open-shell systems for the multiplicity-general
path (Table~\ref{tab:openshell}). At chromophore scale the full-virtual
(uncompressed) eigensolve is a demonstrated capability rather than a
tolerance-converged production result; the FNO-compressed production eigensolve used
here converges to a $<\!5$~meV drift---looser than the $2.3\times10^{-6}~E_h$
benzene/STO-3G class, but two orders of magnitude below the $1.5$~eV $Q$--$B$ gap and
well inside the separation of each bright $^1E_u$ state from its neighbours, so the
state \emph{assignment}---the observable used here---is robust even where the absolute
excitation energy is not tolerance-converged (Sect.~\ref{sec:gpu}).

\paragraph{Gradient as a density functional.}
The excitation-energy gradient is the derivative of the density functional
\begin{equation}
F^{AB}[f,g;t]=\sum_{pq} f_{pq}\,\gamma^{AB}_{pq}
 +\tfrac12\sum_{pqrs} g_{pqrs}\,\Gamma^{AB}_{pqrs}
 \;-\;s_{AB}\,E_{\mathrm{corr}},
\label{eq:Ffunc}
\end{equation}
where $f_{pq}=h_{pq}$ and $g_{pqrs}=(pq|rs)$ are the one- and two-electron
Hamiltonian integrals, $\Gamma^{AB}$ is the two-particle transition density of
Eq.~\eqref{eq:tdm},
$E_{\mathrm{corr}}=\langle 0|\bar H|0\rangle-E_{\mathrm{HF}}$ is the ground-state
correlation energy, and $s_{AB}=\langle L_A|R_B\rangle$ is the interstate overlap
($s_{AB}=1$ for a state gradient, $0$ for a symmetry-distinct NACME). The last
term removes the \emph{disconnected} reference correlation that the cluster
operator injects into the densities; in the normal-ordered representation the
subtracted pieces are $t_1$ in the $ov$ block of $\gamma^{AB}$ and the connected
doubles combination $-(t_{ijab}+t_{ia}t_{jb}-t_{ib}t_{ja})$ in the $oovv$ block of
$\Gamma^{AB}$. At the converged cluster amplitudes $t$ this makes
$F^{AB}=\omega_k$ \emph{in value} (the disconnected piece vanishes; verified to
$3.5\times10^{-9}$) while its amplitude derivative
$\partial F^{AB}/\partial t\neq0$---the non-vanishing derivative is exactly what
the amplitude-response equation of Sect.~\ref{sec:dag} below is constructed to cancel.

\paragraph{Master gradient and NACME.}
The relaxed densities feed one AO-direct contraction,
\begin{equation}
\frac{dE_A}{dx}=\sum_{\mu\nu}\gamma_{\mu\nu}h^x_{\mu\nu}
 +\sum_{\mu\nu}W_{\mu\nu}S^x_{\mu\nu}
 +J^x(\gamma,\gamma)-\tfrac12 K^x(\gamma,\gamma)
 +2\,\mathrm{Tr}[\mathrm{GF}\,U^x]+E^x_{\mathrm{nuc}},
\label{eq:master}
\end{equation}
where $\gamma,\Gamma$ here denote the \emph{relaxed} one- and two-particle
densities ($\gamma\equiv P$ of Eq.~\eqref{eq:dtotal} below, and its two-particle
partner), $W$ is the energy-weighted density (conjugate to the overlap derivative
$S^x$), $\mathrm{GF}$ the generalized Fock and $U^x$ the coupled-perturbed
Hartree--Fock (CPHF) orbital-response
coefficients (the $2\,\mathrm{Tr}[\mathrm{GF}\,U^x]$ term is the orbital Z-vector
contribution), and each energy denominator inside $\gamma,\Gamma,W$ is resolved
by a Laplace grid of quadrature nodes $\{\tau_\alpha,w_\alpha\}$ (the $\tau_\alpha$
are scalar Laplace nodes, unrelated to the excitation operators $\hat\tau_\mu$;
the $\sum_\alpha w_\alpha$ is suppressed here for compactness). Section~\ref{sec:lt} expands this denominator resolution
explicitly and writes Eq.~\eqref{eq:master} in its Laplace-grid form,
Eq.~\eqref{eq:ltgrad}.

The symbols entering Eqs.~\eqref{eq:master} and \eqref{eq:ltgrad} are the
\emph{relaxed} densities. The total one-particle density decomposes additively,
\begin{equation}
P \;=\; D_{\mathrm{GS}}(T,\Lambda)\;+\;D_{\mathrm{EE}}(T,L,R)\;+\;D^{\zeta}(T,\zeta),
\label{eq:dtotal}
\end{equation}
into the ground-state ($\Lambda$-relaxed) density, the EOM transition density
built from $L,R$, and the amplitude-response density $D^{\zeta}$ obtained by
evaluating the ordinary $\Lambda$-linear coupled-cluster response density with the
multiplier $\zeta$ of Sect.~\ref{sec:dag} below; $P$ is thus the $\gamma$ of
Eq.~\eqref{eq:master} once relaxation is included. The energy-weighted density is
$W_{\mu\nu}=-\sum_{pq}\varepsilon_p\,P_{pq}\,C_{\mu p}C_{\nu q}$ (in Laplace form
$\varepsilon_p\!\to\!\varepsilon_p e^{\varepsilon_p\tau_\alpha}$; only its
symmetric part contributes to the symmetric $S^x_{\mu\nu}$), and
the generalized (non-symmetric) Fock matrix is
$\mathrm{GF}_{pq}=\sum_r h_{pr}P_{rq}+\sum_{rst}(pr|st)\,\Gamma_{qrst}$ (free index
$p$ on the one- and two-electron integrals, contracted against the relaxed
densities), whose occupied--virtual block is the right-hand side of the orbital
Z-vector (CPHF) equation that produces the orbital-response coefficients $U^x$.
Because $L\neq R^{\dagger}$, the two-particle transition density $\Gamma^{AB}$
additionally carries the blocks $\gamma^{(2)}|_{ovoo}$ and $\gamma^{(2)}|_{vvov}$
that are absent from any symmetric $L=R$ treatment.

Analytic EOM-CC nonadiabatic and derivative couplings have themselves been
formulated and validated against multireference references by several
groups~\cite{Tajti2009,Faraji2018}, including the biorthonormal CCSD
construction of the interstate left/right elements~\cite{Kjonstad2023} closest to
the non-Hermitian build used here. The quasidiabatic interstate NACME follows by
the off-diagonal Hellmann--Feynman route~\cite{IGS2009} with the
\emph{transition} densities $\gamma^{AB},\Gamma^{AB}$ in place of state
densities,
\begin{equation}
\lambda^{AB}_x=\frac{\langle L_A|\,\partial_x\bar H\,|R_B\rangle}{E_B-E_A},
\label{eq:nacme}
\end{equation}
the $1/(E_B-E_A)$ pole being the conical-intersection (Berry-connection)
singularity. Equation~\eqref{eq:nacme} is the \emph{quasidiabatic} coupling: $L,R$
are frozen at the reference geometry, which removes the eigenvector-response terms
$(E_A-E_B)\,\partial_x\langle L_A|R_B\rangle$. For different-irrep pairs
$s_{AB}=\langle L_A|R_B\rangle=0$, so the $E_{\mathrm{corr}}$ subtraction and the
nuclear-repulsion term drop out; same-irrep pairs ($s_{AB}\neq0$) carry an
additional gap term that the present implementation guards by assertion rather
than evaluates. Note $\lambda^{AB}_x\neq\lambda^{BA}_x$ in general; the symmetrized
value we report is the geometric mean $\sqrt{\lambda^{AB}_x\lambda^{BA}_x}$, formed
in post-processing. The device kernel returns only the one-sided coupling
numerator $\langle L_A|\partial_x\bar H|R_B\rangle$; the $1/(E_B-E_A)$ gap division
and the geometric-mean symmetrization are applied afterward by the caller in an
FP64 island, not on-device.
Both gradient and NACME reduce to the same AO-direct J/K kernel of
Sect.~\ref{sec:lt} fed different densities.

%% ---------------------------------------------------------------------
\subsection{The AO-direct Laplace-transform J/K kernel}
\label{sec:lt}
Equations~\eqref{eq:master} and \eqref{eq:nacme} reduce every gradient and NACME
to density contractions in which each energy denominator must be resolved. We now
develop the kernel that performs that resolution entirely in the atomic-orbital
(AO) basis, never forming a four-index molecular-orbital (MO) tensor. This is the
formalism that keeps the calculation within an 8\,GB memory budget; we present it
as a derivation, from the bottleneck to the master Laplace-grid gradient.

\paragraph{The denominator bottleneck.}
A correlated gradient contracts amplitudes and densities against two-electron
integrals divided by orbital-energy denominators
$D$ (e.g.\ $D=\varepsilon_a+\varepsilon_b-\varepsilon_i-\varepsilon_j$ for a
double excitation, $\varepsilon_a-\varepsilon_i$ for a single; $\varepsilon$ the
canonical orbital energies, $i,j$ occupied and $a,b$ virtual). Carried out
conventionally, this forms and stores four-index MO tensors, of which the
all-virtual ($vvvv$) block dominates memory and scales as the fourth power of the
virtual-space dimension. On commodity hardware this block alone exceeds the
available memory well before the molecules of interest are reached.

\paragraph{The Laplace identity.}
The H\"aser--Alml\"of identity~\cite{Almlof1991,HaserAlmlof1992} replaces the
reciprocal denominator by an integral, approximated on a short minimax
(best-$L^\infty$ exponential-sum) quadrature grid
$\{\tau_\alpha,w_\alpha\}$~\cite{Takatsuka2008,Hackbusch2019},
\begin{equation}
\frac{1}{D}=\int_0^\infty e^{-D\tau}\,d\tau
 \;\approx\;\sum_{\alpha=1}^{n_\tau} w_\alpha\,e^{-D\tau_\alpha},
\label{eq:laplace}
\end{equation}
with $n_\tau$ of order ten points sufficing for sub-microhartree accuracy; the
grid is generated for a target accuracy over the spanned denominator range
$[D_{\min},D_{\max}]$, over which the minimax error decays nearly exponentially in
$n_\tau$~\cite{Takatsuka2008}, controlled by $n_\tau$ together with the ratio
$D_{\max}/D_{\min}$, not by $n_\tau$ alone. The
decisive property is that the exponential of the (additive) denominator
factorizes into one-index factors,
\begin{equation}
e^{-D\tau_\alpha}
 =e^{+\varepsilon_i\tau_\alpha}\,e^{+\varepsilon_j\tau_\alpha}\,
  e^{-\varepsilon_a\tau_\alpha}\,e^{-\varepsilon_b\tau_\alpha},
\label{eq:factorize}
\end{equation}
so each orbital index carries its own scalar weight. The contraction can then run
index by index in the AO basis, and no four-index MO tensor is ever built.

\paragraph{Laplace-scaled coefficients and the AO density.}
The per-index factors of Eq.~\eqref{eq:factorize} are absorbed directly into the
MO coefficients, defining Laplace-scaled occupied and virtual coefficients at
each grid point $\alpha$,
\begin{equation}
\tilde U^{\alpha}_{\mu i}=C_{\mu i}\,e^{+\varepsilon_i\tau_\alpha},\qquad
\tilde V^{\alpha}_{\mu a}=C_{\mu a}\,e^{-\varepsilon_a\tau_\alpha}.
\label{eq:scaledcoeff}
\end{equation}
Contracting these scaled coefficients over the orbital indices yields an AO-basis
density at grid point $\alpha$,
\begin{equation}
\mathcal D^{(1;\alpha)}_{\mu\nu}=\sum_{ia}\tilde U^{\alpha}_{\mu i}\,
 \tilde V^{\alpha}_{\nu a}.
\label{eq:aodens}
\end{equation}
Equation~\eqref{eq:aodens} is the level-1 density. For correlation order $n$ the
construction iterates to a depth $L=L(n)$ ($L=n-1$ for $n=2,3$,
$L=\lfloor n/2\rfloor$ for $n\ge4$), one Laplace grid per denominator, the
level-$k$ density carrying the order-$k$ amplitude residual $\mathcal R^{(k)}_{ia}$
between the scaled coefficients,
\begin{equation}
\mathcal D^{(k;\alpha_1,\dots,\alpha_k)}_{\mu\nu}
 =\sum_{ia}\tilde U^{\alpha_k}_{\mu i}\,
  \mathcal R^{(k;\alpha_1,\dots,\alpha_{k-1})}_{ia}\,
  \tilde V^{\alpha_k}_{\nu a},
\label{eq:aodensk}
\end{equation}
with $\mathcal R^{(k)}_{ia}$ the occupied--virtual projection of the residual
produced by the forward J/K cascade through the preceding grid points
(Eqs.~\eqref{eq:jkbuild}--\eqref{eq:laplacefock}). The ket that enters the gradient
adds to this forward density the backward density $\hat{\mathcal D}^{(L)}$
constructed as the DAG transpose (Sect.~\ref{sec:dag} below),
\begin{equation}
\Xi^{(L;\boldsymbol\alpha)}=\mathcal D^{(L;\boldsymbol\alpha)}
 +\hat{\mathcal D}^{(L;\boldsymbol\alpha)},
\label{eq:xidef}
\end{equation}
which carries the relaxation ($\Lambda$ / Z-vector / $\zeta$) contribution. The
forward cascade (Eqs.~\eqref{eq:scaledcoeff}--\eqref{eq:aodensk}) is
lines~\ref{ln:fwdstart}--\ref{ln:fwdend} of Algorithm~\ref{alg:ltao}.

\paragraph{The J and K builds.}
Each AO density is contracted with the two-electron integrals through the Coulomb
and exchange builds
\begin{equation}
J(\mathcal D)_{\mu\nu}=\sum_{\kappa\rho}(\mu\nu|\kappa\rho)\,\mathcal D_{\kappa\rho},
\qquad
K(\mathcal D)_{\mu\nu}=\sum_{\kappa\rho}(\mu\kappa|\nu\rho)\,\mathcal D_{\kappa\rho},
\label{eq:jkbuild}
\end{equation}
which are the standard AO-direct Coulomb/exchange operations~\cite{Hohenstein2015}
evaluated with GPU Gaussian-integral engines~\cite{Wang2024}. Resolving a
correlated denominator directly in the AO basis through such builds, with no
AO$\to$MO transform, follows the AO-direct Laplace MP2 gradient of Schweizer,
Doser, and Ochsenfeld~\cite{Schweizer2008}; the present kernel extends that
pseudodensity construction to the non-Hermitian transition densities below.
The gradient requires the AO-direct \emph{derivative} (skeleton) builds, in which
the integral derivatives $(\mu\nu|\kappa\rho)^{x}$ are contracted against two
(generally distinct) density arguments,
\begin{equation}
J^{x}(A,B)=\!\!\sum_{\mu\nu\kappa\rho}\!\!(\mu\nu|\kappa\rho)^{x}A_{\mu\nu}B_{\kappa\rho},
\quad
K^{x}(A,B)=\!\!\sum_{\mu\nu\kappa\rho}\!\!(\mu\kappa|\nu\rho)^{x}A_{\mu\nu}B_{\kappa\rho}.
\label{eq:jkderiv}
\end{equation}
These derivative builds are the J/K kernel of this work; here $A,B$ denote the two
AO density arguments (not the EOM Jacobian $A$ of Eq.~\eqref{eq:jacobian}).
Equation~\eqref{eq:jkderiv} is the ordered build whose non-symmetry,
$J^x(A,B)\neq J^x(B,A)$, the EOM transition densities force (Sect.~\ref{sec:jk} below). The undifferentiated forms of
Eq.~\eqref{eq:jkbuild} additionally enter the generalized (Laplace) Fock
back-transform,
\begin{equation}
\tilde F^{\alpha}_{pq}=\sum_{\mu\nu}C_{\mu p}\,
 \big[2J(\mathcal D^{(1;\alpha)})-K(\mathcal D^{(1;\alpha)})\big]_{\mu\nu}\,C_{\nu q},
\label{eq:laplacefock}
\end{equation}
which returns each grid contribution to the MO basis for the response steps of
Sect.~\ref{sec:dag} below.

\paragraph{The master LT-AO gradient.}
Assembling the per-grid builds gives the explicit Laplace-grid form of the master
gradient, Eq.~\eqref{eq:master}, in its relaxed AO-direct realization,
\begin{equation}
\frac{dE_A}{dx}=\sum_{\boldsymbol\alpha}\prod_k w_{\alpha_k}\,
 \Big[2J^{x}\!\big(\mathcal D^{(1;\alpha_1)},\Xi^{(L;\boldsymbol\alpha)}\big)
     -K^{x}\!\big(\mathcal D^{(1;\alpha_1)},\Xi^{(L;\boldsymbol\alpha)}\big)\Big]
 +\sum_{\mu\nu}\big[P_{\mu\nu}h^{x}_{\mu\nu}+W_{\mu\nu}S^{x}_{\mu\nu}\big],
\label{eq:ltgrad}
\end{equation}
where $P$ and $W$ are the relaxed one-particle and energy-weighted densities and
$\Xi^{(L;\boldsymbol\alpha)}$ is the combined forward+backward AO density of
Eq.~\eqref{eq:xidef}. The per-grid build is written here as $2J^x-K^x$ on the
per-spatial (half) densities $\mathcal D^{(1;\alpha_1)},\Xi^{(L;\boldsymbol\alpha)}$;
this is the same closed-shell two-electron build as the $J^x-\tfrac12K^x$ of the
master gradient Eq.~\eqref{eq:master} on the spin-summed total density
($2J-K=2(J-\tfrac12K)$, the factor $2$ being the spin sum), with $J^x,K^x$ the bare
builds of Eq.~\eqref{eq:jkderiv}. The multi-index
$\boldsymbol\alpha=(\alpha_1,\dots)$ runs over the Laplace grids of all
denominators, and $\prod_k w_{\alpha_k}$ collects their quadrature weights.

\paragraph{Cost.}
Each reciprocal denominator $1/D$ is thus replaced by a short sum of J/K builds,
one per grid point, each of cost $\mathcal{O}(N^2)$ in the AO dimension~$N$ for a
direct (integral-recomputing) scheme, the asymptotic cost once Schwarz screening
is applied (the formal cost of one undifferentiated build is $\mathcal{O}(N^4)$).
The four-index $vvvv$ block is never formed: because each build
(Eq.~\eqref{eq:jkbuild}) contracts the integrals against a \emph{two-index} AO
density, the all-virtual MO summation is never assembled---it is replaced by the
AO-direct build. At the MP2 level the conventional $\mathcal{O}(N^5)$ denominator
contraction becomes $n_\tau$ AO-direct J/K builds; at the doubles level and
above, the all-virtual block that dominates conventional memory is eliminated
outright. This replacement of stored four-index tensors by recomputed
two-index AO builds is what makes the excited-state gradients of this work fit
within an 8\,GB budget.

\paragraph{EOM-CC specialization.}
The excited-state denominators of EOM-CC carry the excitation energy $\omega_k$,
$1/(D_k-\omega_k)$, where $D_k$ is the orbital-energy denominator of the
excitation class (a single $\varepsilon_a-\varepsilon_i$, a double
$\varepsilon_a+\varepsilon_b-\varepsilon_i-\varepsilon_j$, etc.). The same
quadrature applies with a single, $\omega$-dependent reweighting of the grid,
\begin{equation}
\frac{1}{D_k-\omega}\approx\sum_{\alpha}\widetilde w_\alpha(\omega)\,e^{-D_k\tau_\alpha},
\qquad \widetilde w_\alpha(\omega)=w_\alpha\,e^{\omega\tau_\alpha},
\label{eq:eomshift}
\end{equation}
valid for $\omega<\min_k D_k$, i.e.\ the regime set by the HOMO--LUMO gap; for
higher (Rydberg/charge-transfer) roots approaching $\omega\to D_k$ a contour shift
in the complex-$\tau$ plane keeps the integral finite, or the IP-EOM treatment
applies. Only
the per-grid weight changes: the AO-direct J/K kernel of
Eqs.~\eqref{eq:jkbuild}--\eqref{eq:jkderiv} and the grid points $\tau_\alpha$ are
unchanged. The EOM-CC excited-state gradient therefore uses the \emph{same} LT-AO
kernel as the ground state, fed the transition densities of
Eq.~\eqref{eq:tdm} in place of the state densities.

\paragraph{The procedure, end to end.}
Algorithm~\ref{alg:ltao} collects the complete gradient/NACME procedure---the
forward density cascade of this section, the reverse-mode relaxation derived in
Sect.~\ref{sec:dag} below, and their assembly through the J/K kernel of
Sect.~\ref{sec:jk}---into one numbered pass, with each step cross-referenced to its
working equation. The remaining two subsections supply the content the algorithm
invokes by name: \emph{why} the reverse pass (lines~\ref{ln:revstart}--\ref{ln:revend})
is guaranteed to be the transpose of the forward pass (Sect.~\ref{sec:dag}), and
\emph{why} a single ordered kernel covers every build the algorithm calls
(Sect.~\ref{sec:jk}).

\begin{algorithm}[htbp]
\caption{EOM-CC excited-state gradient via the AO-LT contraction DAG.
Lines~\ref{ln:fwdstart}--\ref{ln:fwdend} are the forward density cascade
(Sect.~\ref{sec:lt}); lines~\ref{ln:revstart}--\ref{ln:revend} are its reverse-mode
transpose (Sect.~\ref{sec:dag}); the closing builds reuse the single ordered J/K
kernel (Sect.~\ref{sec:jk}). The contraction DAG that condenses, differentiates,
and orders these steps is compiled once and reused across the grid and across
methods.}
\label{alg:ltao}
\begin{algorithmic}[1]
\Require geometry; converged CC amplitudes $T$ ($\langle\mu|\bar H|0\rangle=0$); target root $k$ (state $A\equiv k$)
\State Solve the two-sided EOM eigenproblem for the biorthonormal $\hat R_k,\hat L_k$ \eqcite{Eqs.~\eqref{eq:hbar}, \eqref{eq:jacobian}}
\State Form the non-symmetric transition densities $\gamma^{AB},\Gamma^{AB}$ \eqcite{Eq.~\eqref{eq:tdm}}
\State Subtract the disconnected $E_{\mathrm{corr}}$ pieces ($t_1$; connected doubles) \eqcite{Eq.~\eqref{eq:Ffunc}}
\State Generate the Laplace grid $\{\tau_\alpha,w_\alpha\}$ for the spanned denominator range \eqcite{Eq.~\eqref{eq:laplace}}
\For{each grid multi-index $\boldsymbol\alpha=(\alpha_1,\dots,\alpha_L)$} \label{ln:fwdstart}
  \State Form Laplace-scaled coefficients $\tilde U^{\alpha},\tilde V^{\alpha}$ \eqcite{Eq.~\eqref{eq:scaledcoeff}}
  \State Build the level-1 AO density $\mathcal D^{(1;\alpha_1)}$ \eqcite{Eq.~\eqref{eq:aodens}}
  \For{$k=2,\dots,L$} \Comment{level-$k$ cascade}
    \State $J/K$-build $\to$ Laplace--Fock back-transform $\to$ residual $\mathcal R^{(k)}$ \eqcite{Eqs.~\eqref{eq:jkbuild}, \eqref{eq:laplacefock}}
    \State Build the level-$k$ AO density $\mathcal D^{(k;\boldsymbol\alpha)}$ \eqcite{Eq.~\eqref{eq:aodensk}} \label{ln:fwdend}
  \EndFor
\EndFor
\State Reverse pass = DAG transpose: seed $\bar F=1$, adjoint each forward edge $\to$ source $\xi=\partial F^{AB}/\partial t$ \label{ln:revstart} \eqcite{Eq.~\eqref{eq:zeta}}
\State Solve the amplitude response $A^{\!\top}\zeta=-\xi$; build the backward density $\hat{\mathcal D}^{(L)}$ \eqcite{Eq.~\eqref{eq:zeta}}
\State Combine the ket $\Xi^{(L;\boldsymbol\alpha)}=\mathcal D^{(L;\boldsymbol\alpha)}+\hat{\mathcal D}^{(L;\boldsymbol\alpha)}$ \label{ln:revend} \eqcite{Eq.~\eqref{eq:xidef}}
\State Assemble the relaxed $P,W$ and orbital ($U^x$) response \eqcite{Eqs.~\eqref{eq:master}, \eqref{eq:dtotal}}
\State \Return $\displaystyle\frac{dE_A}{dx}=\sum_{\boldsymbol\alpha}\!\prod_k w_{\alpha_k}\big[2J^{x}(\mathcal D^{(1)},\Xi)-K^{x}(\mathcal D^{(1)},\Xi)\big]+\!\sum_{\mu\nu}\!\big[P_{\mu\nu}h^{x}_{\mu\nu}+W_{\mu\nu}S^{x}_{\mu\nu}\big]$ \eqcite{Eq.~\eqref{eq:ltgrad}}
\Statex \textbf{NACME variant:} run lines 1--15 with mixed $(\hat L_A,\hat R_B)$ densities and the $1/(E_B-E_A)$ factor \eqcite{Eq.~\eqref{eq:nacme}}
\end{algorithmic}
\end{algorithm}

%% ---------------------------------------------------------------------
\subsection{Relaxation as the contraction-DAG transpose}
\label{sec:dag}
The two foregoing subsections fixed \emph{what} must be contracted---the
transition densities and the gradient functional (Sect.~\ref{sec:eom})---and
\emph{how} every energy denominator inside them is resolved into AO-direct J/K
builds (Sect.~\ref{sec:lt}). What remains is the relaxation: the response of the
amplitudes to nuclear motion, traditionally the most laborious part of an
analytic gradient. We now make the ``DAG is the spine'' thesis concrete and show
that the relaxation equations fall out as the transpose of the forward
density-build graph, with no per-state re-derivation.

\begin{figure}[htbp]\centering
\resizebox{\linewidth}{!}{%
\begin{tikzpicture}[>=Latex, font=\normalsize, every node/.style={inner sep=5pt},
  ms/.style  ={draw,rounded corners=3pt,fill=orange!14,align=center,text width=118mm,minimum height=9mm},
  uni/.style ={draw,line width=1.5pt,rounded corners=3pt,fill=blue!12,align=center,text width=118mm,minimum height=11mm},
  res/.style ={draw,rounded corners=3pt,fill=red!18,align=center,text width=90mm,minimum height=9mm},
  feed/.style={draw,dashed,rounded corners=3pt,fill=gray!7,align=center,font=\small,text width=52mm,minimum height=12mm},
  dag/.style ={draw,dashed,line width=1.1pt,rounded corners=3pt,fill=violet!12,align=center,text width=122mm,minimum height=9mm},
  flow/.style={-{Stealth[length=3.4mm]},line width=1.3pt,black!78},
  feedarr/.style={-{Stealth[length=3mm]},line width=1pt,black!60}]
  \node[dag] (dag) at (0,0)     {\textbf{Contraction-DAG compiler} (\emph{structural invariant}):
        condense shared intermediates \,$\cdot$\, transpose $\Rightarrow$ S4 \,$\cdot$\, optimize path once, reuse over grid $\times$ methods};
  \node[ms]  (s1) at (0,-1.95)  {\textbf{S1--S2} \;\;reference $+$ amplitudes / coefficients:\;
        $T^{(n)}$\,(MP$n$) \,$\vert$\, $T,L,R$\,(CC/EOM) \,$\vert$\, CI\,$\mathbf c$ / MPS\,(MR)};
  \node[ms]  (s3) at (0,-3.75)  {\textbf{S3} \;\;one-\,/\,two-particle density matrices $\gamma,\Gamma$ \;(state, or transition $\gamma^{AB},\Gamma^{AB}$)};
  \node[ms]  (s4) at (0,-5.55)  {\textbf{S4} \;\;relaxation $\Rightarrow$ \emph{non-symmetric} $\gamma,\Gamma,W$ \;
        ($\Lambda$/Z \,$\vert$\, $\zeta$ \,$\vert$\, Z$+$CP-MCSCF \,$\vert$\, none)};
  \node[feed] (lap) at (-3.55,-7.2) {Laplace\;\; $\tfrac{1}{D}{=}\!\int_{0}^{\infty}\! e^{-D\tau}\,d\tau$};
  \node[feed] (int) at ( 3.55,-7.2) {skeleton integrals\;\; $h^{x},\,S^{x},\,(\mu\nu|\kappa\rho)^{x}$};
  \node[uni] (s5) at (0,-9.2) {\textbf{S5\;\; UNIVERSAL AO-LT J/K KERNEL}\\[3pt]
        $\displaystyle\sum_{\alpha} w_\alpha\,[\,2J^{x}-K^{x}\,](\mathcal D,\mathcal D')\;+\;\sum h^{x}\gamma\;+\;\sum S^{x}W$};
  \node[res] (g)  at (0,-11.0)  {\textbf{gradient} $dE_A/dx$\qquad or\qquad \textbf{NACME} $\lambda^{AB}_x$};
  \draw[flow] (dag)--(s1); \draw[flow] (s1)--(s3); \draw[flow] (s3)--(s4);
  \draw[flow] (s4)--(s5);  \draw[flow] (s5)--(g);
  \draw[feedarr] (lap.south) -- ([xshift=10mm]lap.south |- s5.north);
  \draw[feedarr] (int.south) -- ([xshift=-10mm]int.south |- s5.north);
  \draw[dashed,-{Stealth[length=2.6mm]},violet!70,line width=1.1pt]
        (dag.east) to[out=0,in=0,looseness=0.8]
        node[midway,right=2pt,font=\small,text=violet!55!black,align=left]{transpose\\$\Rightarrow$ S4} (s4.east);
  \draw[decorate,decoration={brace,amplitude=7pt,mirror},black!50,line width=1pt]
        (-6.35,-1.35) -- (-6.35,-6.15);
  \node[font=\small\itshape,text=black!70,rotate=90,anchor=south] at (-6.9,-3.75) {method-specific (S1--S4)};
\end{tikzpicture}}
\caption{The universal gradient/NACME pipeline. Stages S1--S4 (orange) are
method-specific---they decide \emph{which} densities are built and \emph{how}
they are relaxed; for EOM-CC, S2 supplies the biorthonormal $T,L,R$, S3 the
non-symmetric transition densities, and S4 the self-consistent $\zeta$ response.
Stage S5 (blue) is the same kernel for every method: a Laplace-resolved,
non-symmetric J/K contraction. Spanning the pipeline, the contraction-DAG
compiler (violet) generates the S4 relaxation equations as its transpose (dashed
edge).}
\label{fig:pipeline}
\end{figure}
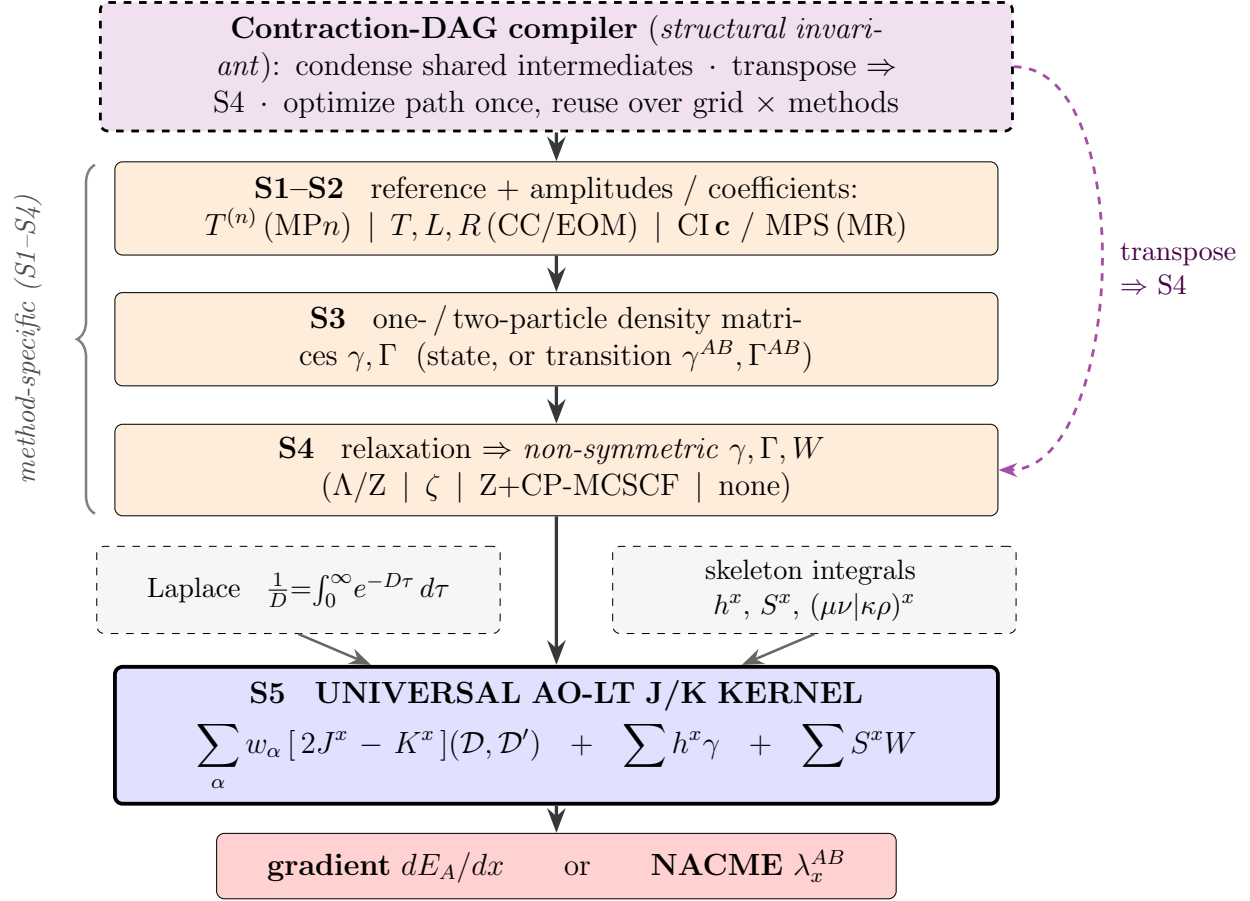

\paragraph{The five-slot template.}
We read any correlation method as five interchangeable parts; the first four are
method-specific, the fifth universal. \textbf{S1}~reference and orbital partition
(a single Hartree--Fock determinant for EOM-CC); \textbf{S2}~amplitudes and
coefficients (the cluster operator $\hat T$ and the biorthonormal left/right
eigenvectors $\hat L_k,\hat R_k$ for EOM-CC); \textbf{S3}~one- and two-particle
densities (the \emph{non-symmetric} transition densities $\gamma^{AB},\Gamma^{AB}$
for couplings); \textbf{S4}~relaxation/response (the self-consistent $\zeta$
amplitude-response, developed below); \textbf{S5}~the universal atomic-orbital
Laplace-transform (LT-AO) J/K kernel of Sect.~\ref{sec:lt}. Only S1--S4 change
between methods; S5 and the contraction-DAG compiler
that condenses, differentiates, and orders it are shared --- which is why the
same engine is designed to specialize across MP2 through CASPT2.
That generality is a design property of the template; in this work we develop and
validate only the EOM-CC case.
The labels S1--S5 index the stages of Fig.~\ref{fig:pipeline}.

The density build and its contraction through the LT-AO kernel
(Fig.~\ref{fig:pipeline}) are never manipulated term by term: the working
expressions are compiled once into a single contraction DAG $G_L$ whose nodes are
intermediate tensors and whose edges are multilinear contractions. Three
properties of that one object carry the leverage of this work, and each maps onto
a step the algorithm above takes for granted.

\emph{Condensation.} The hundreds of contractions in a truncated EOM-CC density
share many subexpressions --- the same half-contracted integral block, the same
$T_2$-dressed intermediate --- that a literal term-by-term evaluation would rebuild
repeatedly. Compiling to a DAG merges identical subexpressions into shared nodes,
so each is computed once and reused by every term that needs it. The amortization
compounds along the two axes the algorithm loops over: a shared node is computed
once and reused across all Laplace-grid points $\boldsymbol\alpha$ (the
$\omega$-reweighting of Eq.~\eqref{eq:eomshift} touches only the per-grid scalar
weight, not the node), and once a method's leaves are swapped in (S1--S4 of the
template) the same condensed skeleton serves MP$n$, CC, and EOM builds.

\emph{Transpose.} The reverse of $G_L$ is the relaxation
($\Lambda$ / Z-vector / $\zeta$) equation, read off by reverse-mode without
per-state re-derivation (developed below). Because the transpose is taken on the
graph rather than on a hand-written term list, gradients and NACMEs --- and every
EE/IP/EA variant at every truncation level --- inherit it as \emph{one} mechanism
rather than a family of derivations.

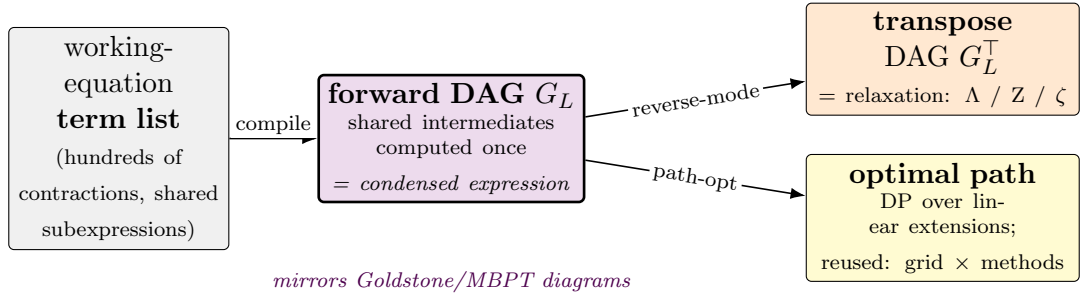
\begin{figure}[htbp]\centering
\begin{tikzpicture}[>=Latex, font=\small, every node/.style={inner sep=3pt},
  tl/.style ={draw,rounded corners=2pt,fill=gray!12,align=center,minimum height=11mm,text width=27mm},
  fw/.style ={draw,line width=1pt,rounded corners=2pt,fill=violet!14,align=center,minimum height=13mm,text width=33mm},
  outp/.style={draw,rounded corners=2pt,fill=yellow!22,align=center,minimum height=12mm,text width=34mm},
  rel/.style ={draw,rounded corners=2pt,fill=orange!18,align=center,minimum height=12mm,text width=34mm},
  arr/.style ={->,semithick,>=Latex}]
  \node[tl]  (tl)  at (0,0) {working-equation \textbf{term list}\\ \scriptsize(hundreds of contractions, shared subexpressions)};
  \node[fw]  (fw)  at (4.4,0) {\textbf{forward DAG} $G_L$\\ \scriptsize shared intermediates computed once \\ \scriptsize\emph{= condensed expression}};
  \node[rel] (rel) at (10.9,1.05) {\textbf{transpose} DAG $G_L^{\top}$\\ \scriptsize $=$ relaxation: $\Lambda$ / Z / $\zeta$};
  \node[outp](opt) at (10.9,-1.05) {\textbf{optimal path}\\ \scriptsize DP over linear extensions; \\ \scriptsize reused: grid $\times$ methods};
  \draw[arr] (tl)--node[above,font=\scriptsize,fill=white,inner sep=1.5pt]{compile}(fw);
  \draw[arr] (fw)--node[sloped,font=\scriptsize,fill=white,inner sep=1.5pt,pos=0.5]{reverse-mode}(rel);
  \draw[arr] (fw)--node[sloped,font=\scriptsize,fill=white,inner sep=1.5pt,pos=0.5]{path-opt}(opt);
  \node[font=\scriptsize\itshape,text=violet!60!black,align=center] at (4.4,-1.9)
       {mirrors Goldstone/MBPT diagrams};
\end{tikzpicture}
\caption{The contraction-DAG leverage. The per-method term list compiles into one
forward DAG that condenses shared intermediates; its transpose yields the
relaxation equations for free (for EOM-CC: the $\zeta$ amplitude-response and the
interstate Z-vector), and its once-optimized path is reused across the Laplace
grid and across methods. The same DAG drives the GPU code generation of
Sect.~\ref{sec:jk}.}
\label{fig:dag}
\end{figure}

\emph{Path.} The contraction order is optimized once (dynamic programming over the
linear extensions of $G_L$) and that single schedule is reused across the Laplace
grid and across methods (Fig.~\ref{fig:dag}), so the optimization cost is paid once
and amortized over every grid point and every downstream method.

Compiling a complex computation into a DAG and exploiting its graph-theoretic
structure is the canonical strategy of optimizing compilers, where a basic block
is represented as a DAG whose shared nodes are common subexpressions and the
compiler performs common-subexpression elimination, dead-code elimination, and
instruction scheduling by topological ordering of that graph~\cite{Aho2007}.
Within quantum chemistry, this compile-to-graph-then-condense-and-order strategy
was established for many-body methods by the Tensor Contraction
Engine~\cite{Hirata2003}, which compiles coupled-cluster and many-body equations
into a contraction graph, factors out common subexpressions, and optimizes the
contraction order --- the direct ancestor of the Condensation and Path properties
below. The
three properties above are exactly these established operations specialized to the
density build: \emph{Condensation} is common-subexpression elimination,
\emph{Path} is instruction scheduling (a topological order of $G_L$), and
\emph{Transpose} is reverse-mode
differentiation of the graph, the same graph rewrite that powers
automatic-differentiation and machine-learning frameworks (the backpropagation
invoked above). The same toolkit recurs across disciplines---optimizing
compilers, automatic-differentiation frameworks, and coupled-cluster code
generators~\cite{Rubin2021,Liebenthal2025}. In quantum chemistry specifically,
graphs already drive automated GPU code generation: the automated code engine of
Song, Wang, and Mart\'{\i}nez represents a single integral-evaluation program as
a dataflow graph and transforms it to enumerate store-versus-recompute code
variants, autotuned for a target device~\cite{Song2016}. A complementary strand
applies the same idea one level down, at the recurrence relations themselves: a
layered (topological-level) code generator emits optimized implementations of
arbitrary recurrences---orthogonal polynomials, special functions, and the
molecular-integral recurrences (McMurchie--Davidson, Obara--Saika, Rys, and the
Boys function)---and is reported to exceed expert hand-optimization~\cite{Guerrero2026}.
That recurrence-level codegen is the lineage of the per-quartet HGP-OS recurrence
kernels emitted by the code generator of \S3 (Fig.~\ref{fig:cdrecur}). In both
cases the graph organizes the evaluation of a fixed integral kernel; it is a
different object from the contraction spine we use here. Our DAG \emph{is} the many-body
density-contraction itself, and its graph-theoretic structure is the
method---shared-node condensation, one reusable contraction path, and a
reverse-mode transpose that, because the graph builds the response functional
$F^{AB}$, \emph{is} the relaxation, generating the response equations ($\zeta$
amplitude response, interstate Z-vector) rather than re-deriving them. This
machinery
is inherited and mature; we claim no novelty for DAGs, common-subexpression
elimination, scheduling, reverse-mode differentiation, or graph-driven GPU code
generation. Our contribution is
their specialization to the correlated excited-state response setting --- the
EOM-CC gradient and NACME --- and specifically the identity that the DAG
transpose \emph{is} the relaxation, realized through Algorithm~\ref{alg:ltao}.

These properties are not incidental to coupled-cluster structure. $G_L$ mirrors
the structure of the Goldstone/MBPT diagrammatic expansion: each diagram is a node,
each contracted internal line an edge, and a shared diagram fragment maps to a
shared DAG node --- so condensation is the diagrammatic factorization made
automatic, the transpose is the conjugate (response) diagram set generated
mechanically, and the optimized path is a contraction ordering of the diagrams.
$G_L$ is the diagrammatic structure made executable, differentiable, and
optimizable.

\paragraph{Relaxation is the DAG transpose.}
\emph{Intuition:} computing a gradient by hand means deriving a separate
``relaxation'' (Lagrange-multiplier) equation for every state --- tedious and
error-prone. This is exactly the situation automatic differentiation solves: if
the energy is a computation graph (a DAG), running that graph \emph{backwards}
mechanically produces all the response equations, just as backpropagation
produces gradients in machine learning. This view is now standard in
differentiable quantum chemistry --- from automatic differentiation through
Hartree--Fock~\cite{Tamayo2018} and the PySCFAD framework~\cite{Zhang2022} to
the reverse-mode coupled-cluster response of Zhang \emph{et~al.}, who obtain the
relaxed (Lagrangian) response density from the backward pass \emph{instead of}
solving the Z-vector equation explicitly~\cite{Zhang2024}. We exploit the same
principle, specialized to the non-Hermitian EOM-CC response: the relaxation
equations are the transpose of the density-build DAG, generated once by the
machine rather than per state by hand.

Standard EOM-CC analytic gradients
(Stanton and Gauss~\cite{Stanton1993,StantonGauss1995}; the spin-conserving and
spin-flip implementation of Krylov and co-workers~\cite{Levchenko2005}) solve
a left-state $\Lambda$-like Lagrange equation (with $\Lambda$ the ground-state
Lagrange multiplier) plus a separate orbital Z-vector (the single linear solve
that yields all orbital-response contributions), each re-derived per state. Here
the amplitude response is a \emph{single self-consistent} multiplier $\zeta$
obtained as the transpose of the forward density-contraction DAG --- identical in
form across EE/IP/EA and across truncation level --- and the orbital Z-vector is
the same J/K build with a different density argument. Our contribution is not a
new physical quantity but a mechanization: all relaxation equations are one
reverse-mode pass over the forward DAG, never re-derived per state. The reverse
pass is lines~\ref{ln:revstart}--\ref{ln:revend} of Algorithm~\ref{alg:ltao}.

Concretely, holding $L,R$ frozen, the amplitudes $T$ still respond to nuclear
motion; that response solves
\begin{equation}
A^{\!\top}\zeta = -\,\xi,\qquad \xi_\mu=\frac{\partial F^{AB}}{\partial t_\mu},
\label{eq:zeta}
\end{equation}
where $A$ is the EOM Jacobian of Eq.~\eqref{eq:jacobian} and $\xi$ is the source
vector $\partial F/\partial t$, built by an analytic term-list adjoint (the
reverse-mode pass) and validated against reverse-mode automatic differentiation
to machine precision. It is \emph{self-consistent} because $\xi$ is the gradient
of the very functional $F$ whose densities feed the build, unifying the
ground-state $\Lambda$ and the excited-state response into one multiplier over the
full amplitude space---replacing the per-state Z-vector prescription.

A subtlety here is load-bearing for correctness: $\xi$ is the $t$-derivative of
the \emph{density functional} $F^{AB}$ of Eq.~\eqref{eq:Ffunc}, \emph{not} of the
connected matrix element $\tilde\omega(t)=\langle L|A(t)|R\rangle$. The two differ
by $\sum_\mu(\partial\Omega_\mu/\partial t)\langle L|R\,\hat\tau_\mu\rangle$ (with
$\Omega_\mu$ the cluster residuals; nonzero only in the single-excitation
$t_1$/$ov$ channel of Eq.~\eqref{eq:Ffunc}, the only disconnected channel
surviving at convergence), and this
is precisely the disconnected $t$-dependence that Eq.~\eqref{eq:Ffunc} was built
to carry. Using the connected form instead leaves an $\mathcal O(10^{-3})$ error
in the gradient---choosing the density-functional derivative is what closes the
finite-difference gate to $\sim\!10^{-8}$.

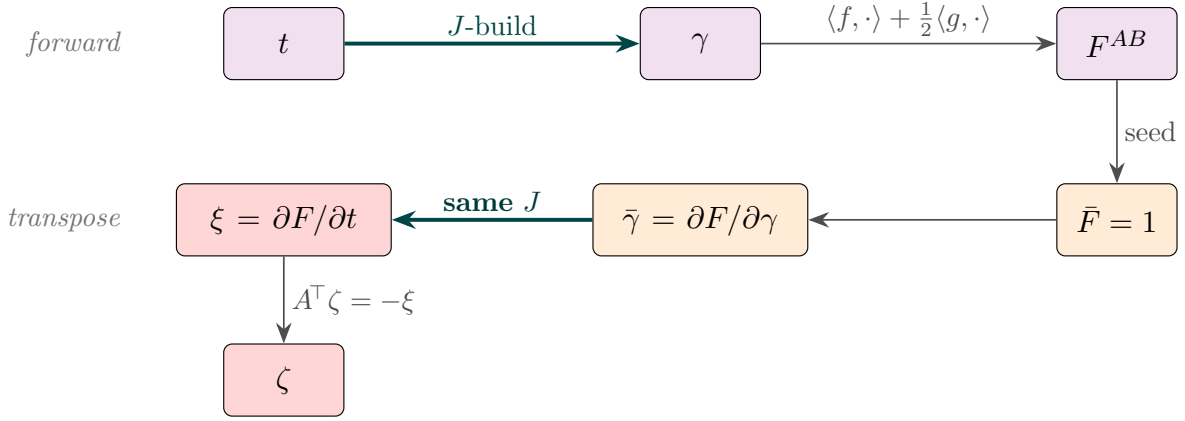
\begin{figure}[htbp]\centering
\resizebox{0.95\linewidth}{!}{%
\begin{tikzpicture}[>=Latex, font=\small, every node/.style={inner sep=4pt},
  bx/.style ={draw,rounded corners=3pt,minimum height=9mm,align=center},
  sm/.style ={bx,minimum width=15mm},
  wd/.style ={bx,text width=24mm},
  fwd/.style={fill=violet!12}, rev/.style={fill=orange!16}, res/.style={fill=red!16},
  je/.style ={-{Stealth[length=3mm]},line width=1.4pt,teal!55!black},
  e/.style  ={-{Stealth[length=2.6mm]},semithick,black!70},
  lbl/.style={font=\footnotesize,fill=white,inner sep=1.5pt},
  rl/.style ={font=\footnotesize\itshape,text=black!55,anchor=east}]
  %% forward row (top)
  \node[sm,fwd] (t) at (0,1.5)    {$t$};
  \node[sm,fwd] (g) at (5.2,1.5)  {$\gamma$};
  \node[sm,fwd] (F) at (10.4,1.5) {$F^{AB}$};
  %% transpose row (bottom)
  \node[wd,res] (xi) at (0,-0.7)    {$\xi=\partial F/\partial t$};
  \node[wd,rev] (gb) at (5.2,-0.7)  {$\bar\gamma=\partial F/\partial\gamma$};
  \node[sm,rev] (Fb) at (10.4,-0.7) {$\bar F=1$};
  \node[sm,res] (z)  at (0,-2.7)    {$\zeta$};
  %% forward edges
  \draw[je] (t)--node[lbl,above=1pt]{$J$-build}(g);
  \draw[e]  (g)--node[lbl,above=1pt]{$\langle f,\cdot\rangle+\tfrac12\langle g,\cdot\rangle$}(F);
  %% transpose edges
  \draw[e]  (F)--node[lbl,right=1pt]{seed}(Fb);
  \draw[e]  (Fb)--(gb);
  \draw[je] (gb)--node[lbl,above=1pt,text=teal!45!black]{\textbf{same}~$J$}(xi);
  \draw[e]  (xi)--node[lbl,right=1pt]{$A^{\!\top}\zeta=-\xi$}(z);
  %% row labels
  \node[rl] at (-1.9,1.5)  {forward};
  \node[rl] at (-1.9,-0.7) {transpose};
\end{tikzpicture}}
\caption{The relaxation is automatic differentiation applied to a density build:
reading the forward graph backwards produces the response equations
mechanically, the same way backpropagation produces gradients in machine
learning. \emph{Forward}
(violet): the build $t\!\to\!\gamma\!\to\!F^{AB}$. \emph{Transpose} (orange): its
reverse-mode pass seeds $\bar F\!=\!1$, propagates
$\bar\gamma=\partial F/\partial\gamma$, and reaches the source
$\xi=\partial F/\partial t$ through the \emph{same} $J$-build kernel (teal, both
directions) with the adjoint density as argument; the amplitude response then
solves $A^{\!\top}\zeta=-\xi$. Identical kernel signatures forward and backward
are why one GPU kernel (Sect.~\ref{sec:jk}) serves both.}
\label{fig:transpose}
\end{figure}

Equation~\eqref{eq:zeta} is not a separate derivation: it \emph{is} the forward
density DAG run backwards (Fig.~\ref{fig:transpose}). What makes this mechanical,
rather than a per-case insight, is that every node of $G_L$ is a single multilinear
contraction, and multilinear contraction has one adjoint rule. Each forward term
$y \mathrel{+}= c\,\mathrm{einsum}(\text{subs};x)$ has the adjoint
$\bar x \mathrel{+}= c\,\mathrm{einsum}(\text{subs}^{\!\ast};\bar y)$ obtained by
swapping the input and output subscript strings and accumulating into the input
adjoint --- the one operational rule that turns ``transpose $=$ relaxation'' into a
reproducible, mechanizable pass. Applied edge by edge to $G_L$ it sends every
forward operation to its adjoint ($J^{\top}\!=\!J$, $K^{\top}\!=\!K$,
back-transform $\leftrightarrow$ forward-transform, projection self-adjoint) and
turns the density build into the $\zeta$ solve. The result is \emph{structurally
guaranteed}, not fortuitous: reverse-mode differentiation of any DAG returns the
transposed Jacobian-vector product of that exact DAG, so the multiplier equation it
emits is necessarily $A^{\!\top}\zeta=-\xi$ with the same $A$ that the forward
build linearizes. The same construction runs unchanged for the interstate Z-vector
of the NACME (mixed $\hat L_A,\hat R_B$ in place of $\hat L_k,\hat R_k$) and for
EE/IP/EA at any truncation level, because all of them are the same forward graph
with different leaf tensors --- one reverse pass, never a family of hand
derivations.

That the adjoint rule applies cleanly term by term is a deliberate consequence of
how the working equations are emitted. The residuals are generated symbolically
(Sect.~\ref{sec:compdetails} below) as flat lists of single contraction terms,
each carrying its coefficient, integral tags, and amplitude tags. We disable the
compressed (paired) antisymmetrizers and keep only the plain transposition
$P(p,q)$, which the translator expands as $1-\text{swap}$ on the external axes.
Every emitted term is then one sign-explicit \texttt{einsum}, so the adjoint of
Eqs.~\eqref{eq:zeta} above follows from the subscript-swap rule alone, with no
antisymmetrizer sign convention to decode. Generated in this form, the same
single-term structure that builds the forward density DAG and the residuals
$\hat\Omega$ transposes, term for term, into the relaxation solve; emitting
transpositions rather than compressed antisymmetrizers is what keeps that
transpose mechanical end to end.

This term-level adjoint rule also delivers the single-kernel payoff that the title
asserts. Reading the rule on a J-contraction shows that a forward edge building an
intermediate from a density becomes a reversed edge building the adjoint density by
the \emph{same} contraction; only the density argument is swapped. Every relaxation
J/K build therefore shares the exact kernel signature of its forward counterpart,
so one ordered AO-LT J/K kernel (Sect.~\ref{sec:jk}) serves the forward densities
and the relaxation alike --- the kernel-reuse claim is a consequence of the adjoint
rule, established here by construction (its measured GPU throughput is deferred to
Sect.~\ref{sec:gpu}).

\begin{figure}[htbp]\centering
\resizebox{0.97\linewidth}{!}{%
\begin{tikzpicture}[>=Latex, font=\small, every node/.style={inner sep=3pt},
  data/.style={draw,rounded corners=2pt,fill=gray!12,align=center,minimum height=7mm},
  amp/.style ={draw,rounded corners=2pt,fill=blue!12,align=center,minimum width=11mm,minimum height=7mm},
  eig/.style ={draw,rounded corners=2pt,fill=teal!14,align=center,minimum height=7mm},
  dens/.style={draw,rounded corners=2pt,fill=green!16,align=center,minimum height=7mm},
  zv/.style  ={draw,dashed,rounded corners=2pt,fill=orange!16,align=center,minimum height=8mm},
  ker/.style ={draw,line width=1pt,rounded corners=2pt,fill=violet!14,align=center,minimum height=8mm},
  grad/.style={draw,rounded corners=2pt,fill=red!16,align=center,minimum height=7mm},
  a/.style ={-{Stealth[length=2.4mm]},semithick,black!72},
  ta/.style={-{Stealth[length=2.4mm]},dashed,semithick,orange!60!black},
  ll/.style={font=\scriptsize\itshape,text=black!55,anchor=east}]
  \node[data] (h)  at (-2.8,0) {$h,\,(pq|rs),\,h^x,\,S^x$};
  \node[data] (ep) at (2.8,0)  {$\varepsilon$};
  \node[amp] (t1) at (-4.8,-1.6) {$T_1$};
  \node[amp] (t2) at (-1.6,-1.6) {$T_2$};
  \node[amp] (t3) at (1.6,-1.6)  {$T_3$};
  \node[amp] (t4) at (4.8,-1.6)  {$T_4$};
  \node[eig] (R) at (-2.9,-3.2) {$\hat R_k=(r_1,r_2,r_3,r_4)$};
  \node[eig] (L) at (3.1,-3.2)  {$\hat L_k=(l_1,l_2,l_3,l_4)$};
  \node[dens] (d) at (0,-4.7) {$\gamma^{AB},\,\Gamma^{AB}$ \;(non-symmetric)};
  \node[zv] (z) at (-4.5,-6.1) {$\zeta=(\zeta_1,\dots,\zeta_4)$\\[1pt]\scriptsize$A^{\!\top}\zeta=-\xi$};
  \node[ker] (k) at (1.4,-6.1) {AO-LT J/K kernel\\[1pt]$\sum_\alpha w_\alpha[2J^x{-}K^x]$};
  \node[grad] (g) at (1.4,-7.7) {gradient $dE_k/dx$ \;/\; NACME $\lambda^{AB}_x$};
  \draw[a] (h)--(t1); \draw[a] (h)--(t2); \draw[a] (ep)--(t3); \draw[a] (ep)--(t4);
  \foreach \x in {t1,t2,t3,t4}{\draw[a] (\x)--(R); \draw[a] (\x)--(L);}
  \draw[a] (R)--(d); \draw[a] (L)--(d);
  \draw[ta] (d)--node[above=1pt,font=\scriptsize,fill=white,inner sep=1pt,text=orange!55!black]{transpose}(z);
  \draw[a] (d)--(k);
  \draw[ta] (z)--(k);
  \draw[a] (k)--(g);
  \node[ll] at (-7.3,0)    {integrals};
  \node[ll] at (-7.3,-1.6) {CCSDTQ $T$};
  \node[ll] at (-7.3,-3.2) {EOM $L,R$};
  \node[ll] at (-7.3,-4.7) {densities};
  \node[ll] at (-7.3,-6.1) {transpose / kernel};
  \node[ll] at (-7.3,-7.7) {gradient / NACME};
\end{tikzpicture}}
\caption{Computation graph (DAG) for the EOM-CCSDTQ excited-state gradient and
NACME. Solid edges are the forward pass; orange dashed edges are the reverse-mode
transpose that produces the amplitude response $\zeta$ (Fig.~\ref{fig:transpose}).
The CCSDTQ amplitudes $T_1$--$T_4$ are converged once ($D_kT_k=-R_k[T]$); for each
target state the biorthonormal $\hat R_k,\hat L_k$ follow from the native two-sided
non-Hermitian Davidson (Sect.~\ref{sec:eom}). The non-symmetric transition densities
$\gamma^{AB},\Gamma^{AB}$ and their transpose $\zeta$ both feed the \emph{single}
AO-LT J/K kernel, which assembles the gradient or NACME; every J/K build---forward
or transpose, symmetric or non-symmetric---reuses the one kernel of
Sect.~\ref{sec:jk}.}
\label{fig:eomdag}
\end{figure}
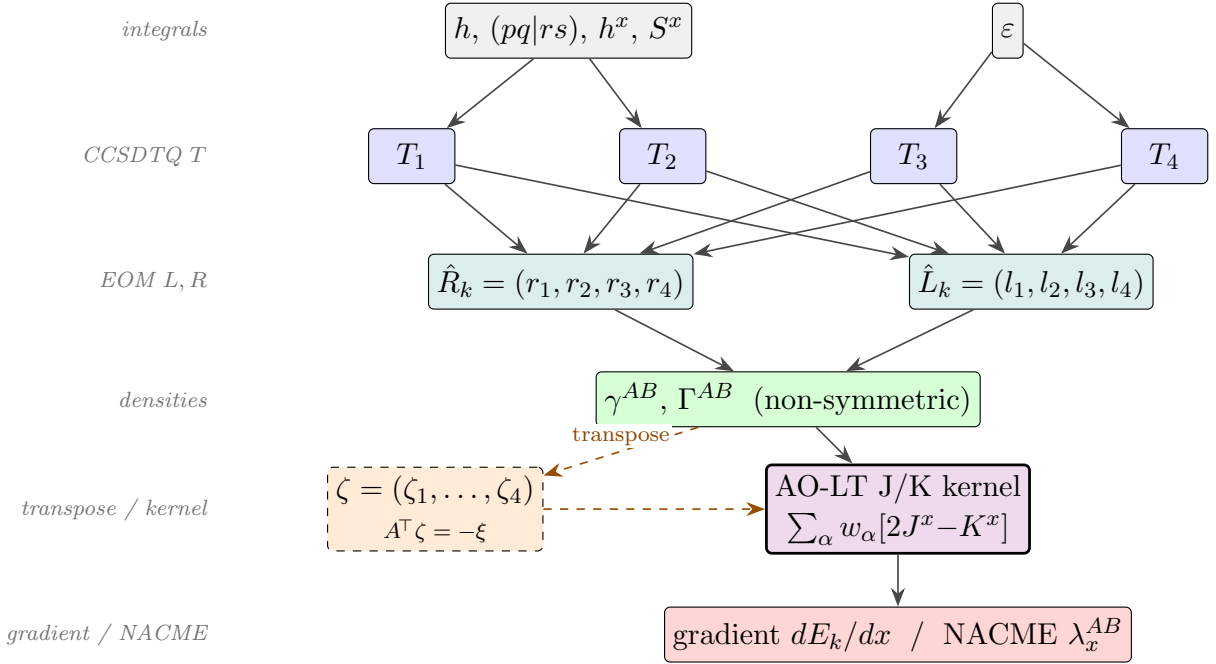

Figure~\ref{fig:eomdag} is Algorithm~\ref{alg:ltao} drawn as the full EOM-CCSDTQ
computation graph: the forward pass (solid) is lines~\ref{ln:fwdstart}--\ref{ln:fwdend},
the reverse-mode transpose (dashed) is lines~\ref{ln:revstart}--\ref{ln:revend},
and both feed the single AO-LT J/K kernel that returns the gradient or NACME.

%% ---------------------------------------------------------------------
\subsection{Non-symmetric J/K for transition densities}
\label{sec:jk}
The LT-AO kernel of Sect.~\ref{sec:lt} is written for two distinct density
arguments because the EOM transition densities demand it. We close the theory by
making that requirement explicit and reducing the two-particle contraction to the
one-particle J/K builds the kernel actually performs.

The two-particle gradient term is
$\tfrac12\sum_{pqrs}\Gamma^{AB}_{pqrs}\,(pq|rs)^x$. To carry it through the
one-particle kernel we cumulant-decompose the two-particle (transition) density
into its separable (mean-field) part and the connected cumulant,
\begin{equation}
\Gamma^{AB}_{pqrs}
 =\underbrace{\gamma^{AB}_{pq}\gamma^{AB}_{rs}
   -\tfrac12\,\gamma^{AB}_{ps}\gamma^{AB}_{rq}}_{\gamma\wedge\gamma}
 \;+\;\Gamma^{\mathrm{c},AB}_{pqrs},
\label{eq:cumulant}
\end{equation}
written in the spatial closed-shell convention of the Notation (the $\tfrac12$
exchange coefficient is the closed-shell spin-traced form of the unit-coefficient
spin-orbital cumulant). The separable part collapses \emph{exactly} onto the
ordered one-particle derivative builds of Eq.~\eqref{eq:jkderiv}, with the
exchange and functional factors kept explicit,
\begin{equation}
\tfrac12\sum_{pqrs}(\gamma\wedge\gamma)_{pqrs}\,(pq|rs)^x
 = \tfrac12 J^x(A,B)-\tfrac14 K^x(A,B)
 = \tfrac12\big[J^x(A,B)-\tfrac12K^x(A,B)\big],
\label{eq:cumcollapse}
\end{equation}
where the leading $\tfrac12$ is the functional prefactor of
$\tfrac12\sum\Gamma^{AB}(pq|rs)^x$, the inner $\tfrac12$ on $K^x$ is the exchange
coefficient, and $J^x,K^x$ are the bare builds of Eq.~\eqref{eq:jkderiv}. The two
density factors $A,B$ here are spin-summed \emph{total} densities; the same
separable build appears as $2J^x-K^x$ on the per-spatial (half) densities
$\mathcal D^{(1)},\Xi$ of the Laplace-grid gradient Eq.~\eqref{eq:ltgrad} (the
factor $2$ is the closed-shell spin sum, $2J-K=2(J-\tfrac12K)$), and as
$J^x(\gamma,\gamma)-\tfrac12K^x(\gamma,\gamma)$ on the total relaxed density
$\gamma$ in the master gradient Eq.~\eqref{eq:master}. This single chain fixes the
normalization across Eqs.~\eqref{eq:cumulant}, \eqref{eq:cumcollapse},
\eqref{eq:master}, and \eqref{eq:ltgrad}. The decomposition leaves only a small
active-only cumulant remainder
$\tfrac12\sum_{pqrs}\Gamma^{\mathrm{c},AB}_{pqrs}(pq|rs)^x$, so a single J/K kernel
covers the separable part of every method's two-particle term, the active-only
remainder handled separately.
The two one-particle factors $A,B$ in Eq.~\eqref{eq:cumcollapse}
are \emph{distinct} objects---in the Laplace-grid assembly they are the level-1
forward density $\mathcal D^{(1;\alpha_1)}$ (bra) and the combined
forward+backward ket $\Xi^{(L;\boldsymbol\alpha)}$ of Eq.~\eqref{eq:xidef}---and
each is itself non-symmetric because $L\neq R^{\dagger}$. Hence $A\neq B$,
$J^x(A,B)\neq J^x(B,A)$, and neither the bra/ket swap nor the antisymmetric part
$\tfrac12(A-A^{\top})$ may be dropped: a symmetric-only kernel retains
$\tfrac12(A+A^{\top})$ and silently discards the rest, corrupting the gradient.
The symmetric ground-state J/K is the special case $A=B$. At the MP2 level
($L=1$) the two factors coincide and Eq.~\eqref{eq:cumcollapse} reduces to the
symmetric $J^x(\mathcal D,\mathcal D)$ form of Ref.~\citenum{Hohenstein2015};
only at $L\ge2$ (MP3/CCSD and above), and for every EOM transition density, is the
genuinely ordered $J^x(\mathcal D^{(1)},\Xi^{(L)})$ build required---which is why a
tuned, ordered kernel is the object of the GPU work below.

%% =====================================================================
\section{Computational Details}
\label{sec:compdetails}
\paragraph{DAG-driven, spill-bounded GPU chunking.}
A naive GPU port of the non-symmetric J/K spills registers and shared memory and
falls far below the roofline (the peak throughput achievable for a kernel of its
arithmetic intensity)---or exceeds the 8\,GB budget---on commodity hardware.
The contraction DAG resolves this: its graph structure is used to split each
kernel into chunks sized to the device register and shared-memory limits, so the
build stays compute/bandwidth-bound rather than spill-bound. The same DAG that
emits the working equations and their transpose (Sect.~\ref{sec:dag}) thus also
emits the contraction schedule and its chunking---one abstraction, both payoffs.

Concretely, the HGP-OS ERI and derivative recurrence is itself a DAG (a
vertical-recurrence tower feeding horizontal raises), and two codegen regimes
are keyed to its size. For low angular momentum ($s/p/d$) the DAG is
\emph{unrolled}---fully inlined---which is fast but explodes the live register
set: a $(dd|dd)$ quartet is $7623$ nodes / $1296$ outputs and spills
$\sim\!96$\,KB per thread, while an $(ffff)$ quartet ($\sim\!78{,}000$ nodes,
$10.4$\,MB of source) times out \texttt{ptxas} ($>1500$\,s) and is effectively
uncompilable. For $f$ shells and above the kernel is instead \emph{looped}: the
ladder structure of the DAG is encoded as integer index tables walked by runtime
loops, so source size is $O(\text{tables})$ rather than $O(\text{nodes})$ and the
result is machine-$\varepsilon$ equal to the unrolled path through $(ffff)$.
Where the unrolled register pressure still dominates, the output cone is
partitioned. Output-tiling cuts the DAG into sub-DAGs capped at $\sim\!2048$
outputs ($(dd|dd)$ spill $96\!\to\!34$\,KB, $\sim\!1.55\times$); frontier
(antichain) chunking bounds the per-chunk live frontier but the $(dd|dd)$ peak
simultaneous-live antichain ($\sim\!2172$) is irreducible, so the per-chunk
frontier is genuinely \emph{not} the peak antichain and output-tiling wins. The
heaviest cc-pVTZ-class buckets exceed the $512$\,KB per-thread local limit; their
working set is placed in an explicit HBM scratch slab indexed by a fixed
thread-pool slot, with no \texttt{\_\_syncthreads} and no grid synchronization,
so the threads stay independent and the kernel scales linearly (constant
$\sim\!220$\,MB scratch and $\sim\!50\,\mu$s/quartet as $n_{\text{quartet}}$ runs
$512\!\to\!4096$). NVIDIA Nsight Compute showed these $f$ and gradient
global-scratch kernels to be occupancy-starved ($2$--$4\%$ of SM issue), not
compute-bound; a shared $\sim\!3$\,GB scratch pool with a small block size and
grid-stride slot reuse then gave $2.2$--$4.5\times$. The \emph{limiter} reported
by Nsight, not \texttt{ptxas}+wall-clock, drove the partitioning choice
throughout.

Each thread evaluates one contracted ERI quartet $(\mu\nu|\kappa\rho)$ through
the HGP-OS recurrence and scatters it into the Fock matrices it contributes to;
in the $J/K$ build of Eq.~\eqref{eq:jkbuild} this per-quartet scatter is
\begin{equation}
J_{\mu\nu}\mathrel{+}=w_{\mu\nu\kappa\rho}\,(\mu\nu|\kappa\rho)\,\mathcal D_{\kappa\rho},
\qquad
K_{\mu\kappa}\mathrel{+}=w_{\mu\nu\kappa\rho}\,(\mu\nu|\kappa\rho)\,\mathcal D_{\nu\rho},
\label{eq:quartet_scatter}
\end{equation}
where $w_{\mu\nu\kappa\rho}$ is the canonical-fold multiplicity of the quartet
and the accumulation is by \texttt{atomicAdd} into the global Fock matrices, so
threads need neither cooperate nor synchronize.

\begin{figure*}[htbp]\centering
\includegraphics[width=\linewidth]{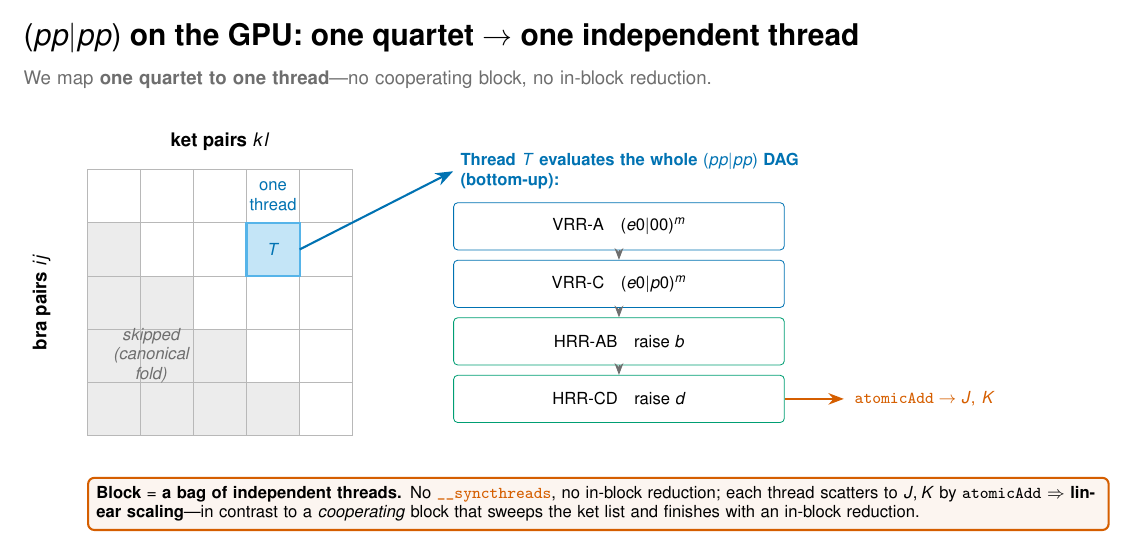}
\caption{The HGP-OS recurrence kernel: one CUDA thread evaluates one quartet's
full DAG (VRR-A~$\to$~VRR-C~$\to$~HRR-AB~$\to$~HRR-CD) over a grid-stride pool with
no barriers, so threads never synchronize and the kernel scales linearly. Low
angular momentum ($s/p/d$) is inlined into registers with output-tiling; high
angular momentum ($f$ and up) is a table-walked loop whose per-thread working set
lives in a dedicated HBM global-scratch slot. Each thread evaluates the
per-quartet scatter of Eq.~\eqref{eq:quartet_scatter}. Only the realized
(bra~$\times$~pivot) buckets are launched. This contrasts with the
cooperating-block, ket-sweeping map of the TeraChem engine~\cite{Wang2024}
(Mart\'inez group), on which the AO-direct $J/K$ formulation here builds.
Measured on production RHF/CD-RHF workloads; the identical
non-symmetric transition-density path (\texttt{compute\_densities\_cd\_nonsym}
with the \texttt{k\_G3combine\_nonsym}/\texttt{k\_G2combine\_nonsym} combine
kernels) serves the EOM-CC response
(Sect.~\ref{sec:gpu}).}
\label{fig:cdrecur}
\end{figure*}

The thread-to-work map underlying these regimes (Fig.~\ref{fig:cdrecur})
distinguishes this kernel from prior GPU integral engines: where those map a
cooperating thread block to one quartet, the present kernel maps \emph{one thread
to one quartet} over a grid-stride pool, externalizing each thread's working set
to an HBM global-scratch slab so the threads never cooperate or synchronize, with
high-$L$ relief from the two DAG-size-keyed codegen regimes above (unrolled with
output-tiling for $s/p/d$, looped integer-index-table walk with external scratch
for $f$ and up) rather than from per-output kernel fission (the execution-model
contrast with per-quartet GPU integral kernels is drawn in full below). The
absence of \texttt{\_\_syncthreads} in the looped path is what makes the threads
fully independent and the kernel scale linearly, and it is the same graph
abstraction that, transposed, emits the response skeletons of Sect.~\ref{sec:dag}.

\begin{figure*}[htbp]\centering
\includegraphics[width=\linewidth]{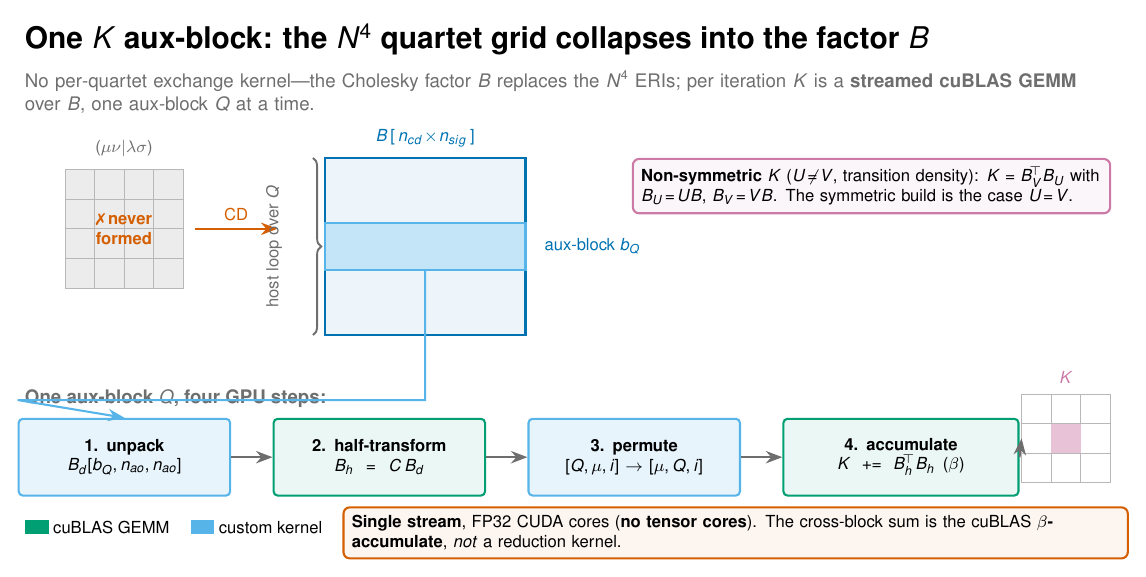}
\caption{Coulomb ($J$) and exchange ($K$) energy build as cuBLAS
contraction-DAGs over the compact Cholesky factor $B$, with the $N^4$ ERI tensor
never formed. $J$ is two bandwidth-bound FP64 GEMVs over $B$; $K$ streams the
auxiliary index in $\sim\!256$\,MB blocks as an FP32 GEMM cascade
(unpack~$\to$~half-transform~$\to$~permute~$\to$~$K\!\mathrel{+}=\!B_h^{\!\top}B_h$).
The transition-density (EOM, $L\ne R$) variant uses two distinct half-transforms
$B_U=UB$, $B_V=VB\Rightarrow K=B_V^{\!\top}B_U$, of which the symmetric
ground-state build is the special case $U=V$. The panel evaluates
Eqs.~\eqref{eq:cdfact}--\eqref{eq:cdknonsym} ($J$, Eq.~\eqref{eq:cdj}; symmetric
$K$, Eq.~\eqref{eq:cdk}; non-symmetric $K$, Eq.~\eqref{eq:cdknonsym}). Markers:
\textbullet{}~symmetric / ground-state density, $\blacktriangle$~non-symmetric /
transition density; orange denotes FP64, blue FP32.
Measured on production RHF/CD-RHF workloads; the identical
non-symmetric transition-density path (\texttt{compute\_densities\_cd\_nonsym}
with the \texttt{k\_G3combine\_nonsym}/\texttt{k\_G2combine\_nonsym} combine
kernels) serves the EOM-CC response
(Sect.~\ref{sec:gpu}).}
\label{fig:cdenergy}
\end{figure*}

\paragraph{J and K as the shared computational primitive.}
The Coulomb ($J$) and exchange ($K$) builds and their nuclear gradients are the
universal primitive of this backend: the \emph{same} kernels that assemble the
Hartree--Fock reference (the SCF) also assemble the skeleton/derivative builds
$J^x(A,B)$, $K^x(A,B)$ of Eq.~\eqref{eq:jkderiv} that the response gradients
contract. Characterizing the DAG-generated $J/K$ kernel on a production RHF
workload therefore directly characterizes the same kernel object used for the
EOM-CC response skeletons; this is what licenses the RHF and Cholesky timing and
robustness data below to stand in for the response build. In particular the
production launcher exposes \texttt{build\_K\_screened} (the symmetric $L=R$ case,
the $U=V$ special case of Fig.~\ref{fig:cdenergy}) and the non-symmetric
transition-density path \texttt{compute\_densities\_cd\_nonsym} (with the
\texttt{k\_G3combine\_nonsym} (\texttt{\_blk}/\texttt{\_blk\_f}) and
\texttt{k\_G2combine\_nonsym} combine kernels) for the
transition-density ($L\ne R$) exchange,
$K[\mu\nu]=\sum_{Qm}\tilde U_t[\mu,Qm]\,\tilde V_t[\nu,Qm]$ with distinct
left/right factors $U\ne V$ (the symmetric build is the $U=V$ special case)---the
GPU realization of the ordered $J^x(A,B)\ne J^x(B,A)$ build derived in
Sect.~\ref{sec:jk}, here measured.
We state the scope explicitly: the GPU data in this section characterize the
\emph{shared $J/K$ kernel object} as exercised on production RHF and CD-RHF
workloads at the $\sim\!730$-AO scale---representative of the chromophore systems
for which this backend was engineered---whereas the EOM-CC response builds invoke
the \emph{identical} kernel entry points
(\texttt{compute\_densities\_cd\_nonsym} and its
\texttt{k\_G3combine\_nonsym}/\texttt{k\_G2combine\_nonsym} combine kernels), so the kernel
under test is the same object in both settings. As a kernel-level in-method
characterization---distinct from a full excited-state run---we wall-time the
non-symmetric transition-density gradient \emph{assembly} on the same RTX~4060,
driven by a surrogate (model) transition density rather than by amplitudes from a
converged chromophore-scale EOM solve; this is reported in
Sect.~\ref{sec:gpu} and times the assembly kernels themselves, not an end-to-end
EOM-CCSD gradient (that full gradient is delivered at chromophore scale in
Sect.~\ref{sec:deveom}). It anchors the
characterization here at the level of the kernels actually invoked.

A second design choice underlies these builds: rather than form the four-index
ERI tensor per quartet, the backend Cholesky-decomposes the two-electron operator
into a compact factor $B$ and reads $J$ and $K$ off $B$ as cuBLAS contraction
DAGs (Fig.~\ref{fig:cdenergy})---the $N^4$ tensor is never materialized. The
factorization writes the ERI as
\begin{equation}
(\mu\nu|\kappa\rho)\approx\sum_Q B_{Q,\mu\nu}\,B_{Q,\kappa\rho},
\label{eq:cdfact}
\end{equation}
and the Coulomb build then reads off $B$ as two contractions over the auxiliary
index $Q$,
\begin{equation}
J_{\mu\nu}=\sum_Q B_{Q,\mu\nu}\,\gamma_Q,\qquad
\gamma_Q=\sum_{\kappa\rho}B_{Q,\kappa\rho}\,\mathcal D_{\kappa\rho},
\label{eq:cdj}
\end{equation}
while the symmetric exchange build factors through the half-transform
$B^{h}_{Q,\mu i}=\sum_\kappa C_{\kappa i}\,B_{Q,\mu\kappa}$ onto the occupied MO
coefficients $C$ (an $n_{\mathrm{ao}}\!\times\!n_{\mathrm{occ}}$ block),
\begin{equation}
K_{\mu\nu}=\sum_{Q,i}B^{h}_{Q,\mu i}\,B^{h}_{Q,\nu i}.
\label{eq:cdk}
\end{equation}
For the non-symmetric (transition/interstate, $L\ne R$) density the left and
right half-transforms differ, $B^{U}=U\!\cdot\!B$ and $B^{V}=V\!\cdot\!B$, giving
\begin{equation}
K_{\mu\nu}=\sum_{Q,m}B^{U}_{Q,\mu m}\,B^{V}_{Q,\nu m},
\label{eq:cdknonsym}
\end{equation}
of which the symmetric build of Eq.~\eqref{eq:cdk} is the special case $U=V$.
This compact-factor-$B$ machinery adapts the GPU $J/K$-over-pseudodensity
construction of Hohenstein \emph{et~al.}~\cite{Hohenstein2015}, part of the broader
low-rank two-electron-factorization program of Hohenstein, Parrish, and
Mart\'inez~\cite{Hohenstein2012}: reading $J$ and $K$ off a low-rank factor of the
two-electron operator rather than from the explicit $N^4$ tensor is the
methodological link between that Stanford-lineage machinery and the present
excited-state-response application, in which the same factored contraction is driven
by left/right transition densities $U\ne V$. What the present work contributes on
top of that inheritance is the
ordered, non-symmetric build itself: the \emph{same} cuBLAS GEMM/GEMV schedule
serves the symmetric SCF $J/K$ ($U=V$), the non-symmetric transition-density build
$J^x(A,B)\ne J^x(B,A)$ that the non-Hermitian response densities force, and
(below) the gradient---all within $8$\,GB.

\paragraph{Hardware, software, and the FP32-mixed boundary.}
All GPU measurements were performed on a single consumer NVIDIA GeForce RTX~4060
(Ada, \texttt{sm\_89}, $8$\,GB), whose FP64 throughput is $\approx\!1/64$ of its
FP32 rate; the $J/K$ kernels are generated as CUDA from the contraction DAG and
evaluate the AO Gaussian integrals with established GPU integral
methods~\cite{Hohenstein2015,Wang2024}. Kernels are built with
\texttt{-arch=sm\_89 -O3 -lineinfo -use\_fast\_math}; register usage and
occupancy were read from \texttt{ptxas -v} and NVIDIA Nsight Compute. The
reference (CPU) verification stack uses a spin-orbital implementation and a C
backend driven from Python. Precision is mixed along a deliberate boundary. The
integral-derivative and exchange builds run in FP32, while the
Cholesky/CD-vector build, the two-center gradient term, the gradient densities,
and the two-center metric stay FP64. The Coulomb ($J$) build is
\emph{genuinely} FP64: it is bandwidth-bound (the roofline of
Sect.~\ref{sec:gpu}), so FP32 would buy nothing. ``FP32-mixed'' thus denotes $K$ and the three-center gradient
in FP32 with $J$, the two-center gradient, and all densities in FP64. This
density-precision statement holds for the symmetric ground-state path; the
non-symmetric packed-response build is the one exception---there the
transition-density V-transform GEMMs (the two exchange $K$-sandwiches and the
$B\!\cdot\!W$, $B^{\!\top}\!\cdot\!\mathrm{bg}K$ contractions) run FP32, and only
the metric core (the pivots $B_{\text{piv}}$, the pseudo-inverse $J^{+}$, the
half-transform $W$, the metric-folded $\Phi/\Psi$, and the two-center assembly)
stays FP64. A further enabling rule, given the $1/64$ FP64 penalty, evaluates the
Boys special function in FP32 inside an otherwise-FP64 kernel. Because the finite-difference gates of
Table~\ref{tab:fd} exercise the dense FP64 reference contraction rather than this
FP32-mixed GPU path, the precision boundary is validated separately: the
FP32-mixed kernels reproduce their own FP64 build to machine-$\varepsilon$ and
agree with our direct four-center Cartesian ERIs to $\le10^{-7}$ (the same
bit-exact gate detailed in the screening-robustness paragraph below), so the
boundary is measured, not assumed.

\paragraph{FP32 is the precision floor.}
FP32 is also the \emph{floor}: the TF32 tensor-core path sits below it and we do
not use it for the amplitudes the response is built on. Running the ground-state
CCSD particle-particle ladder---the $O(o^2v^4)$ GEMMs whose converged amplitudes
seed the EOM-CC amplitude response and the interstate Z-vector---in TF32 rather
than FP32 perturbs the correlation energy of Mg-porphine (def2-SVP) by
$1.5\times10^{-3}$\,Ha, at the edge of chemical accuracy, and the doubles by
$7\times10^{-4}$ in relative Frobenius norm (Section~S8). The relative
perturbation is $\approx\!4.3\times10^{-4}$ and, across the two systems we tested,
nearly size-independent, so the absolute error tracks $|E_{\mathrm{corr}}|$ and
reaches chemical accuracy only at chromophore scale---a small-molecule proxy would
hide it. The energy is a necessary-but-not-sufficient witness of amplitude quality: it
is blind to the virtual-tail amplitudes TF32 truncates first and that gradients
and small-gap NACMEs weight most heavily, where its data-dependent rounding could
inject geometry-dependent jitter into the potential surface (argued, not scanned;
Section~S8). We therefore keep these amplitudes in FP32 and use no tensor cores. The
temptation is real and portable: cuBLAS does not auto-dispatch the FP32 tensor
path here, but an explicit CUTLASS tensor-op ladder GEMM runs $\sim\!1.3$--$1.7
\times$ faster on this $4060$ ($\sim\!10\%$ end-to-end) and several-fold faster on
datacenter parts (A100/H100), where TF32 is the first optimization a reader would
reach for---which is precisely why we document that the banked seed amplitudes are
where it must \emph{not} be applied.

\paragraph{Bounded-footprint memory discipline.}
Device memory is claimed and released in small increments, so resident usage
tracks a bounded sawtooth even for the largest systems ($\sim\!730$ AO on the
$8$\,GB card). This is the composition of aux-block chunking that bounds the
transient dense slab to $\sim\!256$\,MB, a growable bump-pool reused (not churned)
across buckets, phase-boundary releases (the $J/K$ buffers are freed after the SCF
and before the gradient; the four-center-ERI build pool is freed once the
Cholesky factor exists), JIT cubin release after the build, and never-dense
invariants---only the buckets actually launched are materialized, the
packed/screened layout is $O(n_{\text{sig}}\,n_{\text{aux}})$ rather than
$O(n_{\text{ao}}^2 n_{\text{aux}})$, and $K$ unpacks one aux-block at a time so
the dense $[n_{\text{aux}},n_{\text{ao}},n_{\text{ao}}]$ tensor is never resident.
For the $730$-AO system the peak vs.\ resident footprint is $\sim\!4.95$ vs.\
$\sim\!4.83$\,GB. This ceiling is for the symmetric ground-state gradient; the
non-symmetric response gradient carries a higher constant---the packed
transition-density relaxed density $G_3^{\,p}[n_{\text{sig}},n_{\text{cd}}]$ plus a
per-block scratch footprint $\sim\!2\times$ the symmetric path (its two
$K$-sandwiches and a $\Phi$-sandwich each need their own buffers, halving the
aux-block). Even with this higher constant the response-gradient \emph{assembly}
is feasible for all eleven benchmark systems on the $8$\,GB card, because the
device footprint tracks the Cholesky/integral working set---the maximum angular
momentum $L_{\max}$, the Cholesky rank $n_{\mathrm{cd}}$, and the significant-pair
count $n_{\mathrm{sig}}$---rather than the raw AO count. The largest system by AO
count, chlorophyll~a/pcseg-0 ($730$~AO but $s,p$-only, $L_{\max}=1$, and
CD-sparse), therefore carries a \emph{smaller} device peak than the f-shell-dense
decamethylferrocene/def2-SVP ($486$~AO, $L_{\max}=3$, $n_{\mathrm{sig}}=163{,}954$),
the tightest case. The two largest by AO count (chlorophyll~a and
dna\_strand\_4mer, $752$~AO at $\delta=10^{-3}$) fit through the device-resident
$G_3^{\,p}$-eviction path that
host-snapshots the FP64 $B$ and its FP32 mirror to free the device for the
per-pivot kernels. This assembly-feasibility question---the footprint of the
gradient build---is distinct from the ground-state CCSD-\emph{solve} $t_2$
residency ceiling (Sect.~\ref{sec:gpu}; SI Section~S8), which only 6 of the 11
systems clear. Both are fixed memory limits, not algorithmic ones.

\begin{figure*}[htbp]\centering
\includegraphics[width=\linewidth]{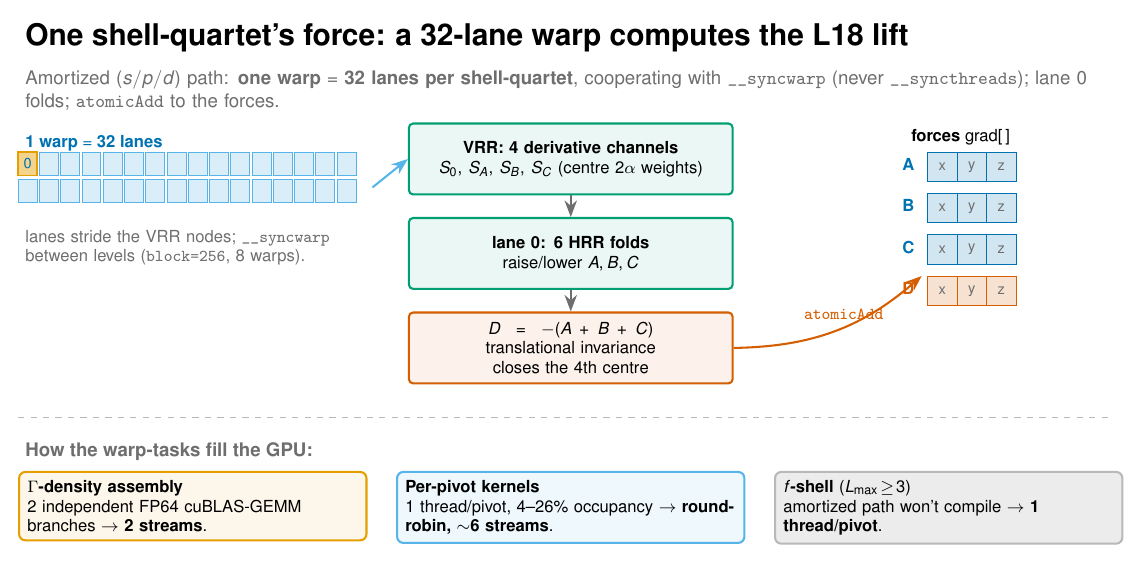}
\caption{Gradient execution. The relaxed-density assembly is a forward
($\gamma\!\to\!c\!\to\!h\!\to\!g$) plus energy-weighted
($W_q\!\to\!U\!\to\!V$) DAG of cuBLAS GEMMs whose two topologically
independent branches run on concurrent streams; in the production FP32-mixed path
the transition-density V-transform GEMMs run FP32 while the metric core (pivots,
pseudo-inverse, half-transform, and two-center assembly) stays FP64. Their output
feeds derivative-ERI kernels built by the L18 single-index lift (channels
$S_0,S_A,S_B,S_C$; the fourth centre closed by $D=-(A+B+C)$,
Eq.~\eqref{eq:l18}), with the three-center derivative in FP32 and the two-center
derivative in FP64. Production evaluates these through a warp-per-quartet
\emph{amortized} 3c/2c kernel (one warp per quartet, \texttt{\_\_launch\_bounds\_\_(256,3)},
$\sim\!50\%$ occupancy), $6.4\times$/$3.8\times$ over the earlier grid-underfilled
per-pivot launch and bit-identical in the forces; the per-pivot $\sim\!6$-stream
launcher is retained as the FP64 parity oracle. The build evaluates the relaxed
two-particle contraction Eq.~\eqref{eq:cdgrad} and reuses the $J^x/K^x$ kernel
object of Eq.~\eqref{eq:jkderiv}, so the symmetric and non-symmetric
($J^x(A,B)\ne J^x(B,A)$) gradients differ only in the densities fed to the shared
derivative kernel---the non-symmetric density assembly itself differs (the
$\tfrac12(G_3\!+\!G_3^{\!\top})$ symmetrization for the canonical-pair fold and
two metric-folded $K$-sandwiches). Measured on production RHF/CD-RHF workloads;
the identical non-symmetric transition-density path
(\texttt{compute\_densities\_cd\_nonsym} with the
\texttt{k\_G3combine\_nonsym}/\texttt{k\_G2combine\_nonsym} combine kernels) serves
the EOM-CC response
(Sect.~\ref{sec:gpu}).}
\label{fig:cdgrad}
\end{figure*}

\paragraph{Gradient execution on the same spine.}
The nuclear gradient runs as a two-stage DAG on the same kernel object
(Fig.~\ref{fig:cdgrad}). A forward ($\gamma\!\to\!c\!\to\!h\!\to\!g$) and an
energy-weighted ($W_q\!\to\!U\!\to\!V$) relaxed-density branch assemble the
relaxed densities; being topologically independent in the DAG, the two branches
run on concurrent CUDA streams. In the production FP32-mixed path the
V-transform GEMMs of the transition-density build---the two exchange
$K$-sandwiches and the $B\!\cdot\!W$ and $B^{\!\top}\!\cdot\!\mathrm{bg}K$
contractions---run in FP32, while the metric core (the Cholesky pivots
$B_{\text{piv}}$, the pseudo-inverse $J^{+}$, the half-transform $W$, the
metric-folded $\Phi/\Psi$, and the two-center assembly) stays FP64. Their output
feeds derivative-ERI kernels generated by the L18 single-index lift (four VRR
channels $S_0,S_A,S_B,S_C$, with the fourth centre closed by $D=-(A+B+C)$); the
three-center derivative kernel is FP32 while the two-center derivative and the
two-center metric stay FP64. The production launcher evaluates these through a
warp-per-quartet \emph{amortized} 3c/2c kernel---one warp per quartet, the VRR
body run lane-parallel---under \texttt{\_\_launch\_bounds\_\_(256,3)} ($3$
blocks/SM, $\sim\!85$ registers/thread, $\sim\!50\%$ occupancy), which fills the
grid that the earlier per-pivot launch left underfilled. The $N\!=\!3$ bound is a
sweep optimum (the three-center kernel runs $9.05\!\to\!7.75$\,s, $-14\%$;
$N\!\ge\!6$ hits a register-spill cliff), giving $6.4\times$ on the three-center
build ($31.2\!\to\!4.85$\,s) and $3.8\times$ on the two-center build at the
$f$-shell scale (ferrocene/def2-SVP), bit-identical in the forces to the per-pivot
path. The per-pivot $\sim\!6$-stream launcher is
retained, now as the FP64 parity oracle: FP64 gradients run per-pivot, FP32
production runs amortized. The two-particle part of this build is the relaxed contraction
\begin{equation}
\frac{dE}{dX}=\sum_{\mu\nu\kappa\rho}\Gamma_{\mu\nu\kappa\rho}\,
 \frac{\partial(\mu\nu|\kappa\rho)}{\partial X},
\label{eq:cdgrad}
\end{equation}
with $\Gamma$ the relaxed two-particle density, assembled through the ordered
$J^x/K^x$ skeleton builds of Eq.~\eqref{eq:jkderiv}. The centre derivatives
$\partial(\mu\nu|\kappa\rho)/\partial X$ are supplied by the L18 single-index
lift, which raises one bra/ket index at a time to give the three independent
centre channels, the fourth closed by translational invariance,
\begin{equation}
\frac{\partial}{\partial D}=-\Big(\frac{\partial}{\partial A}
 +\frac{\partial}{\partial B}+\frac{\partial}{\partial C}\Big).
\label{eq:l18}
\end{equation}
The gradient thus reuses the contraction-graph spine end to end: the same
$J^x/K^x$ builds of Eq.~\eqref{eq:jkderiv}, fed the relaxed densities, with the
$\Gamma$-density GEMM DAG supplying their arguments.

\paragraph{Geometry-frozen caches for per-step reuse.}
The pivoted-metric pseudo-inverse $J^{+}=(B_{\text{piv}}^{\!\top}B_{\text{piv}})^{+}$
is a function of the Cholesky pivots alone---independent of the transition
densities---so it is held device-resident and reused, by a single
device-to-device copy, across the many transition densities of one nuclear
geometry. This skips a cuSOLVER eigensolve that is the single dominant per-call
GPU cost ($\approx\!7.7$\,s/call on Mg-porphine). The cache is shared between the
symmetric and non-symmetric paths (the same pivots give a bit-identical $J^{+}$)
and is invalidated whenever the Cholesky factor is rebuilt---once per geometry---so
it is geometry-optimization- and dynamics-safe; cache-on reproduces cache-off
bit-identically, being a copy of the same buffer. A second, host-side cache reuses
the geometry-frozen Schwarz/pivot screening, a marginal add-on by comparison. The
combined repeated-call speedup is $1.716\times$ (Mg-porphine) and $1.799\times$
(dna nucleoside), almost entirely from the metric cache. This is the regime an
excited-state molecular-dynamics trajectory or a multi-state response runs in---many
transition densities at one fixed geometry against the same factor $B$---and is
the cache ladder of Fig.~\ref{fig:response}(b).

\paragraph{Device-resident CCSD amplitudes on the same card.}
The response above consumes converged ground-state CC amplitudes, whose own
$O(o^2v^4)$ solve is the heaviest single object the pipeline touches. We run the
closed-shell CD-RCCSD solve \emph{fully device-resident} on the same $8$\,GB card
under one organizing rule: \emph{never materialize a dense $O(n_{\mathrm{mo}}^4)$
intermediate}. Only the rank-3 CD factors $B^Q$ ($\sim\!1.6$\,GB) and the doubles
$t_2[o^2v^2]$ ($\sim\!3.7$\,GB, FP32) stay resident; no dense
\texttt{oooo}/\texttt{ovov}/\texttt{oovv}/\texttt{ovvv}/\texttt{vvvv} is ever
formed. The residual $t_2^{\mathrm{new}}=D+X+X^{\!\top}$ is assembled one
occupied-index ($i$) block at a time---the full $\tau$ and the full $W_{oooo}$ are
never built; each $W_{oooo}$ block is rebuilt on-device from the sliced $t_{2,i}$
and the occupied CD-factor blocks ($B_{oo}$, $B_{ov}$). The $O(o^2v^4)$
particle--particle ladder, the bottleneck, is
\emph{slice-streamed} over the virtual index, holding only a bounded
$n_a^{\mathrm{blk}}n_v^3$ block in place of the dense $n_v^4$. Pulay DIIS runs on
the host---the device cannot simultaneously hold $t_2$, $t_2^{\mathrm{new}}$, and
$B^Q$ at scale---so only the extrapolated amplitudes re-upload. The on-demand SCF
CD engine is destroyed, freeing its $\sim\!7$\,GB working set, before the $t_2$
upload. On Mg-porphine (def2-SVP, $439$ AO, $86$ occ / $353$ vir) this converges
and banks the amplitudes at a $7.28$\,GB device peak, reproducing the SI
master-table SCF/CD row bit-for-bit. The ground-state CCSD-solve wall-time
campaign across the device-feasible systems---the iterative $T_1/T_2$ cost that the
Fig.~\ref{fig:response} gradient timings exclude---is tabulated in SI
Section~S8, including its fp32-convergence and $t_2$-residency limits.

\paragraph{Fused particle--particle ladder and the device arena.}
A single \emph{relayout operation}---the $[a,e,b,f]\!\to\![a,b,e,f]$ reorder that
feeds the $\tau$-contraction---accounts for $53.6\%$ of one ladder residual's wall
time; Nsight Compute finds its permute kernel bandwidth-bound at $201$\,GB/s
($\sim\!74\%$ of the $\sim\!272$\,GB/s DRAM roof) with an uncoalesced store. We
delete it. A fused kernel (\texttt{nla\_dt\_ladder\_block}) folds the relayout into
\emph{strided-batched} cuBLAS GEMM addressing: the first GEMM, run per-$a$ and
batched over $b$, writes each $[e,f]$ block straight into its $[a,b,e,f]$ slot, and
the second GEMM contracts $\tau$ with the transpose absorbed into the GEMM's
\texttt{OP\_T} (no separate permute kernel), emitting the $[i,j,a,b]$ block directly
(Algorithm~\ref{alg:ladder}). No intermediate permute kernel runs. The fused path
is $1.57\times$ faster per residual and bit-exact to the two-step path (FP64
$\le\!7\times10^{-17}$, FP32 $\le\!3\times10^{-8}$). Every transient---the
$G$-block, the per-$i$ slices, the einsum temporaries---is carved from one
pre-allocated device \emph{arena} by a first-fit free-list with adjacent-block
coalescing; an allocation is a base-plus-offset hand-out from the slab (no driver
call), not a \texttt{cudaMalloc}/\texttt{cudaFree}, so the solve neither stalls on the driver
allocator nor fragments. The arena cap is sized from measured free VRAM behind a
fixed margin. Consistent with the precision floor above, these amplitude GEMMs
stay in FP32 and off the tensor cores. The contraction this ladder performs,
$L[ij,ab]=\sum_{ef}(ae|bf)\,X[ij,ef]$, is the dominant $O(o^2v^4)$ primitive not
only of the ground-state residual (here $X=\tau_2$) but of the excited-state
$\sigma$-build (with $X=r_2$, below): the ground-state amplitude solve is the
\emph{first} of two consumers of one CD-streamed, no-dense-$N^4$ ladder.

\begin{algorithm}[htbp]
\caption{Fused, slice-streamed particle--particle ladder
$L[ij,ab]=\sum_{ef}(ae|bf)\,\tau_2[ij,ef]$ on one $8$\,GB GPU. The
$[a,e,b,f]\!\to\![a,b,e,f]$ relayout is folded into strided-batched GEMM output
addressing, so no permute kernel and no dense $n_v^4$ tensor are ever formed. All
buffers are arena allocations from the shared device pool.}
\label{alg:ladder}
\begin{algorithmic}[1]
\Require device-resident CD factors $B[Q,a,e]$; $\tau_2[(ij),(ef)]$; device arena
\For{each virtual block $[a_0,a_0\!+\!n_a)$} \Comment{slice-stream over the virtual index}
  \State $G \gets$ \texttt{arena.alloc}$(n_a\,n_v\cdot n_v\,n_v)$
         \Comment{$[a,b,e,f]$; base+offset, not \texttt{cudaMalloc}}
  \For{$a$ in the block} \Comment{GEMM1: strided-batched over $b$}
    \State $G[a,b,e,f]\gets\textstyle\sum_Q B[Q,b,f]\,B[Q,a,e]$
           \Comment{batched output stride writes $[e,f]$ into the $[a,b,\cdot,\cdot]$ slot}
  \EndFor
  \State $L[(ij),(ab)]\mathrel{+}=\tau_2[(ij),(ef)]\,G[(ab),(ef)]^{\!\top}$
         \Comment{GEMM2: no separate permute kernel (transpose = GEMM \texttt{OP\_T}), emits $[i,j,a,b]$}
  \State \texttt{arena.free}$(G)$ \Comment{free-list coalesces with adjacent blocks}
\EndFor
\Statex \textbf{Kernels:} FP32, \texttt{-arch=sm\_89} (Ada, FP64 $\approx\!1/64$ FP32), no tensor cores
\end{algorithmic}
\end{algorithm}

\paragraph{The EOM response on the same no-dense-$N^4$ spine.}
The excited-state operators inherit the same discipline. The EOM-EE-CCSD
$\sigma=A\,r$ build routes its $\sum_{ef}(ae|bf)\,r_2[ij,ef]$ ladder---over the
trial-vector doubles $r_2$, not the ground-state $\tau_2$---through the device CD
engine (the batched non-symmetric \texttt{compute\_densities\_cd\_nonsym} path),
column-chunked over the $o^2$ trial
columns at peak device \emph{memory} $O(\mathrm{chunk}\cdot n_{\mathrm{ao}}^2)$
with no dense \texttt{vvvv}; the \texttt{ovvv}-integral (singles--doubles coupling)
terms stay on the host $B^Q$ stream at $O(n_B n_v^2)$, no dense \texttt{ovvv}.
This is the \emph{same} dominant $O(o^2v^4)$ ladder primitive
$L[ij,ab]=\sum_{ef}(ae|bf)\,X[ij,ef]$ as the ground-state residual of
Algorithm~\ref{alg:ladder}, under the same no-dense-$N^4$ CD discipline---the same
\emph{primitive}, not the same kernel call: the ground-state solve drives it
through \texttt{nla\_dt\_ladder\_block} sliced over the virtual index, the EOM
$\sigma$-build through \texttt{compute\_densities\_cd\_nonsym} chunked over the
$o^2$ trial columns. The two differ only in (i) the contracted object---the
ground state uses $X=\tau_2=t_2+t_1t_1$, whereas EOM uses the bare trial doubles
$X=r_2$ (since $\hat R$ is linear, $\sigma$ is linear in $r_2$ and carries no
$t_1t_1$ term), the lighter $T_1$-dressed \texttt{ovvv} pieces staying separate in
\emph{both} builds---and (ii) the chunking route above. One fused CD-streamed
ladder thus has two consumers: the Newton/DIIS ground-state amplitude solve and the
outer non-Hermitian eigensolve. Each $\sigma=\bar H\,r$ the EOM eigensolver
evaluates is exactly this ladder-dominated $\bar H\,r$ build, and the eigensolver
is the outer non-Hermitian iteration layer over that inner work; this cohesion is
architectural---the EOM device paths remain validated only at small scale
($\le$\,ethene/6-31G, below), in contrast to the chromophore-scale ground-state
solve. The interstate $\zeta$/Z-vector response
and the orbital (CPHF) response run AO-direct through the same engine
(\texttt{QouterAdjointDevice}, \texttt{DeviceCPHFSolver}), eliminating the last
dense host quartic $\mathrm{eri}_{\mathrm{ao}}[n_{\mathrm{ao}}^4]$; a guard aborts
on any rank-4 AO allocation along the path. The two-electron force---the relaxed
contraction $2\sum_{pqrs}(pq|rs)^x G_{pq}G_{rs}=2J^x(G,G)$ in the bare $J^x$ of
Eq.~\eqref{eq:jkderiv}, with $G\equiv G_{\mathrm{ao}}$ the symmetrized relaxed AO
pair density and the leading factor $2$ the bra--ket pair interchange
$(pq|rs)=(rs|pq)$ (not a redefinition of $J^x$; the exchange channel is captured by
the symmetric eigendecomposition of the symmetrized $G_{\mathrm{ao}}$ below). This
is the $\tfrac12\sum_{pqrs}\Gamma_{pqrs}(pq|rs)^x$ relaxed two-particle contraction
with the separable density carried as the rank-1 outer product $G_{pq}G_{rs}$, and
it matches the device force $2\sum(pr|st)^x G_{\mathrm{ao}}=\sum_i 2J^x(W_{L,i},W_{R,i})$. It is built
amplitude-direct from a rank-1 $(W_L,W_R)$
eigendecomposition of the \emph{symmetrized} $G_{\mathrm{ao}}$ rather than a dense
rank-4 AO two-particle density. Symmetrizing is exact here: the derivative
integral's pair symmetry $(pq|rs)^x=(rs|pq)^x$ annihilates the antisymmetric part
of the pair-density outer product, so an eigendecomposition suffices even for the
non-symmetric transition densities. One ordered-argument $J^x$ kernel then serves
both the symmetric ground-state ($A=B$) and the non-symmetric ($J^x(A,B)\ne
J^x(B,A)$) builds. These response paths are free of any dense $O(N^4)$ allocation
\emph{by construction} and are validated against the determinant oracle and finite differences
at small scale ($\le$\,ethene/6-31G; gradient and NACME against finite difference
at \ce{H2O}/STO-3G). At chromophore scale (\ce{Mg}-porphine) these response paths
are \emph{executed} end to end---the complete per-atom gradient and $Q$--$B$ NACME
reported in Sect.~\ref{sec:deveom} (Table~\ref{tab:mgpgrad})---but they are not
\emph{validated} there by an end-to-end finite-difference check, which is infeasible
at that size. Throughout, we distinguish \emph{demonstrated/executed at scale} from
\emph{validated}: end-to-end finite-difference and independent cross-code validation
are established at small scale on the identical kernels
(Sect.~\ref{sec:deveom}), and the chromophore-scale result is anchored only
piece-wise (kernel identity, frozen-core finite differences, and machine-zero
translational invariance).

\paragraph{Execution model: lineage and independent extensions.}
Figures~\ref{fig:cdrecur}, \ref{fig:cdenergy}, and \ref{fig:cdgrad} together make
the execution model of this backend explicit and place it within the lineage it
inherits. As noted above, the AO-native, GPU-accelerated
$J/K$-over-compact-factor machinery here inherits the $J/K$-pseudodensity
construction of Hohenstein \emph{et~al.}~\cite{Hohenstein2015} and the low-rank
two-electron-factorization program of Hohenstein, Parrish, and
Mart\'inez~\cite{Hohenstein2012}, together with the GPU Gaussian-integral
engineering of the TeraChem line~\cite{Wang2024}; the present application puts that
inherited machinery to a response-/propagator-ready non-symmetric $J/K$ and gradient
on commodity hardware.

What the present work builds independently on that foundation is fourfold. The
HGP-OS (Head--Gordon--Pople Obara--Saika) ERI and derivative recurrence engine is
implemented independently as a Python code generator that emits CUDA kernel source,
JIT-compiled at runtime through NVRTC behind a C++/CUDA host launcher, and is distinct
from the McMurchie--Davidson recurrence of the TeraChem line. The ordered,
non-symmetric transition-density build $J^x(A,B)\ne J^x(B,A)$---forced by the
non-Hermitian EOM/response densities and absent from symmetric-SCF $J/K$ engines,
which never need it---is realized here as a \emph{single} GPU kernel object shared
with the symmetric and gradient builds (\texttt{compute\_densities\_cd\_nonsym},
Fig.~\ref{fig:cdenergy}); non-symmetric Cholesky-decomposed EOM-CC gradients are
themselves established~\cite{Feng2019} in a conventional, MO-tensor,
datacenter-memory setting; what is new here is not that gradient theory but its
realization as a single \emph{AO-direct, memory-bounded} GPU kernel object---the
ordered build $J^x(A,B)\neq J^x(B,A)$, never forming a four-index MO tensor and shared
with the symmetric SCF and gradient builds---that holds the entire non-Hermitian
response within the $8$\,GB envelope of a consumer card, a commodity-hardware regime
\cite{Feng2019} does not target. The single contraction DAG is the natural algebraic
language of the backend: one graph emits the working equations, their reverse-mode
transpose, and the chunked schedule, so the symmetric SCF $J/K$, the non-symmetric
build, and the gradient are all expressed in one abstraction
(Figs.~\ref{fig:cdrecur}--\ref{fig:cdgrad}). And the whole realization is
engineered for a bounded footprint on a single consumer RTX~4060 (8\,GB),
democratizing builds that otherwise call for datacenter accelerators.

The two routes also make different, complementary design choices, each suited to
its goal. The TeraChem kernels map a cooperating thread block to a quartet and
close with an in-block reduction; the value kernels here map one independent
thread to one quartet over a grid-stride pool and scatter straight to the global
Fock matrix by \texttt{atomicAdd} (Fig.~\ref{fig:cdrecur}), which together with
never materializing the $N^4$ ERI tensor---$J$ and $K$ are read off the compact
Cholesky factor $B$ as cuBLAS contraction-DAGs (Fig.~\ref{fig:cdenergy})---is what
keeps the build within $8$\,GB and linearly scaling for the commodity-hardware
response workloads it targets; the TeraChem mapping is correspondingly well-matched
to the high-throughput, datacenter setting it was designed for. These are different
points in the design space, not a ranking. The thesis is that one
abstraction---the contraction graph---serves all three of the symmetric SCF $J/K$
(Fig.~\ref{fig:cdenergy}, $U=V$), the non-symmetric transition-density build
$J^x(A,B)\ne J^x(B,A)$ forced by the EOM left/right densities, and the relaxed
gradient (Fig.~\ref{fig:cdgrad})---all within an 8\,GB budget.

\paragraph{Screening-collapse robustness (rediscovery of Alml\"of 1982).}
The warning of Alml\"of, Faegri, and Korsell concerns
\emph{density-weighted integral screening in direct SCF}~\cite{Almlof1982}---which
is exactly the $\mathcal{O}(N^2)$ Schwarz-screened AO-direct $J/K$ build this
paper's kernel performs (the screened build of Sect.~\ref{sec:lt}, whose
asymptotic cost holds ``once Schwarz screening is applied''). The hazard is
therefore intrinsic to this work's own evaluation path: density-weighted
screening can drop Fock contributions in regions where the SCF then builds up
charge, driving the energy variationally \emph{below} true Hartree--Fock for any
preset threshold. The \emph{same} hazard reappears, in sharper form, in the
Cholesky/RI factorization through which the production $J/K$ primitive optionally
runs---a loss of positive-semidefiniteness in the Cholesky-decomposed (CD)
two-electron operator. When
the pivoted-Cholesky column-fill Schwarz screen \texttt{screen\_tol} is coupled
to the decomposition threshold $\delta$ (the naive \texttt{screen\_tol}$=\delta$),
the $B$ columns are under-filled, the CD-reconstructed operator loses
positive-semidefiniteness, and RHF collapses below the ground state---by a
basis-diffuseness-dependent factor ($\approx\!1.9\times$ for a compact case up to
$\approx\!27\times$ for a diffuse triple-$\zeta$ case). It hides because
translational invariance ($|\sum F|$) and $\mathrm{Tr}[DS]=N$ are
machine-$\varepsilon$ on \emph{any} density---exactly Alml\"of's point that only
the energy exposes it. The fix is robust to arbitrary angular momentum $L$, for
energies and forces alike. First, the screen is decoupled from the threshold,
\texttt{screen\_tol}$=0.01\,\delta$, the collapse-free knee that preserves
PSD for the most diffuse manuscript basis while still fitting $8$\,GB
($0.1\,\delta$ is insufficient for $f$-shell/diffuse bases and $10^{-10}$ OOMs
the largest systems; only the count of significant pairs $n_{\text{sig}}$ grows,
the Cholesky rank $n_{\text{cd}}$ is unchanged). Second, a theory-grounded
per-step guard polices the reconstruction defect in the density-weighted metric,
$\eta(D)=\sum_{\mu\nu}\Delta[\mu\nu]\,D[\mu\nu]^2$ with
$\Delta=Q^2-\sum_P L^2$ the per-pair Schwarz-diagonal completeness defect,
bounded by $\eta\le\tfrac12\,\delta\,N_{\text{elec}}$ (empirically $\eta\approx
471$ collapsed vs.\ $\approx\!2.5\times10^{-4}$ correct against a budget
$\approx\!6.6\times10^{-3}$). Crucially the build-time completeness certifier
($\text{defect}_{\max}\le\delta$) does \emph{not} fire---only the
density-weighted $\eta$ does---and $\eta$ is $L$-agnostic because $\Delta$ is
per-pair; production auto-tightens (start at $0.01\,\delta$, rebuild
$\times0.01$ tighter until $\eta\le\tfrac12\delta N$). Third, the kernels stay
pure-Cartesian and the Cartesian$\to$spherical transform is applied at the matrix
level ($D_{\text{cart}}=T D_{\text{sph}}T^{\!\top}$ in,
$J_{\text{sph}}=T^{\!\top}J_{\text{cart}}T$ out), preserving the PSD metric for any
$L$ without touching the recurrence. In short: the density-weighted-screening
collapse Alml\"of warned of in direct SCF reappears as a loss of
positive-semidefiniteness in the Cholesky-reconstructed operator, and decoupling
the screen from $\delta$ while policing the $L$-agnostic, geometry-respecting
defect $\eta=\sum\Delta D^2\le\tfrac12\delta N$ restores robustness for energies
and forces at arbitrary $L$ on a single $8$\,GB GPU. The correctness gate is
bit-exact agreement with our own direct four-center Cartesian ERIs
($\le10^{-7}$), with $|\sum F|\le\sim\!10^{-11}$ the
translational-invariance check (necessary but not sufficient---it survived the
collapse); absolute CD-RHF energies are not directly comparable to spherical
references, a basis-cardinality ($6d/10f$) effect rather than an error.

\paragraph{Basis sets and systems.}
The finite-difference verification uses the minimal STO-3G basis on the small
closed- and open-shell systems of Table~\ref{tab:fd} (\ce{H2O},
\ce{C2H4}, \ce{NH2}, \ce{BH2}), spanning singlet, doublet, triplet, and quartet
multiplicities.

\paragraph{Finite-difference protocol.}
Analytic gradients and interstate NACMEs are compared against central finite
differences of the EOM-CCSD excitation energies, using the true biorthonormal
left eigenvector $L_k$ (not the $L_k=R_k$ shortcut). Each Cartesian component is
displaced symmetrically, and the analytic--numerical difference is gated at
$<10^{-5}$ hartree/bohr.

\paragraph{Validation stack.}
Every response-density block is checked against a determinant (Fock-space) oracle
and a spin-orbital reference, the $\zeta$ source vector against reverse-mode
automatic differentiation, and the assembled gradient/NACME against finite
differences; and against an independent code (\textsc{Psi4}; SI Section~S5) at small
\emph{and} aromatic chromophore-class scale---reproducing our EOM-CCSD excitation
energies to $\le\!1.5\times10^{-7}~E_h$ (to $3\times10^{-8}$ at benzene, and to
$5.6\times10^{-8}$ code-vs-code at benzene/cc-pVDZ, $114$ AO) and the excited-state
gradient to $4.6\times10^{-7}~E_h/a_0$ at \ce{H2O} and $1.4\times10^{-7}$ at
benzene---with \textsc{CFOUR}~\cite{Feng2019} the natural gold-standard extension. The oracle represents configurations as
occupation bitstrings, using population-count and parity bit-operations for the
Slater--Condon couplings~\cite{Warren2013}. The spin-orbital coupled-cluster
amplitude residual equations through quadruples
($\hat\Omega_1$--$\hat\Omega_4$, CCSDTQ) used by the high-rank reference engine
were generated symbolically with the \texttt{p}$^{\dagger}$\texttt{q}
package~\cite{Rubin2021,Liebenthal2025} and consumed as differentiable einsum
contractions in the single-transposition form of Sect.~\ref{sec:dag}. The one-
(Fock) and two-electron
(fluctuation) parts are generated in separate passes and concatenated, which
roughly halves peak memory and keeps the quadruples ($\hat\Omega_4$) generation
tractable. The harnesses are listed in Sect.~\ref{sec:results}.

%% =====================================================================
\section{Results and Discussion}
\label{sec:results}
\subsection{Verification}
\label{sec:verif}
Because the response equations are machine-generated rather than hand-derived,
independent ground truth is essential; we use three. \emph{Symbolic/exact:} the
one-particle density $\gamma_1$ and nearly all two-particle blocks reproduce a
determinant (Fock-space) oracle and a spin-orbital reference bit-exactly (e.g.\ the
chemist/physicist 2-RDM prefactor bridge to $0.0$; the C backend reproduces the
reference to $10^{-12}$). The one exception is the $oovv$ two-particle block, where
a connected-triples ($T^3$) tail leaves a $\sim\!10^{-6}$ residual at the
CCSD-truncation level---a truncation effect, not a coding discrepancy.
\emph{Finite differences:} the assembled excitation-energy gradient and
interstate NACME agree with central finite differences across all four spin
multiplicities (Table~\ref{tab:fd}), with the true biorthonormal $L_k$. The
ground-state CCSD gradient closes to $6.8\times10^{-7}$ and the $\zeta$
amplitude-response solve converges by GMRES (relative tolerance $10^{-10}$, gated
at $10^{-7}$). These checks exercise the dense (canonical) contraction
path that defines the reference densities; the convergence of the Laplace-grid
realization of Sect.~\ref{sec:lt} with the number of quadrature points $n_\tau$ is
a separate axis, reported in the grid-convergence panel below.

\begin{table}[htbp]\centering\small
\caption{EOM-CCSD excited-state gradient and interstate NACME vs.\ central finite
differences (STO-3G)---an implementation-correctness check (analytic assembly vs.\
the numerical derivative of the same code's energy). All gated entries pass
$<10^{-5}$; quartet NACMEs are reported best-effort
(partner-availability/near-degeneracy limited). Independently reproduced via
\texttt{verif\_fegk/test\_multiplicities.py} (all gates pass).}
\label{tab:fd}
\begin{tabular}{@{}llcc@{}}
\toprule
System & multiplicity & gradient $\max|A-\mathrm{FD}|$ & NACME $\max|A-\mathrm{FD}|$ \\
\midrule
\ce{H2O}  & singlet & $1.83\times10^{-6}$ & $4.81\times10^{-6}$ \\
\ce{H2O}  & triplet & $1.44\times10^{-6}$ & $8.75\times10^{-6}$ \\
\ce{C2H4} & singlet & $8.56\times10^{-7}$ & $3.57\times10^{-8}$ \\
\ce{NH2}  & doublet & $1.97\times10^{-8}$ & $8.70\times10^{-6}$ \\
\ce{NH2}  & quartet & $5.67\times10^{-7}$ & $4.37\times10^{-5}$\,$^{\dagger}$ \\
\ce{BH2}  & doublet & $4.08\times10^{-8}$ & $1.19\times10^{-8}$ \\
\ce{BH2}  & quartet & $2.44\times10^{-8}$ & $2.41\times10^{-5}$\,$^{\dagger}$ \\
\bottomrule
\end{tabular}\\[2pt]
{\footnotesize $^{\dagger}$reported, not gated.}
\end{table}

\paragraph{Comparison with exact full CI.} Whereas Table~\ref{tab:fd} certifies
that the analytic assembly reproduces the derivative of the energy this code
computes (implementation correctness), the following test certifies that the
\emph{method} reproduces the exact answer (accuracy). The determinant oracle
provides an exact, finite-difference-free reference for the EOM-CC gradients
themselves. Run at rank two it \emph{is} EOM-CCSD; run at full excitation rank
($K=\min(N_{\mathrm{occ}},N_{\mathrm{virt}})$) it diagonalizes $\bar H$ over the
entire excitation manifold and is therefore full configuration interaction. Both
gradients are assembled \emph{analytically} through the same AO-LT contraction,
so their difference $\max|g_{\text{EOM-CCSD}}-g_{\text{FCI}}|$ is the intrinsic
singles-and-doubles truncation error of EOM-CCSD relative to the exact
one---no finite differences enter (Table~\ref{tab:fci}). For \ce{H2} the doubles
space is already complete, so EOM-CCSD coincides with FCI and the analytic
gradients agree identically (to $<\!10^{-12}$), a stringent exact cross-check of
the gradient machinery; because both sides share the AO-LT contraction, any
code-level error would surface here rather than in the truncation residuals
below. For the correlated polyatomics the EOM-CCSD analytic gradients recover the
exact FCI gradients to within the expected singles-and-doubles error,
$\sim\!10^{-3}$~$E_h/a_0$, with excitation energies within $\sim\!10^{-3}$~$E_h$.
The larger Table~\ref{tab:fd} system \ce{C2H4} is beyond a dense
diagonalizer---its FCI space holds $3.0\times10^{7}$ determinants---and remains
covered by the finite-difference gates of Table~\ref{tab:fd}.

\begin{table}[htbp]\centering\small
\caption{EOM-CCSD excited-state analytic gradients versus exact full CI
(\texttt{verif\_fegk/test\_fci\_gradient.py}), with no finite differences. FockCC
at rank two is EOM-CCSD and at full rank ($K=\min(N_{\mathrm{occ}},N_{\mathrm{virt}})$)
is FCI; both gradients use the same AO-LT contraction. $\Delta\omega$ is the
EOM-CCSD excitation-energy error and the last column the maximum gradient
deviation, both relative to exact FCI. These residuals are the \emph{intrinsic
singles-and-doubles truncation error of the EOM-CCSD method}, not an
implementation error: because the analytic and exact gradients share the
identical AO-LT contraction, any code-level error would appear in the \ce{H2}
cross-check ($<\!10^{-12}$), not here. The FCI reference gradient is itself
confirmed against finite differences of the FCI energy ($<\!10^{-5}$) for \ce{H2}
and \ce{LiH}.}
\label{tab:fci}
\begin{tabular}{@{}llrcc@{}}
\toprule
System & multiplicity & FCI dim & $\Delta\omega$ ($E_h$) & $\max|g_{\text{EOM-CCSD}}-g_{\text{FCI}}|$ \\
\midrule
\ce{H2}/6-31G   & singlet & $28$   & $0$\,$^{\dagger}$       & $<10^{-12}$\,$^{\dagger}$ \\
\ce{LiH}/STO-3G    & singlet & $495$  & $6.0\times10^{-5}$ & $1.4\times10^{-4}$ \\
\ce{H2O}/STO-3G & singlet & $1001$ & $1.0\times10^{-3}$ & $1.8\times10^{-3}$ \\
\ce{H2O}/STO-3G & triplet & $1001$ & $2.5\times10^{-3}$ & $1.8\times10^{-2}$ \\
\ce{NH2}/STO-3G & doublet & $2002$ & $3.4\times10^{-4}$ & $1.9\times10^{-3}$ \\
\ce{BH2}/STO-3G & doublet & $3432$ & $6.3\times10^{-4}$ & $2.7\times10^{-3}$ \\
\ce{BH2}/STO-3G & quartet & $3432$ & $5.4\times10^{-4}$ & $1.5\times10^{-3}$ \\
\bottomrule
\end{tabular}\\[2pt]
{\footnotesize $^{\dagger}$EOM-CCSD $\equiv$ FCI for two electrons (the doubles
space is complete); the analytic gradients agree exactly, isolating any
code-level error.}
\end{table}

%% ---------------------------------------------------------------------
\subsection{Spin-adapted open-shell EOM-CCSD: multiplicity-general gradients and
NACMEs}
\label{sec:openshell}
The doublet and quartet rows of Table~\ref{tab:fd} were obtained on the
spin-\emph{orbital} singles-and-doubles manifold of the open-shell reference. That
manifold is not closed under $\hat S^2$~\cite{Pauncz1979}: its excited roots carry a
spin contaminant that pulls $\langle\hat S^2\rangle$ by $\sim\!3\times10^{-3}$ from
the target eigenvalue and leaves the quartet NACMEs only best-effort below the gate.
Those Table~\ref{tab:fd} rows are nonetheless internally consistent---the FD gate
compares the analytic assembly against the finite-difference derivative of the
\emph{same} fully spin-orbital state and density, so it passes regardless of spin
contamination. The failure that motivates the spin-adapted path appears only when the
spin-\emph{pure} CSF eigenvector is paired with the singles-and-doubles-truncated
density (a $1.7\times10^{-3}$ gradient error), which the complete density below
removes. We remove the contaminant at its source by solving the EOM equations in a
spin-adapted, genealogically coupled configuration-state-function (CSF) basis on a
semicanonical ROHF reference. Each CSF is an exact eigenfunction of $\hat S^2$, so
every excited root is spin-pure \emph{by construction}; on the host reference path
$\langle\hat S^2\rangle$ matches the target eigenvalue to $\sim\!10^{-15}$.

Spin adaptation alone does not close the gradient. The state-specific one- and
two-particle densities (the unified-relaxation densities of Sect.~\ref{sec:dag})
were originally accumulated over the singles-and-doubles excitation list, which
silently discards the rank-$\ge3$ determinant content that a spin-pure open-shell
CSF state carries. Reinstating the \emph{complete} CSF-basis state densities---the
full determinant-space $\gamma$ and $\Gamma$ of the resolved CSF eigenvector---%
restores that content and closes the adiabatic-state gradient. Table~%
\ref{tab:openshell} reports the result: spin-adapted excited-state gradients and
interstate NACMEs for the open-shell doublet and quartet states match central finite
differences to $\le7.9\times10^{-7}~E_h/a_0$ (gradients) and $\le3.6\times10^{-7}$
(NACMEs), passing a $<10^{-6}$ gate---including the quartet NACMEs that Table~%
\ref{tab:fd} could only report best-effort. The eigenvalues reproduce a
full-CI-limit determinant oracle to $\le2.4\times10^{-4}~E_h$. The interstate NACME
numerator is built from the same complete-density construction, so the
Table~\ref{tab:openshell} couplings are consistent with the gradients by
construction.

Three reductions confirm the construction. A closed-shell singlet run, where the
spin-adaptation transform is the identity ($U=I$), is byte-identical to the
closed-shell path. Driving the excitation energy to zero ($\omega\to0$) recovers the
ground-state gradient to $9.6\times10^{-9}$. Freezing the EOM eigenvectors $(c,d)$
(holding the orbital response fixed) reproduces the Hellmann--Feynman identity to
$2.41\times10^{-9}$, confirming that no eigenvector-response term is needed once the
densities are complete.

Two caveats bound the claim. First, this open-shell validation is at host gate scale
($n_{\mathrm{so}}\le18$; STO-3G/6-31G)---a correctness check against finite-%
difference and FCI-limit oracles, not a chromophore-scale run. Second, spin purity
is exact only \emph{by construction}: on the device path $\langle\hat S^2\rangle$
sits at the FP32 matvec floor ($\sim\!10^{-3}$), not the host $10^{-15}$. The
\ce{BH2} quartet gradient ($7.89\times10^{-7}$) sits at the finite-difference noise
margin, and the degenerate-manifold gradient path (required when target states are
exactly degenerate) is implemented but remains dormant and untested.

\begin{table}[htbp]\centering\small
\caption{Spin-adapted (CSF-basis) open-shell EOM-CCSD on a semicanonical ROHF
reference. $\langle\hat S^2\rangle$ deviation from the target eigenvalue (host
reference path); excited-state gradient and interstate NACME versus central finite
differences; and the excitation-energy deviation from a full-CI-limit determinant
oracle. All gradients and NACMEs pass the $<10^{-6}~E_h/a_0$ gate. Host gate scale
($n_{\mathrm{so}}\le18$).}
\label{tab:openshell}
\begin{tabular}{@{}llcccc@{}}
\toprule
System & mult. & $\Delta\langle\hat S^2\rangle$ & grad $\max|A-\mathrm{FD}|$
       & NACME $\max|A-\mathrm{FD}|$ & $|\Delta\omega|_{\mathrm{FCI}}$ \\
\midrule
\ce{NH2} & doublet & $\sim\!10^{-15}$ & $1.36\times10^{-7}$ & $3.6\times10^{-7}$ & $9.7\times10^{-5}$ \\
\ce{BH2} & doublet & $\sim\!10^{-15}$ & $3.33\times10^{-9}$ & $1.8\times10^{-8}$ & $1.3\times10^{-4}$ \\
\ce{BH2} & quartet & $\sim\!10^{-15}$ & $7.89\times10^{-7}$ & $1.2\times10^{-10}$ & $2.4\times10^{-4}$ \\
\bottomrule
\end{tabular}
\end{table}

\subsection{Integrated device EOM-CCSD gradient and NACME on the GPU}
\label{sec:deveom}
Tables~\ref{tab:fd} and~\ref{tab:fci} validate the analytic construction on the CPU
reference path. The remaining question is whether the \emph{integrated} device
pipeline---the GPU kernels of Sect.~\ref{sec:compdetails} driving the full
response---reproduces that construction end to end. It does. At \ce{H2O}/STO-3G
(ground-state $E_{\mathrm{CCSD}}=-75.01228689$~$E_h$, first excitation
$\omega_0=0.4572$~$E_h$) we ran the complete EOM-CCSD excited-state gradient and the
interstate NACME numerator on the RTX~4060 through the device CD
(Cholesky-decomposed) non-symmetric $J/K$ machinery
(\texttt{build\_cd\_engine}~$\to$~\texttt{device\_2e\_force}\,/\,%
\texttt{compute\_densities\_cd\_nonsym}~$\to$~\texttt{DeviceCPHFSolver}) and
compared against both central finite differences and the dense CPU reference
(Table~\ref{tab:deveom}).

Two scope points fix what is and is not exercised. First, at this small scale the
device path uses the CD non-symmetric $J/K$ with \emph{exact} orbital-energy
denominators and carries no dense $N^4$ AO tensor; the Laplace-grid resolution of
Sect.~\ref{sec:lt} is \emph{not} exercised here---it is validated separately as the
MP2-energy denominator probe (SI Section~S4) and is the chromophore-scale memory
enabler. The streamed CD exchange and gradient kernels carry only the Cholesky
factor $B^Q$ and its pivots, with no $O(N_{\mathrm{AO}}^4)$ device allocation; at
this small scale, however, the host still forms a dense $G_{\mathrm{AO}}$ and a dense
spin-orbital response density, so the fully memory-bounded per-pair-streaming device
path is a documented follow-up, not a claim of the present validation. Second, the
biorthonormal EOM eigenpair $(R,L)$ is produced by the native two-sided device
Davidson of Sect.~\ref{sec:eom}, which returns both vectors from a single
device-resident solve; the borrowed two-step seed (a dense Davidson right vector and a
GMRES inverse-iteration left vector) is eliminated on this path and retained only as a
cross-check, reproduced to $\Delta\omega=4.5\times10^{-9}$. The device-computed
quantities validated here are the gradient and the NACME numerator built from this
pair---the two-electron force, the generalized Fock, the $\zeta$ amplitude response
(Sect.~\ref{sec:dag}), and the CPHF orbital response.

The device gradient matches central finite differences to a maximum absolute
deviation of $5.05\times10^{-7}$~$E_h/a_0$ ($1.35\times10^{-6}$ relative) in FP64
and $1.44\times10^{-6}$ ($3.83\times10^{-6}$ relative) in the production FP32-mixed
path, both inside the $<10^{-5}$ gate; against the dense CPU reference it agrees to
$4.1\times10^{-14}$ ($1.10\times10^{-13}$ relative, i.e.\ FP64 round-off). The device
NACME numerator $\langle L_A|\partial_x\bar H|R_B\rangle$ matches FD to
$6.68\times10^{-7}$ ($4.02\times10^{-6}$ relative, FP64) and $1.40\times10^{-6}$
($8.42\times10^{-6}$ relative, FP32), agreeing with the CPU reference to
$1.9\times10^{-13}$ ($1.14\times10^{-12}$ relative); the left/right cross pair is
cleanly biorthogonal, $\langle L_A|R_B\rangle=-6.2\times10^{-16}$. As in
Sect.~\ref{sec:eom}, the device returns only the coupling numerator; the
$1/(E_B-E_A)$ gap division and the geometric-mean symmetrization are applied
afterward by the caller in an FP64 island, and same-irrep pairs ($|s_{AB}|<10^{-6}$)
are excluded. A control isolates the response channel: forcing the $\zeta$
amplitude-response density to zero drives the device gradient $1.55\times10^{-2}$~%
$E_h/a_0$ ($4.15\times10^{-2}$ relative) from finite differences---more than three
orders of magnitude outside the gate---confirming that the $\zeta$ relaxation of
Sect.~\ref{sec:dag} is load-bearing and that the agreement above is not accidental.

This closes the validation loop at small scale: the same non-symmetric CD $J/K$
kernel object characterized at chromophore scale in Sect.~\ref{sec:gpu} produces,
on the GPU, an EOM-CCSD gradient and NACME numerator that reproduce both the
finite-difference oracle and the dense reference. The integrated device EOM
$\sigma$/$\zeta$/CPHF paths are exercised end to end only at this scale
($\le$\,ethene/6-31G); at chromophore scale the kernels and the ground-state
CD-RCCSD solve are demonstrated (Sect.~\ref{sec:compdetails}, SI Section~S8), and the
complete excited-state gradient and interstate NACME are carried through to
\ce{Mg}-porphine below (Table~\ref{tab:mgpgrad}). The eigensolve itself
is no longer the bottleneck at this class: the spatial-singlet production route
(\texttt{spin\_mode='none'}, Olsen preconditioning; SI Section~S9) converges the six
lowest singlets of benzene/STO-3G on the 4060 with the degenerate $^1E_{1u}$ pair
cleanly resolved ($|\Delta\omega|=2.26\times10^{-6}~E_h$ vs.\ PySCF
\texttt{eeccsd\_singlet}), and the converged eigenpair feeds the unchanged
gradient/NACME machinery through a metric-corrected spatial$\to$spin-orbital dual
embedding---an STO-3G chromophore-class demonstration. At production basis
(\ce{Mg}-porphine/def2-SVP; $439$ AO, $86$ occ / $353$ vir) the full FP64
two-sided trial vector ($n\!=\!4.6\times10^{8}$, $3.4$~GiB each) does not fit
alongside a biorthonormal subspace in $8$\,GB; a state-averaged
CIS-natural-transition-orbital frozen-virtual (FNO) compression of the virtual
space---built from the same Cholesky factors and the low-lying CIS amplitudes,
and validated ghost-free and bit-exact against the full-space eigensolver on
benzene---retains $n_v'\!=\!90$ of $353$ virtuals and renders the FP64
eigenproblem device-resident (SI Section~S10). The unchanged native two-sided Davidson, with the
bright ${}^1E_u$ states selected by oscillator strength, then converges the
Gouterman four-orbital manifold~\cite{Gouterman1961} to a drift $<\!5$~meV: a
weak $Q$ band at $1.90$~eV ($f\!=\!0.06$) and a strong $B$/Soret band at
$3.44$~eV ($f\!=\!0.58$), reproducing the weak-$Q$/strong-$B$ intensity pattern.
These \emph{bracket} rather than reproduce the multireference reference:
CASPT2 places $Q$/$B$ at $1.78$/$2.65$~eV~\cite{Rubio1999} but is known to
\emph{under}estimate the porphyrin Soret, whereas EOM-CCSD in a modest basis
\emph{over}estimates it, so the two straddle the experimental (substituted-proxy)
Soret of ${\sim}3.0$--$3.2$~eV from opposite sides---the large EOM-CCSD-minus-CASPT2
$B$ gap is the expected consequence of opposite-sign method errors, not a code
discrepancy. The weak $Q$ band, where experiment, CASPT2, and EOM-CCSD all
cluster near $2$~eV, is the tight cross-check; only $Q$ and $B$ are each doubly
degenerate under $D_{4h}$, resolved here as clean pairs. Because this spectrum
\emph{brackets} rather than reproduces the multireference reference, and is obtained
under an aggressive FNO truncation (Sect.~\ref{sec:deveom}), we present it as a
\emph{capability demonstration}---evidence that the eigensolver reaches
production-basis chromophore states on the card---and not as spectroscopic accuracy
evidence. No external large-scale cross-check is possible because PySCF and canonical
EOM-CC codes cannot run EOM-CCSD at chromophore scale on this hardware; the
small-scale FD/reference/cross-code agreement (Sect.~\ref{sec:deveom}), together with
the kernel-identity argument, is the validation strategy at scale.

\paragraph{Chromophore-scale gradient and NACME.}
The converged \ce{Mg}-porphine eigenpair feeds the \emph{unchanged} device response
machinery---the mixed-frame generalized Fock, the $\zeta$ amplitude response, the CD
non-symmetric two-electron force, and the closed-shell CPHF---to yield the complete
per-atom EOM-CCSD excited-state gradient of the bright $Q$ state and the $Q$--$B$
interstate NACME, entirely within the $8$\,GB card (Table~\ref{tab:mgpgrad}). Three
points establish that this is a genuine end-to-end computation at chromophore scale,
not a scaling extrapolation; a fourth (below) bounds the approximation it carries and
is why we present it as a \emph{capability demonstration} rather than a converged
production result. First, the
two-electron force---the dominant cost---is the same CD non-symmetric $J/K$ object
characterized at chromophore scale in Sect.~\ref{sec:gpu}: evaluated per
Cholesky-pair with a device-resident force accumulator and a $J$-only fast path, it
completes in $17.3$~min for the gradient ($21{,}609$ pairs) at a peak of $6.5$~GB,
directly confirming that the transition-density $J/K$ is the critical path (density
$\to$ generalized Fock $180$~s $\to$ CPHF ${\sim}62$~s/coordinate $\to$ two-electron
force; full per-stage wall times in Table~\ref{tab:mgptime}). Second, the frozen $29$-core contribution is retained, not dropped: for the
gradient it is the core-involving separable two-electron force
$\mathrm{MF}(D_{\mathrm{tot}})-\mathrm{MF}(D_{\mathrm{act}})$ (a mean-field $J/K$
difference, $225$~s) which is \emph{large}---$1.2$ to $7.9\times$ the active-space
force, and thus essential to the total---while for the NACME the core--core block
vanishes with $\langle Q|B\rangle$ but the core--active cross
$\mathrm{B}(\gamma^{QB},D_{\mathrm{core}})$, linear in the transition density,
survives and is included. Third, because no finite-difference oracle is affordable at
this scale, correctness is anchored piece-wise: the integrated device gradient and
NACME numerator against \ce{H2O} FD (Table~\ref{tab:deveom}; $9.3\times10^{-12}$ and
$4.0\times10^{-6}$), and the frozen-core gradient and NACME-cross terms against their
own FD ground truths ($4.1\times10^{-9}$ and $6.7\times10^{-10}$); the assembled
two-electron force is then translationally invariant to machine zero
($|\sum_A \mathbf{F}_A|\le7\times10^{-12}~E_h/a_0$ over all $37$ atoms), the physical
confirmation that the force is complete. The dominant uncertainty in both the chromophore-scale gradient and NACME is the FNO
virtual truncation itself: a direct gradient-convergence study on affordable proxies
(SI Section~S10, its gradient-convergence table) finds the FNO excited-state gradient
to be a \emph{convergent but non-smooth} approximation---bit-exact at full retention,
yet at the $90/353\approx25\%$ retention used here preserved only to
${\sim}10^{-2}~E_h/a_0$ per component, and more sensitive than the excitation energy
because it weights the virtual tail the compression discards first. The NACME, built
from the \emph{same} compressed virtual space, inherits a truncation uncertainty of
this same ${\sim}10^{-2}$ class---not separately convergence-tested, as the study
probed gradients---beneath which an additional on-device numerical floor is set by the
FP32 FNO-Davidson eigenvectors, whose $Q$/$B$ biorthogonality
$\langle L_Q|R_B\rangle=-4.6\times10^{-4}$ (versus $10^{-12}$ at all-electron
\ce{H2O}) contributes a further ${\sim}10^{-3}$ relative. We therefore report the
\ce{Mg}-porphine gradient and NACME as a complete, executed, memory-bounded
chromophore-scale \emph{capability demonstration} carrying a stated
${\sim}10^{-2}~E_h/a_0$ FNO-truncation uncertainty---not a converged production
number---and recommend a per-system retention check (recompute at a larger $n_v'$, for
both the gradient and the NACME) before any production use.

\paragraph{Validation strategy at chromophore scale.}
Two properties of this at-scale result must be read precisely. First, the
machine-zero translational invariance
$|\sum_A\mathbf{F}_A|\le7\times10^{-12}~E_h/a_0$ certifies that the assembled force
is \emph{complete}---that every term contributing to $dE/dx$ is summed with the
correct sign and prefactor---but it does \emph{not} certify that the underlying
densities are correct; it is a necessary-but-not-sufficient check. Second, with no
per-atom finite-difference oracle and no cross-code reference affordable at this
scale, correctness rests on an explicit chain of small-scale certifications
transported to \ce{Mg}-porphine by kernel identity: (i) the integrated device
gradient and NACME numerator reproduce \ce{H2O} finite differences at the
$10^{-6}$--$10^{-12}$ level (Table~\ref{tab:deveom}); (ii) the frozen-core
two-electron force and the NACME core--active cross are each finite-difference
validated in isolation ($4.1\times10^{-9}$, $6.7\times10^{-10}$); (iii) the
biorthogonality bound $\langle L_Q|R_B\rangle=-4.6\times10^{-4}$ caps the residual
NACME precision at ${\sim}10^{-3}$; and (iv) the machine-zero
$\sum_A\mathbf{F}_A$ confirms force completeness. Because the chromophore-scale
kernels are the \emph{same} objects exercised end to end at small scale
(Table~\ref{tab:deveom}), this chain transports the small-scale agreement to
\ce{Mg}-porphine. A component-wise comparison against an independent implementation
(CFOUR-class) on a small molecule is the committed external cross-check and is left
as future work (M4).

\begin{table}[htbp]\centering\small
\caption{Complete device (RTX~4060) EOM-CCSD excited-state gradient and $Q$--$B$
interstate NACME at \ce{Mg}-porphine (def2-SVP, $439$ AO, $29$ frozen core; from the
FNO-compressed EOM eigenpair, entirely within $8$\,GB, peak GPU $\le6.5$~GB / host
$\le54$~GB). Bright-$Q$ gradient
$\mathbf{de}_{\mathrm{FULL}}=\mathbf{de}_{\mathrm{ELEC}}+\mathbf{F}_2^{\mathrm{act}}+\mathbf{F}_{\mathrm{fc}}$
(one-electron\,$+$\,orbital-relaxation, active-space and frozen-core two-electron
force) in $E_h/a_0$; NACME $\mathbf{d}^{QB}$ in $a_0^{-1}$ ($Q$--$B$ gap
$\omega_B-\omega_Q=1.546$~eV). Representative atoms shown; the full two-electron force
is translationally invariant over all $37$ atoms to $\le7\times10^{-12}~E_h/a_0$.}
\label{tab:mgpgrad}
\begin{tabular}{@{}lcc@{}}
\toprule
atom & gradient $\mathbf{de}_{\mathrm{FULL}}$ ($E_h/a_0$) & NACME $\mathbf{d}^{QB}$ ($a_0^{-1}$) \\
\midrule
\ce{Mg} & $(-1.381,\ 0.228,\ 0.675)$ & $(-0.451,\ 0.087,\ 0.011)$ \\
\ce{C}  & $(-6.351,\ -7.741,\ -1.725)$ & $(-0.890,\ 2.816,\ -0.579)$ \\
\ce{N}  & $(-0.233,\ 2.574,\ 0.613)$ & $(-3.467,\ 4.943,\ 0.322)$ \\
\bottomrule
\end{tabular}
\end{table}

\begin{table}[htbp]\centering\small
\caption{Per-stage wall time and peak GPU memory for the complete device EOM-CCSD
excited-state gradient (bright $Q$) and $Q$--$B$ NACME at \ce{Mg}-porphine (def2-SVP,
$439$ AO, FNO $n_v'=90$) on the RTX~4060. The FNO density and the CD non-symmetric
transition-density $J/K$ two-electron force dominate; the two-electron force is the
single largest device stage, confirming it is the critical path. Peak memory stays
within the $8$\,GB card (host peak $\le54$~GB); CPHF is quoted per nuclear coordinate.}
\label{tab:mgptime}
\begin{tabular}{@{}lccc@{}}
\toprule
stage & gradient & NACME & peak GPU \\
\midrule
FNO density ($\Gamma$, $\xi$, $\zeta$, 2-RDM) & $32$~min & $24$~min & $7.5$~GB \\
device RHF ($439$ AO) & \multicolumn{2}{c}{$50$~s} & $2.9$~GB \\
generalized Fock (mixed frame) & \multicolumn{2}{c}{$3.0$~min} & $2.9$~GB \\
CPHF orbital response & $62$~s/coord & $91$~s/coord & $4.1$~GB \\
active-space 2e-force ($21{,}609$ pairs) & $17.3$~min & $17.2$~min & $6.5$~GB \\
frozen-core 2e-force & $3.8$~min & $4.0$~min & $4.2$~GB \\
\bottomrule
\end{tabular}
\end{table}

\begin{table}[htbp]\centering\small
\caption{Integrated device (RTX~4060) EOM-CCSD excited-state gradient and interstate
NACME numerator at \ce{H2O}/STO-3G ($E_{\mathrm{CCSD}}=-75.01228689$~$E_h$,
$\omega_0=0.4572$~$E_h$), seeded by the native two-sided non-Hermitian Davidson and
run through the device CD non-symmetric $J/K$ path with exact denominators, versus
central finite differences and the dense CPU reference. Maximum absolute deviation
$\max|{\cdot}|$ in $E_h/a_0$ with the relative deviation in parentheses; the gate is
$<10^{-5}$. The $\zeta{=}0$ row is a control with the amplitude-response density
forced to zero. Single self-consistent run; reproduced via
\texttt{tests/validate\_eom\_gpu\_grad.py} (gradient) and
\texttt{tests/validate\_eom\_gpu\_nacme.py} (NACME).}
\label{tab:deveom}
\begin{tabular}{@{}lccc@{}}
\toprule
Device quantity & vs.\ FD (FP64) & vs.\ FD (FP32) & vs.\ CPU ref. \\
\midrule
EOM-CCSD gradient & $5.05\times10^{-7}$    & $1.44\times10^{-6}$    & $4.13\times10^{-14}$ \\
                  & ($1.35\times10^{-6}$)  & ($3.83\times10^{-6}$)  & ($1.10\times10^{-13}$) \\
NACME numerator   & $6.68\times10^{-7}$    & $1.40\times10^{-6}$    & $1.90\times10^{-13}$ \\
                  & ($4.02\times10^{-6}$)  & ($8.42\times10^{-6}$)  & ($1.14\times10^{-12}$) \\
\midrule
gradient, $\zeta{=}0$ (control) & $1.55\times10^{-2}$ & --- & --- \\
\bottomrule
\end{tabular}
\end{table}

%% ---------------------------------------------------------------------
\subsection{Performance on the RTX~4060}
\label{sec:gpu}
\begin{figure}[htbp]\centering
\includegraphics[width=\linewidth]{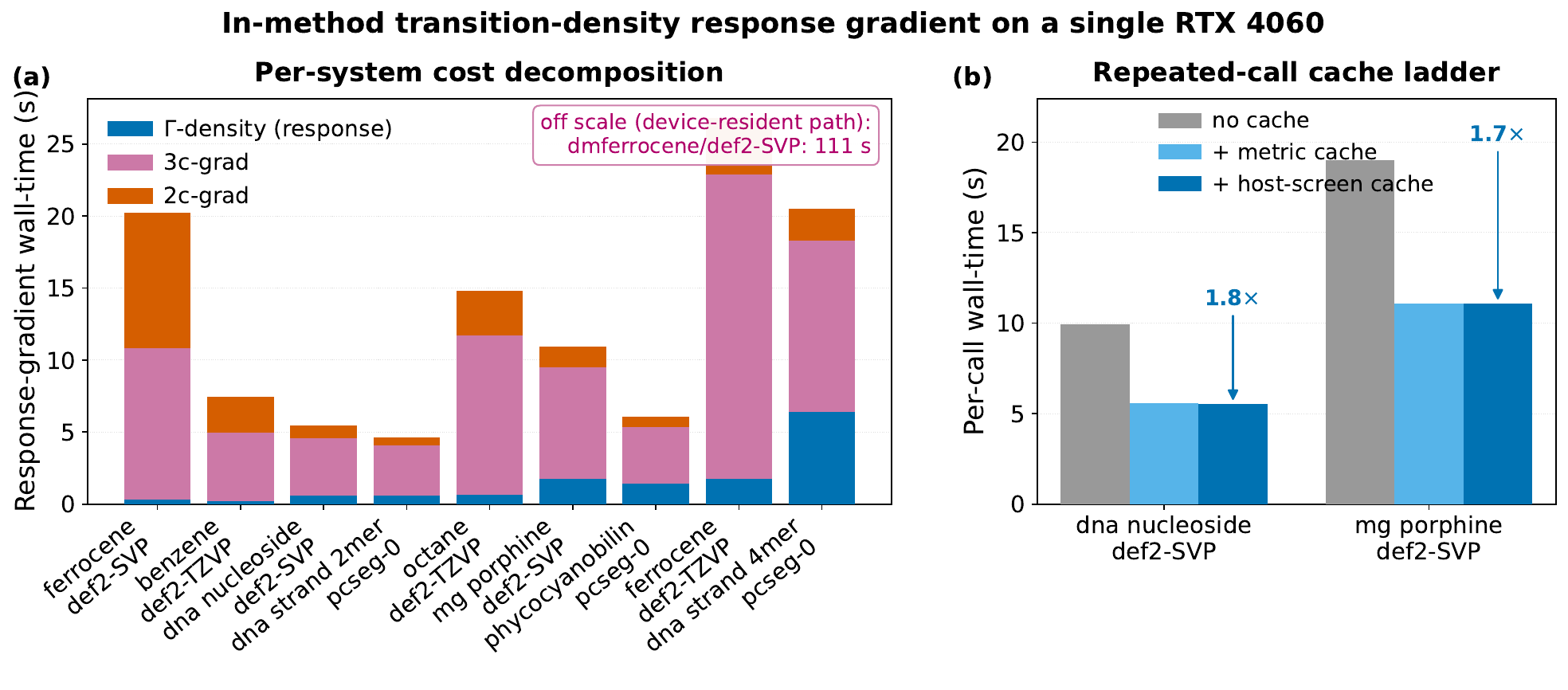}
\caption{In-method transition-density response gradient on the RTX~4060 (the
production FP32-mixed CD path, the \emph{same} 3c/2c kernels as the ground-state
gradient). All eleven response-gradient \emph{assemblies} are feasible on the
$8$~GB device; the two largest by AO count (chlorophyll\_a, $730$~AO;
dna\_strand\_4mer, $752$~AO) and the f-shell decamethylferrocene/def2-SVP
($486$~AO, the tightest case, not the largest) require the device-resident eviction
path. Feasibility tracks the CD footprint ($L_{\max}$, $n_{\mathrm{cd}}$,
$n_{\mathrm{sig}}$), not the AO count, and is distinct from the ground-state
CCSD-\emph{solve} $t_2$ ceiling (6 of 11 systems; SI~S8) that this figure excludes
by construction. (a)~Per-system cost: the response gradient is
three-center-dominated, and the non-symmetric transition-density $\Gamma$ build---the
only kernel specific to the ordered $J^x(A,B)\ne J^x(B,A)$ density---is a thin slice
($0.1$--$1.7$~s). Decamethylferrocene fits only via the eviction path; its fused
3c$+$2c launch runs at $\sim\!3.6\times$ the per-call cost of the comparably sized
ferrocene/def2-TZVP ($\sim\!110$~s, off-scale), and its degenerate per-term split
(ratio $0.757$) is the price of the sub-$1$~GB margin, not comparable term-by-term
to the standard path. (b)~Repeated-call cache ladder: reusing the geometry-frozen
Cholesky metric across the many transition densities of one geometry gives a
$1.7$--$1.8\times$ per-step speedup---the regime of an excited-state AIMD trajectory.
On the standard (non-evicting) path the response gradient matches the ground-state
gradient to within $1\%$ (ratio $0.99$--$1.00$), since one kernel serves both density
symmetries and the contraction cost is density-independent. This is a
cost/kernel-identity assembly timing (a surrogate transition density; resp/ground
$\approx1.00$ by construction), \emph{not} a chromophore-scale EOM-gradient
demonstration (the full end-to-end run is M3); it complements the small-scale
end-to-end device validation of Sect.~\ref{sec:deveom}.}
\label{fig:response}
\end{figure}

We characterize the DAG-generated J/K and gradient kernels on the RTX~4060 at the
scale the backend was engineered for---production Hartree--Fock and CD-RHF runs on
the chromophore systems of Table~\ref{tab:fd} up to $730$ Cartesian AO. Because
the same kernel object serves the SCF reference and the response skeletons
(Sect.~\ref{sec:compdetails}), this characterization is of the very kernels the
EOM-CC gradient invokes; the symmetric builds are the ground-state-density case
and the non-symmetric build is the transition-density case
($J^x(A,B)\ne J^x(B,A)$, Sect.~\ref{sec:jk}). The in-method response-gradient
build---the wall-time of the non-symmetric transition-density gradient, its
per-step (repeated-call) cost under nuclear dynamics, and its parity with the
ground-state gradient---is given in Fig.~\ref{fig:response}; because the
contraction cost is independent of the density \emph{values}, a surrogate
transition density reproduces the wall-time of a converged EOM-CCSD eigenvector.

\begin{figure}[htbp]\centering
\includegraphics[width=0.92\linewidth]{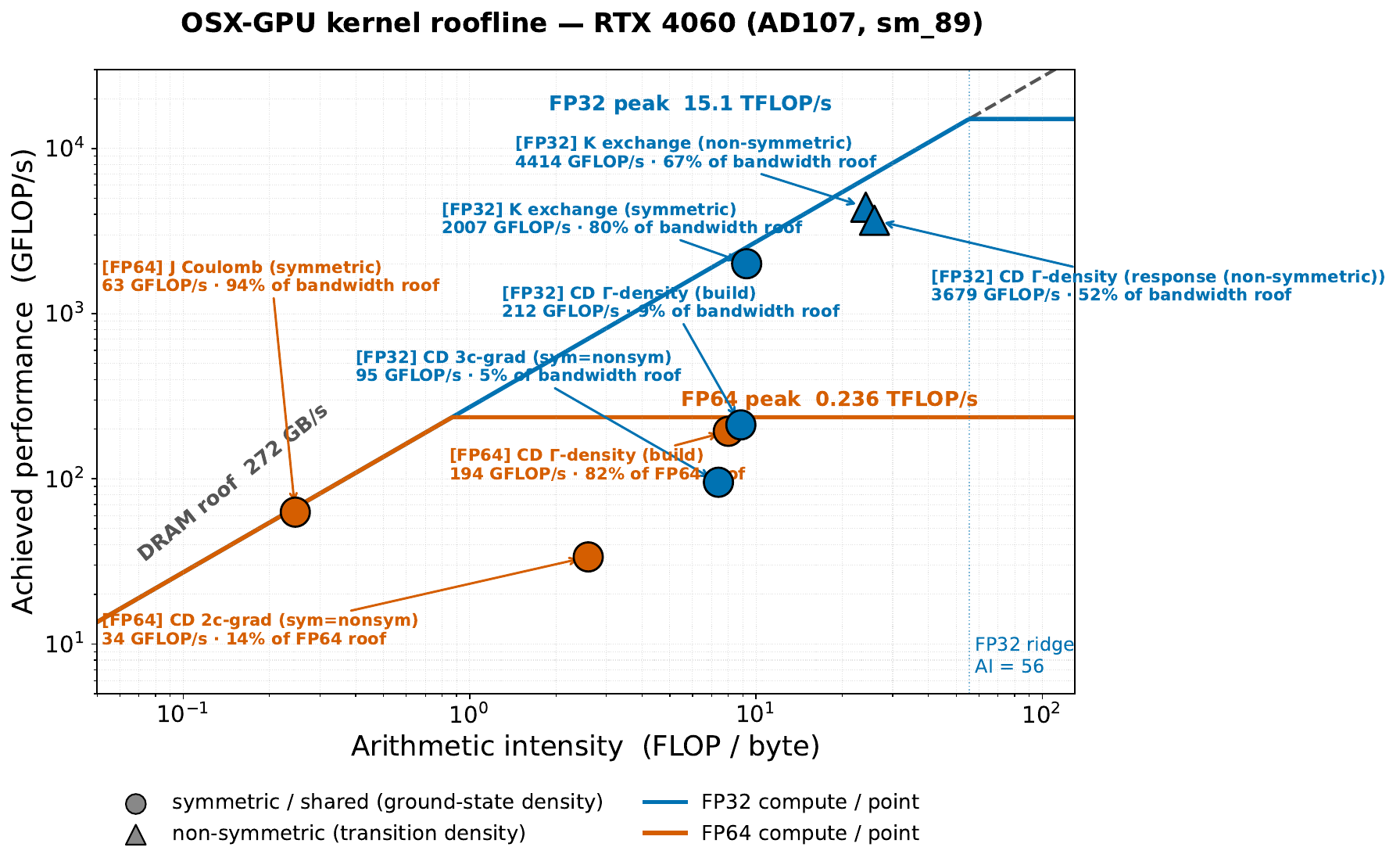}
\caption{Kernel roofline on the RTX~4060 (AD107, \texttt{sm\_89}): achieved
throughput vs.\ arithmetic intensity for the production J/K and gradient kernels,
against the FP32 ($15.1$~TFLOP/s) and FP64 ($0.236$~TFLOP/s) compute peaks and the
$272$~GB/s DRAM roof (FP32 ridge at AI~$=56$). Circles are the symmetric/shared
(ground-state-density) builds; triangles are the non-symmetric transition-density
builds---the GPU realization of the ordered $J^x(A,B)\ne J^x(B,A)$ build of
Sect.~\ref{sec:jk}. The Coulomb build [FP64]
($63$~GFLOP/s, $94\%$ of the bandwidth roof) is bandwidth-bound, which is why it
stays FP64; symmetric K [FP32] ($2007$~GFLOP/s, $80\%$) and the non-symmetric
transition-density K [FP32] ($4414$~GFLOP/s, $67\%$) are likewise bandwidth-bound;
the three-center gradient [FP32] ($95$~GFLOP/s, $5\%$) is latency/occupancy-bound,
and the two-center gradient [FP64] ($34$~GFLOP/s, $14\%$ of the FP64 roof) is the
only compute-leaning kernel. The transition-density $\Gamma$-density build [FP32]
($3679$~GFLOP/s, $52\%$ of the bandwidth roof at AI~$=26$)---the response
gradient's distinctive kernel---carries $\sim\!3\times$ the arithmetic intensity of
the symmetric $\Gamma$-density build (its two $K$-sandwiches vs.\ one), lifting SM
utilization from $15\%$ to $58\%$ and placing it well up the FP32 line; the metric
pseudo-inverse is excluded as it is cached once per nuclear geometry.}
\label{fig:roofline}
\end{figure}

\begin{figure}[htbp]\centering
\includegraphics[width=0.78\linewidth]{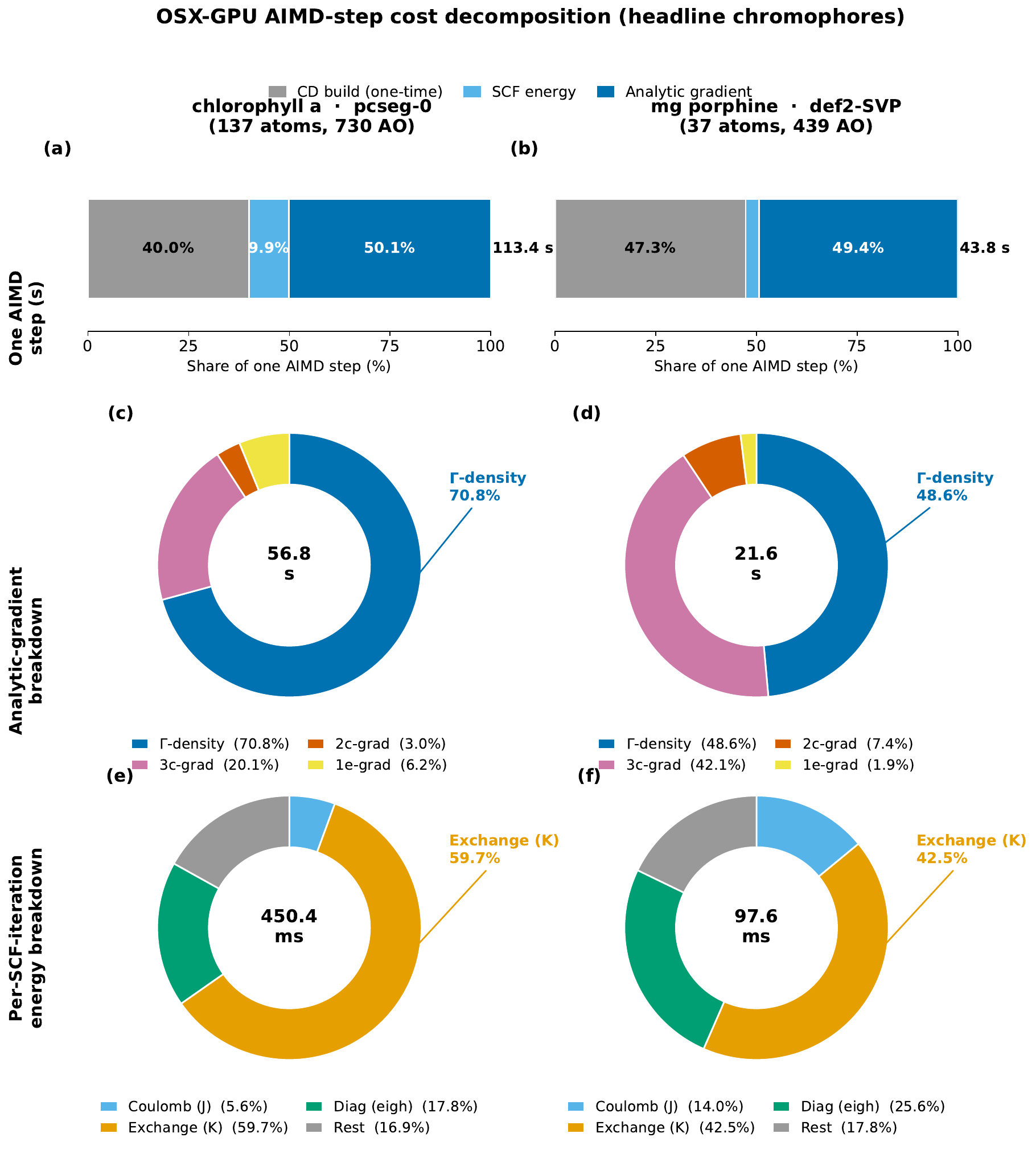}
\caption{Cost decomposition of one Born--Oppenheimer molecular-dynamics step for
the two headline chromophores, chlorophyll~a (pcseg-0, $137$ atoms, $730$~AO) and
Mg-porphine (def2-SVP, $37$ atoms, $439$~AO). (a,b)~the one-time Cholesky build
vs.\ the per-step analytic gradient (the gradient is $80.4\%$ / $73.9\%$ of the
step). (c,d)~the analytic gradient is dominated by the $\Gamma$-density
contraction ($91.8\%$ / $80.7\%$), the three- and two-center gradient builds and
the one-electron term making up the rest. (e,f)~the per-SCF-iteration energy build
is exchange-dominated (K $59.7\%$ / $42.5\%$). These are the same J/K and gradient
kernels the EOM-CC response invokes (Sect.~\ref{sec:compdetails}).}
\label{fig:cake}
\end{figure}

\begin{figure}[htbp]\centering
\includegraphics[width=\linewidth]{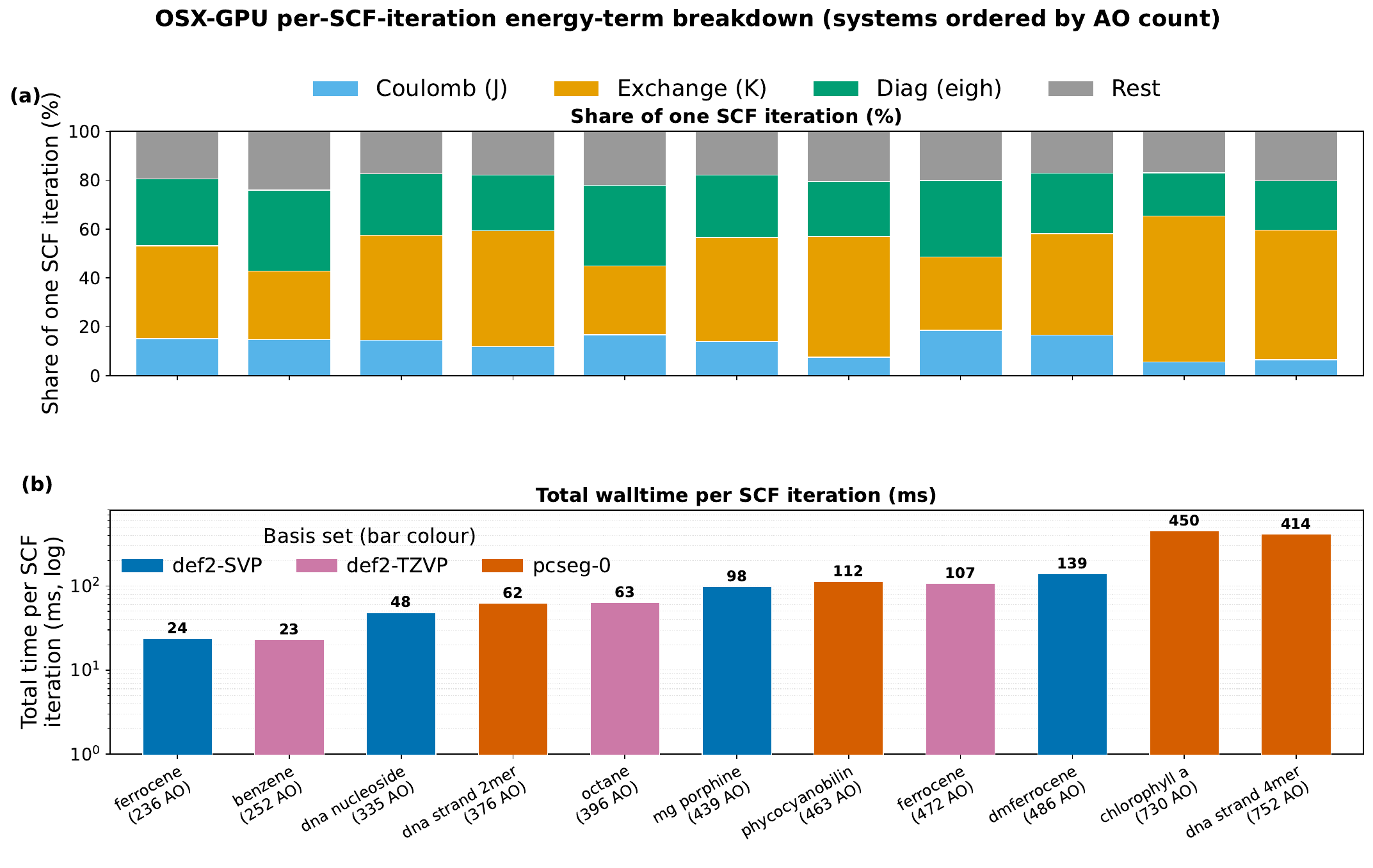}
\caption{Per-SCF-iteration energy-term breakdown across the eleven benchmark
systems, ordered by Cartesian AO count. (a)~relative shares of Coulomb (J),
exchange (K), diagonalization (eigh), and the remaining work; (b)~absolute
wall-time per SCF iteration (log scale; bar color denotes the basis set), from
$24$~ms (ferrocene, $236$~AO) to $450$~ms (chlorophyll~a, $730$~AO).}
\label{fig:stacked}
\end{figure}

\begin{figure}[htbp]\centering
\includegraphics[width=\linewidth]{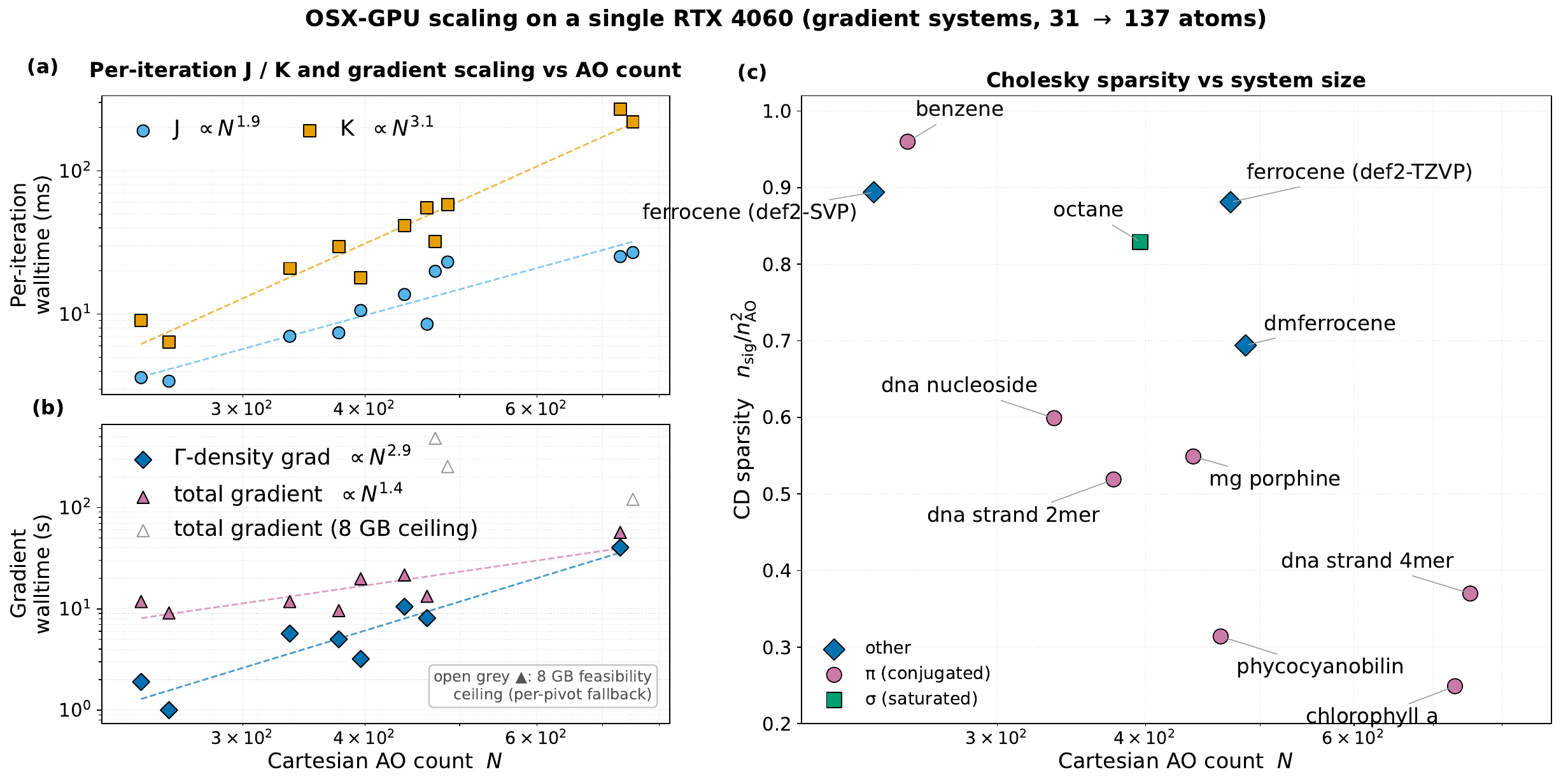}
\caption{Scaling on a single RTX~4060 over the gradient systems ($31$--$137$
atoms). (a)~per-iteration J and K wall-time vs.\ Cartesian AO count $N$, with
fitted exponents ($J\propto N^{1.9}$, $K\propto N^{3.1}$); (b)~gradient wall-time
for the eight systems using the optimized path, the device-resident FP32-mixed
$\Gamma$-density build $\propto N^{2.9}$ and the total amortized gradient
$\propto N^{1.4}$ (the warp-per-quartet 3c/2c kernels now apply at all $L$,
including the f-shell, so the total is no longer dominated by the recurrence
tower); the three largest systems at the $8$~GB feasibility ceiling (per-pivot
fallback) are shown separately as open markers; (c)~the Cholesky sparsity
$n_{\text{sig}}/n_{\text{ao}}^2$ vs.\
system size, contrasting the denser $\pi$-conjugated systems with the sparser
saturated ones.}
\label{fig:scaling}
\end{figure}

\begin{figure}[htbp]\centering
\includegraphics[width=\linewidth]{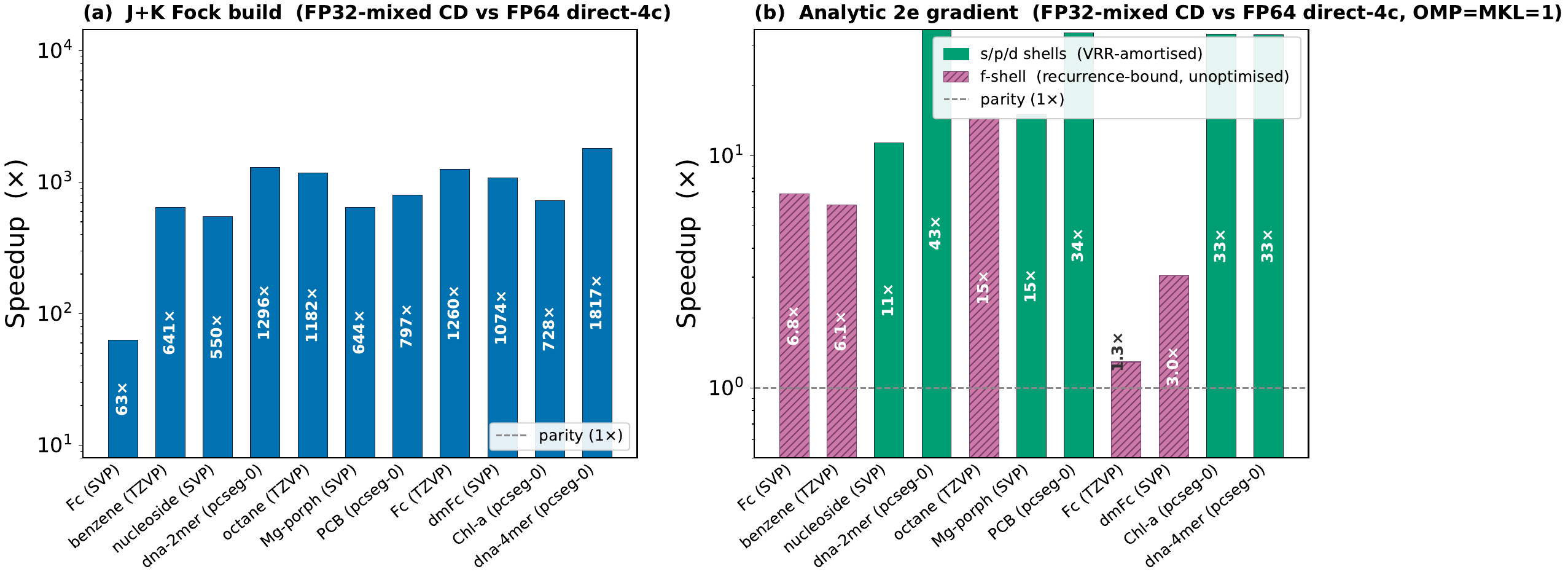}
\caption{Speedup of the GPU backend (OSX-GPU, CD, FP32-mixed) over single-thread
PySCF (FP64 direct ERIs, \texttt{OMP\_NUM\_THREADS}~$=$~\texttt{MKL\_NUM\_THREADS}~$=1$),
the same physical observable on both sides, across the eleven systems. This
single-thread, FP64, direct-ERI baseline confounds three effects---hardware,
density fitting (Cholesky), and FP32-mixed precision---so panel~(a) is a
reference-implementation comparison rather than a like-for-like kernel benchmark;
the defensible same-machine figure is the analytic gradient of panel~(b).
(a)~the per-SCF-iteration J+K Fock build, $63\times$--$1817\times$
(reference-implementation comparison); (b)~the
analytic two-electron gradient, $6\times$--$43\times$ across the eight systems
that fit the $8$~GB device with the optimized path (device-resident FP32-mixed
$\Gamma$-density build $+$ VRR-amortized warp-per-quartet 3c/2c kernels at all
$L$, including the f-shell). The three feasibility-ceiling systems (open markers; two
f-shell---ferrocene/def2-TZVP and decamethylferrocene/def2-SVP---plus one large
pcseg-0 system, dna\_strand\_4mer) exceed device memory with both optimizations
co-resident and retain the per-pivot gradient ($1.3\times$--$33\times$); this is a
fixed $8$~GB-memory limit, not an algorithmic one. Dashed line: parity.}
\label{fig:speedup}
\end{figure}

The roofline (Fig.~\ref{fig:roofline}) places the symmetric and non-symmetric J/K
kernels and the gradient builds against the device ceilings; the per-step and
per-iteration decompositions (Figs.~\ref{fig:cake}--\ref{fig:stacked}), the
scaling exponents (Fig.~\ref{fig:scaling}), and the speedups over single-thread
PySCF (Fig.~\ref{fig:speedup}) characterize the shared kernel across the
chromophore benchmark set. The headline same-machine figure is the analytic
two-electron gradient, $6\times$--$43\times$ (Fig.~\ref{fig:speedup}(b)); the larger
Fock-build ratios ($63\times$--$1817\times$, Fig.~\ref{fig:speedup}(a)) are taken
against a single-thread, FP64, direct-ERI reference and confound hardware, density
fitting, and precision, so we report them as a reference-implementation comparison,
not a headline. The in-method response gradient closes this loop at the
kernel level (Fig.~\ref{fig:response}): the non-symmetric transition-density
gradient \emph{assembly} (driven by a surrogate transition density, not a
chromophore-scale EOM solve, and complementing the small-scale end-to-end device
validation of Sect.~\ref{sec:deveom}), wall-timed on the 4060 across all eleven feasible
response-gradient assemblies, is
three-center-dominated with the transition-density $\Gamma$-density on its critical
path, matches the ground-state gradient to within $1\%$ ($0.99$--$1.00\times$, one
kernel for both density symmetries), and inherits the $6\times$--$43\times$ kernel
speedup of Fig.~\ref{fig:speedup}, with a further $1.7$--$1.8\times$ from reusing
the geometry-frozen Cholesky metric (and, marginally, the screening) across the per-MD-step transition
densities. The Laplace-grid convergence of the AO-LT kernel---gradient/energy
error versus the number of Laplace points $n_\tau$, exercising the Laplace path
that the dense reference checks of Sect.~\ref{sec:verif} do not---is reported in
SI Section~S4 (Fig.~S1). That study probes the shared minimax-Laplace grid through
the LT-MP2 energy as a proxy on the same grid, rather than as a direct
EOM-gradient-versus-$n_\tau$ sweep. The DAG-derived chunking's effect---shifting the
kernel from spill-bound to compute/bandwidth-bound---is evidenced by the
register-pressure and Nsight-Compute occupancy characterization above; a dedicated
chunked-versus-unchunked ablation table (\texttt{ptxas -v} spills, occupancy,
throughput) would isolate it causally and is left as a future refinement (SI
Section~S11).

%% =====================================================================
\section{Conclusions}
\label{sec:outlook}
We have realized the complete non-Hermitian EOM-CCSD excited-state gradient and
interstate NACME within the $8$\,GB of a consumer GPU: validated end to end at small
scale---including an independent cross-code (\textsc{Psi4}) check that reaches the
excited-state gradient itself and holds from \ce{H2O} to aromatic benzene---and
executed at the chlorophyll-core chromophore \ce{Mg}-porphine as a bounded capability
demonstration. The enabling mechanism is a single contraction DAG whose reverse-mode
transpose \emph{is} the EOM-CC relaxation, from which the symmetric, non-symmetric,
and gradient $J/K$ builds are all emitted.

By construction the DAG spine and LT-AO kernel are designed to apply across
MP2--MP6, CCSD--CCSDTQ~\cite{Aikens2003}, and the
CASSCF/CP-MCSCF/DMRG-SCF/CASPT2 families---each filling the same S1--S5 template,
differing only in which densities are built and how they are relaxed. This
generality is a structural property of the shared LT-AO kernel of
Sect.~\ref{sec:lt} and the DAG transpose of Sect.~\ref{sec:dag}; in the present
work it is the EOM-CC case that we develop and validate in full, and we do not
claim the other families as tested here. In the multireference regime in
particular, the cumulant collapse of Eq.~\eqref{eq:cumcollapse} leaves a large
active-only remainder that the single J/K kernel does not absorb, so the
CASSCF/CASPT2 reach is a structural property of the template rather than a
demonstrated capability. The reason this generality is structural
rather than coincidental is that the construction operates on the contraction
graph, which mirrors the Goldstone/MBPT diagrammatic expansion
(Sect.~\ref{sec:dag}): for a property expressible as a \emph{finite, truncated}
sum of diagrams and cast as a differentiable (Lagrangian) functional, the
reverse-mode transpose of the diagram-contraction graph \emph{is} the
corresponding relaxation/Lagrange-multiplier equation, generated rather than
re-derived. This is a design property of the framework, demonstrated here only
for non-Hermitian EOM-CC. It is \emph{not} a statement about infinite
resummations (e.g.\ RPA or ladder series) or frequency-dependent self-energies,
where the response follows from differentiating the underlying self-consistency
or linear-response solve rather than transposing a finite term list; extending
the present mechanism to Green's-function/self-energy formulations is therefore an
open question rather than a corollary.

The reach of the method is bounded by the assumptions of its two pillars. The
shifted Laplace quadrature of Eq.~\eqref{eq:eomshift} requires
$\omega<\min_k D_k$, restricting the excited states to the HOMO--LUMO-gap regime;
states above the first denominator pole fall outside the present grid. The
AO-direct formulation eliminates one power of the virtual-space dimension at every
truncation (the entire $vvvv$ block at the doubles level), which---together with
the spill-bounded chunking---is what makes correlated excited-state gradients and
NACMEs fit and run on commodity hardware.

A natural extension, squarely in the same machinery, is analytic
gradients/NACMEs for ab-initio cavity-QED (QED-EOM-CC) polaritonic states, where
the non-symmetric transition densities and the DAG-transpose response carry over
unchanged.

%% =====================================================================
\section*{Data and code availability}
The validation stack is publicly deposited so that the paper's accuracy claims can be
independently checked against the same references: the verification suites---the
determinant (Fock-space) oracle, the finite-difference gradient/NACME harness
(\texttt{test\_multiplicities.py}), and the full-CI gradient harness
(\texttt{test\_fci\_gradient.py})---together with the independent cross-code
comparison scripts (Section~S5), the molecular geometries of
Tables~\ref{tab:fd}--\ref{tab:fci}, the complete working-equation derivations, and
portions of the PTX code-generator output (the emitted GPU-kernel assembly for the
symmetric, non-symmetric, and gradient builds), are deposited on Zenodo (DOI assigned
on acceptance) and mirrored in a public GitHub repository. These host and
independent-code checks reproduce the small-scale accuracy references without the
proprietary kernels, and Section~\ref{sec:compdetails} describes the device
algorithm---the contraction schedule and its live-set chunking, the shared-memory
lift, the arena allocator, and the fused particle--particle ladder---in sufficient
detail to reimplement the kernels from the paper. The performance-tuned production
CUDA $J/K$ kernels---and hence the timing, roofline, and chromophore-scale execution
data---are proprietary to the funder and available from the corresponding author upon
reasonable request.

\begin{suppinfo}
The Supporting Information is available free of charge at
\url{http://pubs.acs.org}. It contains: the computational environment and
hardware/software for the CPU verification stack and the GPU backend
(Section~S1); full Cartesian molecular geometries for all finite-difference and
full-CI systems (S2); the coupled-cluster, EOM, finite-difference, and full-CI
numerical protocols and tolerances (S3); the Laplace-grid ($n_\tau$) convergence
study (S4, Figure~S1); expanded verification tables (S5); the determinant-oracle
and \texttt{p}$^{\dagger}$\texttt{q} amplitude-generation details (S6); the
complete working-equation derivations (S7); GPU-backend reproducibility, timing,
the device-resident CCSD-solve campaign, and roofline data (S8); an algorithmic
characterization of the non-Hermitian EOM Davidson eigensolver on synthetic small
matrices (S9); the FNO virtual compression for the production-basis eigensolve
(S10); the status of in-method deliverables (S11); and the data- and
code-availability statement (S12).
\end{suppinfo}

\begin{acknowledgement}
We acknowledge financial support and computational resources provided by
NeuroTechNet S.A.S. The validation harnesses, cross-code comparison scripts, and
molecular geometries supporting the findings of this study are publicly deposited
(Zenodo, DOI on acceptance; public GitHub mirror); the production CUDA kernels are
available from the corresponding author upon reasonable request.
\end{acknowledgement}

\noindent The author declares no competing financial interest.

%% =====================================================================
\bibliography{refs}

\end{document}